\documentclass[11pt,a4paper]{article}
\usepackage{amssymb}
\usepackage{graphicx}
\usepackage{amsmath}
\usepackage{makeidx}
\usepackage{indentfirst}
\usepackage[T1]{fontenc}
\usepackage[utf8]{inputenc}
\usepackage{authblk}

\newcounter{resultnum}[section]
\newcounter{conclusionnum}[section]
\newcounter{conditionnum}[section]
\newcounter{conjecturenum}[section]
\newcounter{examplenum}[section]
\newcounter{exercisenum}[section]
\newcounter{lemmanum}[section]
\newcounter{notationnum}[section]
\newcounter{theoremnum}[section]
\newcounter{definitionnum}[section]
\newcounter{corollarynum}[section]
\newcounter{remarknum}[section]
\newcounter{propositionnum}[section]
\newcounter{acknowledgementnum}[section]
\newcounter{algorithmnum}[section]
\newcounter{axiomnum}[section]
\newcounter{casenum}[section]
\newcounter{claimnum}[section]
\newcounter{summarynum}[section]
\newcounter{problemnum}[section]
\begin{document}

\title{\textbf{Einstein Spaces  Modeling Nonminimal Modified Gravity}}
\date{The variant accepted to EPJP on May 16, 2015}

\author[1]{Emilio Elizalde\thanks{elizalde@ieec.uab.es}}%

\author[2]{Sergiu I. Vacaru\thanks{sergiu.vacaru@uaic.ro; sergiu.vacaru@gmail.com}}%

\affil[1]{\small Instituto de Ciencias del Espacio, Consejo Superior de Investigaciones Cient\'{\i}ficas,
\newline ICE-CSIC and IEEC,  Campus UAB,  Torre C5-Parell-2a planta,
\newline 08193 Bellaterra (Barcelona) Spain%
\newline {\qquad }
}

\affil[2]{\small Theory Division, CERN, CH-1211, Geneva 23, Switzerland \footnote{associated visiting research affiliation};\
\newline
Max-Planck-Institute for Physics, Foehringer Ring 6,  Muenchen,  Germany D-80805\footnote{DAAD fellowship  affiliation};
\newline
Inst.  Theor. Phys., Lebiniz Univ. Hannover, Appelstrasse 2, Hannover, Germany, D-30167\footnote{DAAD fellowship  affiliation}; and
\newline
 Rector's Department, Alexandru Ioan Cuza University,\
 Alexandru Lapu\c sneanu street, \newline  nr. 14, UAIC - Corpus R, office 323;\
 Ia\c si,\ Romania, 700057
}%

\renewcommand\Authands{ and }

\maketitle

\begin{abstract}
Off-diagonal vacuum and nonvacuum configurations in Einstein gravity can
mimic physical effects of modified gravitational theories of $f(R,T,R_{\mu
\nu}T^{\mu \nu})$ type. To prove this statement, exact and approximate
solutions are constructed in the review, which encode certain models of
covariant Ho\v rava type gravity with dynamical Lorentz symmetry breaking.
The corresponding FLRW cosmological dynamics with possible nonholonomic
deformations and the reconstruction procedure of certain actions closely
related with the standard $\Lambda$CDM universe are studied. Off-diagonal
generalizations of de Sitter universes are constructed which are generated
through nonlinear gravitational polarization of fundamental physical
constants and which model interactions with non-constant exotic fluids and
effective matter. The problem of possible matter instability for such
off-diagonal deformations in (modified) gravity theories is briefly
discussed.

\vskip0.1cm


\vskip3pt

\end{abstract}




\section{Introduction}

There are several motivations for the study of modified gravity theories. At
short scales it seems clear that General Relativity (GR) needs to be
modified in order to take into consideration quantum effects, which leads to
second- and higher-order terms in the curvature $R$, and these same
modifications might be also very useful to solve problems at large scales,
as the acceleration of the universe expansion and other, as the nature
itself of dark matter (DM) and dark energy (DE). Not the least, it is an
attempt to formulate a self-consistent (in some particular way, at least)
theory of quantum gravity. Moreover, an increasing amount of more and more
accurate and constraining observational data will help to discriminate among
the different, alternative modifications of the gravity theory in the search
for a better description of our universe.

Among the different classes of extensions of general relativity some of the
most popular are $f(R), f(R,T)$, and $f(\mathbf{R},T,F)$---which we will
here generically call $f(R,...)$-modified theories, being $R$ the Ricci
scalar and $T$ the metric torsion. In these approaches, the standard
Lagrangian for GR, namely as $\mathcal{L}=R$, on a pseudo-Riemannian
manifold, $V$---where $R$ is the Ricci scalar curvature for the Levi-Civita
connection, $\nabla $---is modified by the addition of a functional, $%
f(R,...)$, of the Ricci scalar only, in the first case, of $R$ and the
torsion tensor, $T_{\beta \gamma }^{\alpha }$, the energy-momentum tensor
for matter, $T_{\beta \gamma },$ and/or its trace $T=T_{\alpha }^{\alpha }$
(in the second), and of a generalized Ricci scalar $\mathbf{R}$, and a
Finsler generating function, $F$, in the third case (such values may be
defined on the tangent bundle $TV$), etc. Classes of modified theories of
these kinds can be successfully constructed, and also the corresponding
reconstruction procedures, able to mimic the ${\Lambda}$CDM model including
the dark energy epochs and the transitions between the different main stages
of the universe evolution, thus providing a unified description of the
entire cosmological history. For reviews of some of the most important
results along this line, see Refs.~\cite{revfmod,revfmod01,revfmod02,
revfmod03,revfmod04,revfmod05,revfmod06,revfmod07,revfmod08,
revfmod09,revfmod10,revfmod11,revfmod12,revfmod13}.

Several of these model constructions of modified theories are actually
related with different forms of the so-called covariant Ho\v{r}ava gravity
associated with a dynamical breaking of Lorentz invariance \cite{covhl,covhl1,covhl2,covhl3}, and
with further developments, as well, including in particular generic
off-diagonal solutions, Lagrange-Hamilton-Finsler like generalizations,
deformation quantization, A-brane models, and gauge like gravity \cite{vhl,vhl1,vhl2,vhl3}.
In some simplified approaches, theories of this kind can be constructed in a
power-counting renormalizable form or as nonholonomic brane configurations
which correspond to power-law versions of actions of type $f(R,T,R_{\mu \nu
}T^{\mu \nu })$ \cite{odgom,odgom1,odgom2}. In general, the spacetime geometries can be of
Finsler type, with commutative and/or noncommutative parameters, and
off-diagonal metrics for wapred/trapped solutions \cite{voffds,voffds1,voffds2,voffds3,voffds4}). This also
includes effects as Lorentz violations, nonlinear dispersion relations and
locally anisotropic re-scaling, and effective polarizations of constants,
which provide a deeper understanding of the possible connections between
these $f(R,...)$ modified theories, and Ho\v{r}ava-Lifshitz and Finsler
theories (see also Refs.~\cite{finslmavr,finslmavr1,finslmavr2,finslmavr2a,finslmavr3,finslmavr4,finslmavr5,finslmavr6}, in this respect).

Field equations for gravitational and matter field interactions in GR and
various modified theories of the types described above usually consist of
very sophisticated systems of nonlinear partial differential equations
(PDEs). No wonder they request advanced numeric, analytic and geometric
techniques for constructing exact and approximate solutions. The most
important physical solutions, for black hole configurations, observable
cosmological scenarios, etc., have been therefore constructed with the
simplifying diagonalizable ansatz for the corresponding metric (obtained by
appropriate coordinate transformations and frame rotations) with Killing
symmetries. After a series of assumptions of ``high symmetry'' of the
relevant interactions (for spherical, cylindrical or torus ans\"{a}tze, with
a possible additional Lie group interior symmetry), the systems of resulting
nonlinear PDEs are usually transformed into much more simplified systems of
nonlinear ordinary differential equations (ODEs), what is a great advantage,
indeed. In such cases, some classes of exact solutions can be obtained in
explicit form (see the monographs \cite{monexsol,monexsol1} for reviews of some
results in GR). Even then, it actually took more than half a Century to
understand the fundamental physical implications of these solutions; for
instance, of the Schwarzschild and Kerr black hole metrics, and to finally
elaborate the Friedmann-Lema\^{\i}tre-Robertson-Walker (FLRW) cosmological
scenario. At present, it is already possible to construct more general
classes of exact solutions in (modified) gravity theories depending
generically on two or three variables in a ``less symmetric'' form for
four-dimensional (4-d), and extra-dimensional models using advanced
geometric, analytic and numerical methods. We can provide a plausible
physical interpretation when such solutions are defined by certain symmetry
transforms or small deformations of some well-known classes of solutions.
However, to derive a new class of exact solutions of a gravity theory is by
itself not of main interest for physicists, neither to applied
mathematicians, unless such construction does result in new and interesting
classical or quantum-physical effects, or does provide convincing
explanations to cosmological data coming from the newest surveys.

In a series of works \cite{voffds,voffds1,voffds2,voffds3,voffds4}, the so-called anholonomic frame
deformation method (AFDM) for the construction of exact solutions in gravity
has been developed. It provides a general geometric technique which allows
to integrate PDEs for gravitational and matter fields interactions, for
generic off-diagonal metrics with generalized connections, or for the
torsionless Levi-Civita one. Such solutions may depend on all spacetime
coordinates via various classes of generating and integration functions,
commutative and non-commutative parameters, and so on. In particular, the
corresponding metrics and connections can exhibit anisotropic ellipsoidal or
toroidal symmetries, or encode certain generalized solitonic hierarchies,
coming from an effective non-holonomic (with non-integrable constraints)
dynamics with nontrivial topological configurations. It is furthermore
possible to analyze the physical implications of geometric constructions of
this kind, provided they describe certain ``small parameter'' deformations
related to well defined black hole objects in cosmological models, or
particle physics interactions with broken symmetries.

In brief, the AFDM is based on a quite surprising decoupling property of the
vacuum and on certain classes of non-vacuum fundamental field equations in
GR and modified gravities. The main idea is to work with an ``auxiliary''
connection when physically important systems of nonlinear PDE decouple in
certain classes of nonholonomic frames. This allows to integrate systems of
this kind of very general form. Usually, the auxiliary connection which is
needed involves nontrivial nonholonomically induced torsion, which in GR is
determined by certain generic off-diagonal terms of the metric and
corresponding classes of nonholonomic (equivalently, anholonomic, i.e.
non-integrable) constraints on gravitational and matter field dynamics. A
nonholonomically induced torsion is different from that in the
Einstein-Cartan or string gravity theory, where torsion fields are subject
to additional algebraic or dynamical field equations. Having constructed
certain general integral varieties of solutions, we can almost generically
consider certain classes of constraints where the auxiliary connection
transforms into the Levi-Civita one. Here we note that it is important to
impose such zero-torsion constraints after certain classes of generalized
solutions are found in general form, but not before applying the AFDM.
Generic off-diagonal solutions for nonlinear systems can be restricted to
torsionless configurations, provided certain nontrivial solutions have been
already found. If non-holonomic constraints are imposed from the very
beginning (for instance, spherical symmetries and a simplified diagonal
ansatz for the metric), then one excludes from the analysis more general
classes of nonlinear interactions.

The crucial importance of generic off-diagonal solutions in GR and modified
theories is determined by a series of geometric and analytic properties of
the associated systems of nonlinear PDEs which are used for elaborating the
cosmological models and for performing the quantization of the associated
gravity theories. In this respect we should emphasize three key issues: 1) A
number of physical effects and observed cosmological data can already be
explained in GR or in the context of modified gravity theories (for
instance, within $f(R)$ gravity or with nontrivial massive terms, see \cite%
{massgr,massgr1}) or can alternatively be modelled with the help of off-diagonal
interactions in different types of modified theories. 2) Generic
off-diagonal solutions encode configurations with a nontrivial parametric
vacuum and effective matter field interactions, also with gravitational
polarizations of the interaction and a cosmological constant, and for
nontrivial generating and integration functions. For certain well-defined
conditions, such models describe broken fundamental symmetries---as, for
instance, locally anisotropic interactions, terms deploying violation of
local Lorentz symmetry, and warped and trapping configurations---which
points towards new methods of quantization and results in alternative
scenarios of accelerating and anisotropic cosmological theories, with
different solutions for the dark energy and dark matter physical problems.
3) Considering off-diagonal configurations one can model classical and
quantum $f(R)$ modified gravities and the like, and also Ho\v{r}ava-Lifshitz
and Finsler like theories in a suggestive, unified geometric way \cite%
{vhl,vhl1,vhl2,vhl3,voffds,voffds1,voffds2,voffds3,voffds4}.

The aim of this paper is to apply the anholonomic frame deformation method
for the constructing of exact off-diagonal solutions corresponding to
cosmological models of modified gravity of the general form $f(R,T,R_{\mu
\nu }T^{\mu \nu })$, and to study the conditions under which such
configurations can be alternatively modeled as effective Einstein spaces
with nontrivial off-diagonal parametric vacuum and non-vacuum
configurations. The FLRW cosmological dynamics and a reconstruction
procedure of the ${\Lambda}$CDM universe will be investigated. To be noted
is that we will not work with exotic anisotropic fluid configurations as in
\cite{covhl,covhl1,covhl2,covhl3,odgom,odgom1,odgom2}, but rather with off-diagonal deformations of de Sitter
solutions, as in \cite{vhl,vhl1,vhl2,vhl3}. The problem of matter instabilities in modified
and deformed GR theories will be analyzed and solutions will be obtained for
certain classes of non-holonomic configurations.

\section{Off-diagonal interactions in modified gravity and cosmology}

\label{s2}

In this section, we formulate a geometric approach to $f(R,T,R_{\mu \nu
}T^{\mu \nu })$\ gravity and summarize the anholonomic frame deformation
method \cite{voffds,voffds1,voffds2,voffds3,voffds4}. The geometric constructions will be adapted to
nonholonomic distributions with associated nonlinear connection
(N-connection) structure.\footnote{%
It should be noted that in generalized Finsler like theories, the
N-connection structure is given by a set of three fundamental geometric
objects which, for certain models, is completely defined by the so-called
Lagrange/Finsler generating function. We do not study in this work Finsler
like modifications of GR.} The N-connections formalism will be used for
constructing certain classes of N-adapted frames with respect to which the
gravitational and matter field equations decouple in very general forms (see
Sect.~\ref{safdm} below). This is possible for metric compatible linear
connections with non-holonomically induced torsions completely defined by
metric tensors. Imposing additional constraints, we can generalize zero
torsion configurations for the Levi-Civita connection, $\nabla$.

\subsection{Modeling dark energy with off-diagonal metrics}

In order to motivate our approach, we discuss simple FLRW cosmology and dark
energy and dark matter models which are extended for generic off-diagonal
solutions in GR and modifications. Working in a spatially flat spacetime
with diagonal quadratic form%
\begin{equation}
ds^{2}=\mathring{g}_{\alpha }(t)(du^{\alpha })^{2}=\mathring{a}%
^{2}(t)[(dx^{1})^{2}+(dx^{2})^{2}+(dy^{3})^{2}]-dt^{2},  \label{flrwel}
\end{equation}%
for local coordinates $u^{\alpha }=(x^{i},y^{3},y^{4}=t),$ when $i=1,2,$ the
FLRW equations are
\begin{equation*}
\frac{3}{\kappa ^{2}}\mathring{H}^{2}=\mathring{\rho}\mbox{ and }\mathring{%
\rho}^{\diamond }+3\mathring{H}(\mathring{\rho}+\mathring{p})=0,
\end{equation*}%
where $\mathring{\rho}$ and $\mathring{p}$ are, respectively, the total
energy and pressure of a perfect fluid (pressureless or just radiation), $%
\mathring{H}:=\mathring{a}^{\diamond }/\mathring{a}$ for $\mathring{a}%
^{\diamond }:=\partial \mathring{a}/\partial t=\partial _{4}\mathring{a}%
=\partial _{t}\mathring{a}$, and $\kappa ^{2}$ is related to the
gravitational (Newton) constant.\footnote{%
We use a system of notations different from that in standard cosmology, as
this will be convenient for constructing cosmological models with generic
off-diagonal metrics, and also in order to follow the conventions in our
previous works.} To explain the observational data of an accelerating
universe, and dark energy and matter, various models have been studied (see
reviews and references in \cite{revfmod,revfmod01,revfmod02,revfmod03,
revfmod04,revfmod05,revfmod06,revfmod07,revfmod08,revfmod09,revfmod10,
revfmod11,revfmod12,revfmod13}), with effective or exotic matter
with an equation of state (EoS) of phantom kind, $p=\varpi \rho ,$ with $\
\varpi <-1.$ The simplest model of phantom DE is given by
\begin{equation*}
\frac{3}{\kappa ^{2}}H_{DE}^{2}=\rho _{DE}\mbox{ and }\rho _{DE}^{\diamond
}+3H_{DE}(1+\varpi )\rho _{DE}=0,
\end{equation*}%
which for $\varpi <-1$ admits an exact solution
\begin{equation}
H_{DE}=\frac{2}{3(1+\varpi )(t_{s}-t)}.  \label{brip}
\end{equation}%
This solution has a finite-time future singularity (Big Rip) at $t=t_{s}.$

Some models have been considered where the Hubble function $H(t)$ is
determined by a phantom DE coupled with DM, via a coupling constant, $Q,$
which results in the conservation law%
\begin{equation*}
\rho _{DE}^{\diamond }+3H(1+\varpi )\rho _{DE}=-Q\rho _{DE},\ \rho
_{DM}^{\diamond }+3H\rho _{DM}=Q\rho _{DM}.
\end{equation*}%
The solutions of these equations can be expressed as
\begin{equation*}
\rho _{DE}=\ ^{0}\rho _{DE}\ e^{-3(1+\varpi )}e^{-Qt} \mbox{ \ and } \rho
_{DM}a^{3}=Q\ ^{0}\rho _{DE}\int\nolimits^{t}dt^{\prime }e^{-3\varpi
}e^{-Qt},
\end{equation*}
respectively, where $\ ^{0}\rho _{DE}$ is an integration constant and the
EoS is taken to be $p=\varpi \rho _{DE}.$ These functions can be used for
the second FLRW equation,
\begin{equation*}
-\frac{1}{\kappa ^{2}}(2H^{\diamond }+3H^{2})=p.
\end{equation*}%
We have
\begin{equation}
H=-Q/3(1+\varpi )  \label{dssol}
\end{equation}%
for the exact solution of this equation, which corresponds to the evolution
for de Sitter space, $a(t)=a_{0}e^{-Qt/3(1+\varpi )},$ where $a_{0}$ is
determined from $a_{0}^{3(1+\varpi )}=-\frac{3\kappa ^{2}}{Q^{2}}(1+\varpi
)^{2}\varpi $ $\ ^{0}\rho _{DE}.$ The value of $H$ in (\ref{dssol}) is
positive for $\varpi <-1$, what does not mean that the Big Rip singularity
in (\ref{brip}) can be avoided, but just shows that the coupling of the
phantom DE and DM gives a possibility that the universe could evolve as a de
Sitter phase. More than that, the first FLRW equation,
\begin{equation}
\frac{3}{\kappa ^{2}}H^{2}=\rho _{DE}+\rho _{DM},  \label{demeq}
\end{equation}%
imposes the relation $\rho _{DM}=(1+\varpi )\rho _{DE}.$ Considering a de
Sitter solution as an attractor, with $\varpi \sim -4/3,$ we obtain $%
-(1+\varpi )\sim 1/3$, which is almost independent from the initial
condition, i.e., it solves for free the so-called coincidence problem.%
\footnote{%
If DE does not couple with DM, we have $\rho _{DM}\sim a^{-3}$ and $\rho
_{DE}\sim a^{-3(1+\varpi )}$, which do not satisfy the observed $1/3$ ratio
of DE and DM and does result in a coincidence problem.}

Since the DE-DM coupling does not always remove the singularity and there is
no such fluid with constant EoS parameter, models were considered which are
proportional to a power of the scalar curvature, for instance, $%
p_{fluid}\propto R^{1+\epsilon },$ for $\epsilon >0.$ In that case the total
EoS parameter is greater than $-1$ and a Big Rip does not occur for large
curvature. Two variants of theories have been exploited where this kind of
inhomogeneous effective fluid matter is realized, by a conformal anomaly and
other quantum effects or by some modified model of gravity, for instance,
when the gravitational Lagrange density $R\rightarrow f(R)=R+R^{\varkappa }.$
In the case $1<\varkappa <2$, we have that solutions with
\begin{equation*}
Ht\sim -\frac{(\varkappa -1)(2\varkappa -1)}{\varkappa -2}\mbox{ and }%
w_{eff}\sim -1-2H^{\diamond }/3H^{2}>-1
\end{equation*}%
do not result in a Big Rip or any other kind of future singularity. Similar
classical and quantum arguments were considered as motivations to study $%
f(R) $ modified gravity theories \cite{revfmod,revfmod01,revfmod02,revfmod03,revfmod04,
revfmod04,revfmod06,revfmod07,revfmod08,revfmod09,revfmod10,revfmod11,revfmod12,
revfmod13,covhl,covhl1,covhl2,covhl3,odgom,odgom1,odgom2,massgr,massgr1}.

In a series of works \cite{voffds,voffds1,voffds2,voffds3,voffds4,vhl,vhl1,vhl2,vhl3,voffdmgt,voffdmgt1,voffdmgt2,voffdmgt3}, various classes of
off-diagonal solutions were studied which can be constructed by geometric
methods in modified gravity theories. We proved that, for instance, certain
important effects and cosmological models related to $f(R)$ modified
theories and the like can alternatively be explained by nonlinear
off-diagonal gravitational and matter field interactions with respect to
nonholonomic frames. Let us briefly recall the main ideas supporting such an
approach. Off-diagonal ans\"{a}tze for metrics (see, for instance, (\ref%
{qelgenofd})),{\small
\begin{equation}
g_{\underline{\alpha }\underline{\beta }}=\left[
\begin{array}{cccc}
g_{1}+\omega ^{2}(w_{1}^{\ 2}h_{3}+n_{1}^{\ 2}h_{4}) & \omega
^{2}(w_{1}w_{2}h_{3}+n_{1}n_{2}h_{4}) & \omega ^{2}w_{1}h_{3} & \omega
^{2}n_{1}h_{4} \\
\omega ^{2}(w_{1}w_{2}h_{3}+n_{1}n_{2}h_{4}) & g_{2}+\omega ^{2}(w_{2}^{\
2}h_{3}+n_{2}^{\ 2}h_{4}) & \omega ^{2}w_{2}h_{3} & \omega ^{2}n_{2}h_{4} \\
\omega ^{2}w_{1}h_{3} & \omega ^{2}w_{2}h_{3} & \omega ^{2}h_{3} & 0 \\
\omega ^{2}n_{1}h_{4} & \omega ^{2}n_{2}h_{4} & 0 & \omega ^{2}h_{4}%
\end{array}%
\right] ,  \label{ans1a}
\end{equation}%
}where the coefficients are parameterized by functions of type $%
g_{1}=g_{2}\sim e^{\psi (x^{i})}$ and $n_{i}(x^{k})$ (we can fix certain
constants for corresponding classes of generating, $\Phi (x^{k},t),$ and
integration functions), $h_{a}[\Phi (x^{k},t)]\sim h_{a}(t),$ [for $%
a=3,4],w_{i}[\Phi (x^{k},t)]\sim w_{i}(t)$ and $\omega (x^{k},t)\sim \omega
(t),$ were found to generate exact (in general, nonhomogeneous) cosmological
solutions in modified gravity theories. Such generic off-diagonal metrics%
\footnote{%
Which cannot be diagonalized by coordinate transformations.} can be
represented in the form
\begin{equation}
ds^{2}=a^{2}(t)[(e^{1})^{2}+(e^{2})^{2}]+a^{2}(t)\widehat{h}_{3}(t)(\widehat{%
\mathbf{e}}^{3})^{2}+(\widehat{\mathbf{e}}^{4})^{2},  \label{ans1b}
\end{equation}
with respect to so-called N-adapted frames (Eqs.~(\ref{nader}) and (\ref%
{nadif}) are used, in general)
\begin{equation*}
\widehat{\mathbf{e}}^{3}=dy^{3}+n_{i}dx^{i},\widehat{\mathbf{e}}%
^{4}=dt+w_{i}(t)dx^{i}.
\end{equation*}

For certain well-defined conditions (see Sect.~\ref{safdm}), we can consider
off-diagonal deformations $\mathring{g}_{\alpha }(t)\rightarrow g_{%
\underline{\alpha }\underline{\beta }}(x^{k},t)$ $\sim g_{\underline{\alpha }%
\underline{\beta }}(t)$ defining new classes of cosmological models which
mimic contributions from $f(R)$ modified gravity encoded into the data for $%
\omega (t),w_{i}(t)$, etc. The corresponding formulas are nonlinear
functionals relating generating functions to the (effective) matter sources.
Such off-diagonal configurations are equivalently modeled as solutions of
some effective field equations $\mathbf{\check{R}}_{\ \beta }^{\alpha }=%
\check{\Lambda}\delta _{\ \beta }^{\alpha }.$ This way, various classes of
cosmological solutions of modified gravities can be alternatively modeled by
metrics of type (\ref{ans1b}), when the scaling factor $a(t)$ is nonlinearly
determined by the coefficients $w_{i}(t)$ and $h_{a}(t)$ via a generating
function $\Phi (t)$ and an effective source $\Upsilon (t).$ We can model ${%
\Lambda}$CDM cosmology and analogously DE and DM effects with $\rho
_{DE}+\rho _{DM}$ encoded into $\Phi (t)$ and $\Upsilon (t),$ but with
respect to the adapted frames $\widehat{\mathbf{e}}^{a}(t).$ Solutions with
off-diagonal metrics may be interpreted in accordance with observational
data if $a(t)$ is chosen to determine, for instance, an effective $H(t)$\ (%
\ref{dssol}) with cosmological evolution from a spacetime background
encoding $f(R)$-modifications. To prove such results in a rigorous
mathematical way we need to apply advanced methods from the geometry of
nonholonomic manifolds. For our purposes, such manifolds can be considered
as usual pseudo-Riemannian spacetimes, endowed with additional
non-integrable distributions and frame structures.

Both classes of metrics (\ref{flrwel}) and (\ref{ans1b}) can be
characterized, respectively, by scaling factors $\mathring{a}(t)$ and $a(t).$
Let us suppose that we have found a cosmological solution of type (\ref%
{ans1b}) in a given theory of modified gravity and analyze how this metric
can be formally diagonalized for deformations of a small real parameter $%
\varepsilon $ (when $0\leq \varepsilon $ $\ll 1).$ We can consider
``homogeneous'' approximations of type $\widehat{h}_{3}(t)\approx
1+\varepsilon \widehat{\chi }_{3}(t),w_{i}(t)\sim \varepsilon \check{w}%
_{i}(t)$ and $n_{i}\sim \varepsilon \check{n}_{i}.$\footnote{%
On inhomogeneity effects in cosmology, see \cite{ellisgfr}. In a more
general context, it is possible \ to consider also ``small'' local
anisotropic deformations depending on space like coordinates when $\widehat{%
\chi }_{3}(x^{k},t),w_{i}(t)\sim \varepsilon \check{w}_{i}(x^{k},t)$ and $%
n_{i}\sim \varepsilon \check{n}_{i}(x^{k}).$ Some amount of anisotropy is
compatible with observational data in various gravity and cosmological
theories. See \cite{goedel,krasinski}, for reviews of various approaches
related to GR and generalizations of Bianchi, Kasner and G\"{o}del type
configurations; \cite{voffdmgt,voffdmgt1,voffdmgt2,voffdmgt3}, for off--diagonal configurations; and \cite%
{appleby}, for $f(R)$--modified gravity theories. We note also that the
approximation $\widehat{h}_{3}(t)\approx 1+\varepsilon \widehat{\chi }%
_{3}(t) $ can be very restrictive---one can consider more general classes of
solutions with arbitrary $\widehat{h}_{3}(t).$} In explicit form, such
metric, with small off--diagonal deformations on $\varepsilon $ and
rescaling $\mathring{a}(t)\rightarrow a(t)$, can be written as
\begin{equation}
ds^{2}=a^{2}(t)[(e^{1})^{2}+(e^{2})^{2}]+a^{2}(t)[1+\varepsilon \widehat{%
\chi }_{3}(t)](dy^{3}+\varepsilon \check{n}_{i}dx^{i})^{2}+(dt+\varepsilon
\check{w}_{i}(t)dx^{i})^{2}.  \label{ans1bb}
\end{equation}%
See below how it is possible to construct subclasses of off--diagonal
configurations in a $\widehat{\mathbf{f}}(\widehat{\mathbf{R}},\mathbf{...})$
gravity where $\ \widehat{\mathbf{\Upsilon }}$ (\ref{dsours1}) goes into $%
\check{\Lambda}$ (\ref{dsours2}), and $\check{\Phi}^{2}=\check{\Lambda}^{-1}[%
\widehat{\Phi }^{2}|\ \widehat{\mathbf{\Upsilon }}|+\int d\zeta \ \widehat{%
\Phi }^{2}\partial _{\zeta }|\ \widehat{\mathbf{\Upsilon }}|]$ (\ref{aux2b})
results in $\widehat{\mathbf{f}}\rightarrow \mathbf{\check{f}=\check{R}}$,
an effective $\mathbf{\check{R}}_{\ \beta }^{\alpha }=\check{\Lambda}\delta
_{\ \beta }^{\alpha }$ which admits LC--solutions with zero torsion. We will
be able to reproduce the $\Lambda$CDM model provided the metric (\ref{ans1bb}%
) defines certain classes of solutions constructed for a corresponding
effective action in GR, namely
\begin{equation}
S=\frac{1}{\kappa ^{2}}\int \delta ^{4}u\sqrt{|\ ^{\varepsilon }\mathbf{g}%
_{\alpha \beta }|}(\ ^{\varepsilon }\mathbf{\check{R}}-2\check{\Lambda}+\
_{m}\mathcal{L}(\ ^{\varepsilon }\mathbf{g}_{\alpha \beta },\ _{m}\Psi )).
\label{actex}
\end{equation}%
In this action, the Ricci scalar $\ ^{\varepsilon }\mathbf{\check{R}=\check{R%
}(}a,\varepsilon )$ is constructed for $\ ^{\varepsilon }\mathbf{g}_{\alpha
\beta }$ with coefficients of (\ref{ans1bb}), $\check{\Lambda}$ is an
effective cosmological constant used for nonholonomic deformations, and $\
_{m}\mathcal{L}$ is considered for certain effective matter fields with
certain pressure $\ _{m}p$ and energy--density $\ _{m}\rho $. The EoS are
chosen, for simplicity, to correspond to an effective de Sitter
configuration determined by $\check{\Lambda},$ where $\check{\varpi}:=\check{%
p}_{\Lambda }/\check{\rho}_{\Lambda }=-1$, with pressure $\check{p}_{\Lambda
}$ and energy--density $\check{\rho}_{\Lambda }.$

We can describe the theories determined by (\ref{actex}) and (\ref{ans1bb})
with respect to nonholonomic (non-integrable) dual frames $\widehat{\mathbf{e%
}}^{\alpha }=(e^{i},\widehat{\mathbf{e}}^{a}),$ which is convenient for
constructing off--diagonal solutions, or to redefine the constructions with
respect to local coordinate coframes $du^{\alpha }=(dx^{i},dy^{a}),$ where
certain analog of the FLRW metric and $\Lambda$CDM like theories can be
analyzed. For $\varepsilon \rightarrow 0,$ the metric (\ref{ans1bb})
transforms into
\begin{equation}
ds^{2}=a^{2}(t)[(e^{1})^{2}+(e^{2})^{2}+(dy^{3})]^{2}+dt^{2}.  \label{ans1bc}
\end{equation}%
which is just (\ref{flrwel}) but with a re--scaled factor because of the
nonholonomic transformations $\ \widehat{\mathbf{\Upsilon }}$ $\rightarrow $
$\check{\Lambda}$ and $\widehat{\Phi }\rightarrow \check{\Phi}.$

The corresponding Einstein equations with respect to the nonholonomic frames
are%
\begin{eqnarray}
3H^{2} &=&\kappa ^{2}\ _{m}\rho +\check{\Lambda},  \label{enst1c} \\
2H^{\diamond } &=&-\kappa ^{2}(\ _{m}\rho +\ _{m}P+\check{\Lambda}),  \notag
\end{eqnarray}%
where $H^{\diamond }:=a^{\diamond }/a.$ We can express $\ ^{\varepsilon }%
\mathbf{\check{R}}+\ _{m}\mathcal{L}\mathbf{=\ }^{a}\mathbf{\check{R}+}\
_{m}^{0}\mathcal{L+}\varepsilon \ _{m}^{1}\mathcal{L}$, where $\mathbf{\ }%
^{a}\mathbf{\check{R}}$ and $\ _{m}^{0}\mathcal{L}$ are computed for the
metric (\ref{ans1bc}) and $\ _{m}^{1}\mathcal{L}$ include all $\varepsilon $%
--deformations in (\ref{actex}). Such $\ _{m}^{1}\mathcal{L}$ results in the
effective splitting $_{m}\rho =\ _{m}^{0}\rho +\varepsilon \ _{m}^{1}\rho $
and $_{m}p=\ _{m}^{0}p+\varepsilon \ _{m}^{1}p.$ In this way, we can encode
the off--diagonal components as certain additional terms into the matter
source, or either consider them as a polarization of the effective
cosmological constant $\Lambda :=\check{\Lambda}+\varepsilon \ ^{1}\check{%
\Lambda}.$\footnote{%
We do not provide here explicit formulas for the corrections proportional to
$\varepsilon$ because, in the end, we shall take smooth limits $\varepsilon
\rightarrow 0.$ The main constructions for nonholonomic off--diagonal
transforms are based on rescaling $\mathring{a}(t)\rightarrow a^{2}(t)$ $\ $%
generated by the solutions with $\widehat{\mathbf{\Upsilon }}$ $\rightarrow $
$\check{\Lambda}$ and $\widehat{\Phi }\rightarrow \check{\Phi}. $} We also
note that possible small inhomogeneous and locally anisotropic
contributions, and concordance with observational data, can be estimated
similarly to those presented, e.g., in \cite{appleby}. This could be a
ground for further investigations of such ``slightly'' $f(R)$--modified
off--diagonal cosmological models. In this subsection we provide only a few
qualitative estimations, in order to demonstrate that ``realistic'' $\Lambda$%
CDM like cosmological theories can be equivalently modeled both in terms of
nonholononomic frames and of coordinate frames, if small off--diagonal
deformations are considered, only, and then the limit $\varepsilon
\rightarrow 0$ is taken.

In coordinate frames, Eqs.~(\ref{enst1c}) are written as
\begin{eqnarray*}
3H^{2} &=&\kappa ^{2}\ _{m}^{0}\rho +\Lambda , \\
2H^{\diamond } &=&-\kappa ^{2}(\ _{m}^{0}\rho +\ _{m}^{0}P+\Lambda ).
\end{eqnarray*}%
For $\varepsilon \rightarrow 0,$ the diagonalized solutions are determined
by $a$ (and not by $\mathring{a}$ in (\ref{flrwel})) and can be
parameterized to define and effective $\Lambda$CDM like model where $%
a=a_{c}e^{H_{c}t},$ for a positive constant $a_{c}.$ Thus, modified
gravities with equivalent off--diagonal encodings of $f(R)$--modified
gravity seem to result in realistic cosmological models, at least for small
parametric $\varepsilon $--deformations.

The main goal of this work is to study possible nonlinear gravitational and
matter field interactions which result in the encoding of modified gravities
into generic off--diagonal metrics defining effective Einstein spaces,
without certain special assumptions on the linearization of some associated
systems of PDEs and their solutions. Surprisingly enough, the AFDM allows us
to find such ``non-perturbative'' solutions in explicit form, by using
geometrical methods. The values $a(t),$ $\widehat{h}_{3}(t),w_{i}(t)$ and $%
n_{i}$ are nonlinearly determined by generating functions and sources of
type $\Phi (t)$ and $\Upsilon (t).$ In general, such nonlinear modifications
of a ``prime'' $\mathring{a}(t)$ are not small. Even if we can introduce an
effective scaling factor $a(t),$ this value describes a nonlinear and
inhomogeneous evolution with respect to nonholonomic (nonintegrable) dual
frames $\widehat{\mathbf{e}}^{\alpha }=(e^{i},\widehat{\mathbf{e}}^{a}).$
All generic off--diagonal \ cosmological models can be also redefined with
respect to local coordinate coframes $du^{\alpha }=(dx^{i},dy^{a}).$ In
local coordinate form, we are not able to analyze common and different
properties of diagonalizable and non--diagonalizable models only by
comparing the evolutions of $a(t)$ and $\mathring{a}(t).$ From the physical
point of view, we can consider the Universe as an aether with a complex
vacuum and nonvacuum nonlinear structure determined by possible $f(R)$%
--modifications. An observer acquires experimental/observational data with
respect to a local comoving frame $\widehat{\mathbf{e}}^{\alpha }=(e^{i},%
\widehat{\mathbf{e}}^{a})$ where generic off--diagonal gravitational and
matter field interactions are taken into consideration. For certain
parametric resonant dependencies, even the smallest nonlinearities can
result in substantial polarizations of the gravitational vacuum aether, with
possible Lie group or solitonic symmetries, or without any anisotropic
symmetry prescribed in advance. Such cosmological models are described by
more sophisticate geometries than the FLRW cosmology (for references, see
\cite{ellisgfr,krasinski}).

The key idea of our work is that, within certain assumptions, various
possible $f(R)$--nonlinear modifications can be encoded into off--diagonal
terms and some effective $a(t),$ $\widehat{h}_{3}(t),w_{i}(t)$ via nonlinear
interactions. This can be done for more general classes of cosmological
solutions with nonlinear gravitational interactions restructuring the
spacetime aether before considering certain small $\varepsilon $%
--parameters. Such nonlinear cosmological evolution is determined by three
functions of a time like variable, $t$, characterizing a more complex model
then the FLRW one. We get, indeed: 1) a scaling factor $a(t)$; 2) a diagonal
inhomogeneity function $\widehat{h}_{3}(t);$ and 3) off--diagonal
deformations via $w_{i}(t).$ Having constructed a class of off--diagonal
solutions then, with certain additional assumptions for the effective
linearization in terms of $\varepsilon $, one can study possible observable $%
\varepsilon $--small inhomogeneous or locally anisotropic contributions. The
physical effects of small $\varepsilon $--deformations can be compared, for
instance, with those for a scaling factor $\mathring{a}(t)$, although this
will not be the aim of this paper. The main results and conclusions of it
(see Sects. ~4 and 5) will have to do with certain special properties of
off--diagonal nonlinear systems with nonintegrable constraints and with the
exact solutions. Even for the physical interpretations of observable
cosmological data at a fixed time $t=t_{0},$ we can take the limit $%
\varepsilon \rightarrow 0$, where $\widehat{h}_{3}(t)\rightarrow 1$, and $%
w_{i}$ and $n_{i}$ may vanish or result in a nonholonomic frame structure; a
generalized nonlinear cosmological evolution by such generalized solutions
may result in a modified scaling factor $a(t)$ encoding both $f(R)$%
--modifications and off--diagonal nonlinear interactions for $t<t_{0}.$

\subsection{Conventions and geometric preliminaries}

\subsubsection{Nonlinear connections and N-adapted frames}

Let us consider a pseudo-Riemannian manifold $V,$\ $\dim V=n+m,$ ($n,m\geq 2
$). A Whitney sum $\mathbf{N}$ is defined for its tangent space $TV$,
\begin{equation}
\mathbf{N}:\ TV=hTV\oplus vTV.  \label{whitn}
\end{equation}
Conventionally, this states a nonholonomic (equivalently, non-integrable, or
anholonomic) horizontal (h) and vertical (v) splitting, or a nonlinear
connection (\textit{N-connection}) structure. In local form, it is
determined by its coefficients $\mathbf{N}=\{N_{i}^{a}(u)\}, $ when $\mathbf{%
N}=N_{i}^{a}(x,y)dx^{i}\otimes \partial /\partial y^{a}$ for certain local
coordinates $u=(x,y),$ or $u^{\alpha }=(x^{i},y^{a}),$ and $h$-indices $%
i,j,...=1,2,...n$ and $v$-indices $a,b,...=n+1,n+2,...,n+m.$\footnote{%
The Einstein rule on index summation will be applied if the contrary is not
stated. For convenience, ``primed'' and ``underlined'' indices will be used,
and boldface letters to emphasize that an N-connection spitting is
considered on a spacetime manifold $\mathbf{V=(}V,\mathbf{N).}$} Such a
h-v-decomposition can be naturally associated with some N-adapted frame or,
respectively, dual frame structures, $\mathbf{e}_{\nu }=(\mathbf{e}%
_{i},e_{a})$ and $\mathbf{e}^{\mu }=(e^{i},\mathbf{e}^{a}),$
\begin{eqnarray}
\mathbf{e}_{i} &=&\partial /\partial x^{i}-\ N_{i}^{a}(u)\partial /\partial
y^{a},\ e_{a} = \partial _{a}=\partial /\partial y^{a},  \label{nader} \\
\mbox{ and  }e^{i} &=&dx^{i},\ \mathbf{e}^{a}=dy^{a}+\ N_{i}^{a}(u)dx^{i}.
\label{nadif}
\end{eqnarray}
The nonholonomy relations hold
\begin{equation}
\lbrack \mathbf{e}_{\alpha },\mathbf{e}_{\beta }]=\mathbf{e}_{\alpha }%
\mathbf{e}_{\beta }-\mathbf{e}_{\beta }\mathbf{e}_{\alpha }=W_{\alpha \beta
}^{\gamma }\mathbf{e}_{\gamma },  \label{nonholr}
\end{equation}%
with nontrivial anholonomy coefficients $W_{ia}^{b}=\partial
_{a}N_{i}^{b},W_{ji}^{a}=\Omega _{ij}^{a}=\mathbf{e}_{j}\left(
N_{i}^{a}\right) -\mathbf{e}_{i}(N_{j}^{a})$. The coefficients $\Omega
_{ij}^{a}$ define the N-connection curvature.

\subsubsection{Distinguished metric structures}

Any metric structure $\mathbf{g}$ on $\mathbf{V}$ (for physical
applications, we consider pseudo-Euclidean signatures of type $\left(
+,+,+,-\right) $) can be written in two equivalent ways: 1) with respect to
a dual local coordinate basis,
\begin{equation}
\mathbf{g}=\underline{g}_{\alpha \beta }du^{\alpha }\otimes du^{\beta },
\label{m1}
\end{equation}%
where
\begin{equation}
\underline{g}_{\alpha \beta }=\left[
\begin{array}{cc}
g_{ij}+N_{i}^{a}N_{j}^{b}g_{ab} & N_{j}^{e}g_{ae} \\
N_{i}^{e}g_{be} & g_{ab}%
\end{array}%
\right] ,  \label{ansatz}
\end{equation}%
or 2) as a distinguished metric (in brief, \textit{d-metric}, i.e. in
N-adapted form,
\begin{equation}
\mathbf{g}=g_{\alpha }(u)\mathbf{e}^{\alpha }\otimes \mathbf{e}^{\beta
}=g_{i}(x^{k})dx^{i}\otimes dx^{i}+g_{a}(x^{k},y^{b})\mathbf{e}^{a}\otimes
\mathbf{e}^{a}.  \label{dm1}
\end{equation}%
To prove the decoupling of fundamental gravitational equations in modified
gravity is possible for d-metrics and working with respect to N-adapted
frames.

\subsubsection{Distinguished connections}

A linear connection is called distinguished,\textit{\ d-connection,} $%
\mathbf{D}=(hD,vD),$ if it preserves under parallelism a prescribed
N-connection splitting (\ref{whitn}). Any $\mathbf{D}$ defines an operator
of covariant derivation, $\mathbf{D}_{\mathbf{X}}\mathbf{Y}$, for a d-vector
field $\mathbf{Y}$ in the direction of a d-vector $\mathbf{X}.$ We note that
any vector $Y(u)\in T\mathbf{V}$ can be parameterized as a d-vector, $%
\mathbf{Y}=$ $\mathbf{Y}^{\alpha }\mathbf{e}_{\alpha }=\mathbf{Y}^{i}\mathbf{%
e}_{i}+\mathbf{Y}^{a}e_{a},$ or $\mathbf{Y}=(hY,vY),$ with $hY=\{\mathbf{Y}%
^{i}\}$ and $vY=\{\mathbf{Y}^{a}\},$ where the N-adapted base vectors and
duals, or covectors, are chosen in N-adapted form (\ref{nader}) and (\ref%
{nadif}). The local coefficients of $\mathbf{D}_{\mathbf{X}}\mathbf{Y}$ can
be computed for $\mathbf{D}=\{\mathbf{\Gamma }_{\ \alpha \beta }^{\gamma
}=(L_{jk}^{i},L_{bk}^{a},C_{jc}^{i},C_{bc}^{a})\}$ and h-v-components of $%
\mathbf{D}_{\mathbf{e}_{\alpha }}\mathbf{e}_{\beta }:=$ $\mathbf{D}_{\alpha }%
\mathbf{e}_{\beta }$ using $\mathbf{X}=\mathbf{e}_{\alpha }$ and $\mathbf{Y}=%
\mathbf{e}_{\beta }.$\footnote{%
We shall use the terms d-vector, d-tensor, etc. for any vector, tensor
valued with coefficients defined in a N-adapted form with respect to the
necessary types of tensor products of bases, (\ref{nader}) and (\ref{nadif}%
), and necessary $h$-$v$-decompositions.} We can characterize a d-connection
by three fundamental geometric objects: the d-torsion, $\mathcal{T},$ the
non-metricity, $\mathcal{Q},$ and the d-curvature, $\mathcal{R},$
respectively defined by
\begin{eqnarray}
\mathcal{T}(\mathbf{X,Y})&:=&\mathbf{D}_{\mathbf{X}}\mathbf{Y}-\mathbf{D}_{%
\mathbf{Y}}\mathbf{X}-[\mathbf{X,Y}],\mathcal{Q}(\mathbf{X}):=\mathbf{D}_{%
\mathbf{X}}\mathbf{g,}  \label{torsnmcurv} \\
\mathcal{R}(\mathbf{X,Y})&:=&\mathbf{D}_{\mathbf{X}}\mathbf{D}_{\mathbf{Y}}-%
\mathbf{D}_{\mathbf{Y}}\mathbf{D}_{\mathbf{X}}-\mathbf{D}_{\mathbf{[X,Y]}}.
\notag
\end{eqnarray}%
The N-adapted coefficients,
\begin{eqnarray*}
\mathcal{T} &=&\{\mathbf{T}_{\ \alpha \beta }^{\gamma }=\left( T_{\
jk}^{i},T_{\ ja}^{i},T_{\ ji}^{a},T_{\ bi}^{a},T_{\ bc}^{a}\right) \},%
\mathcal{Q}=\mathbf{\{Q}_{\ \alpha \beta }^{\gamma }\}, \\
\mathcal{R} &\mathbf{=}&\mathbf{\{R}_{\ \beta \gamma \delta }^{\alpha }%
\mathbf{=}\left( R_{\ hjk}^{i}\mathbf{,}R_{\ bjk}^{a}\mathbf{,}R_{\ hja}^{i}%
\mathbf{,}R_{\ bja}^{c},R_{\ hba}^{i},R_{\ bea}^{c}\right) \},
\end{eqnarray*}%
of such fundamental geometric objects are computed by introducing $\mathbf{X}%
=\mathbf{e}_{\alpha }$ and $\mathbf{Y}=\mathbf{e}_{\beta },$ and $\mathbf{D}%
=\{\mathbf{\Gamma }_{\ \alpha \beta }^{\gamma }\}$ into the formulas above
(see \cite{voffds,voffds1,voffds2,voffds3,voffds4} for details).

\subsubsection{Preferred d-metric and d-connection structures}

A d-connection $\mathbf{D}$ is compatible with a d-metric $\mathbf{g}$ if
and only if $\mathcal{Q}=\mathbf{Dg}=0.$ Any metric structure $\mathbf{g}$
on $\mathbf{V}$ is characterized by a unique metric compatible and
torsionless linear connection called the Levi-Civita (LC) connection, $%
\nabla .$ It should be noted that $\nabla $ is not a d-connection because it
does not preserve under parallelism the N-connection splitting (\ref{whitn}%
). Nevertheless, such a $h$-$v$ decomposition allows us to define N-adapted
distortions of any d-connection $\mathbf{D,}$
\begin{equation}
\mathbf{D}=\nabla +\mathbf{Z},  \label{distr}
\end{equation}%
with respective conventional ``non-boldface'' and ``boldface'' symbols for
the coefficients: $\nabla =\{\Gamma _{\ \beta \gamma }^{\alpha }\}$ and, for
the distortion d-tensor, $\mathbf{Z}=\{\mathbf{Z}_{\ \beta \gamma }^{\alpha
}\}. $

This stands for any prescribed $\mathbf{N}$ and $\mathbf{g}=h\mathbf{g}+v%
\mathbf{g,}$ but alternatively to $\nabla $, on $\mathbf{V}$, we can work
with the so-called canonical d-connection, $\widehat{\mathbf{D}},$ when
\begin{equation*}
(\mathbf{g,N})\rightarrow
\begin{array}{cc}
\mathbf{\nabla :} & \mathbf{\nabla g}=0;\ ^{\nabla }\mathcal{T}=0; \\
\widehat{\mathbf{D}}: & \widehat{\mathbf{D}}\mathbf{g}=0;\ h\widehat{%
\mathcal{T}}=0,v\widehat{\mathcal{T}}=0,hv\widehat{\mathcal{T}}\neq 0;%
\end{array}%
\end{equation*}%
are completely defined \textit{by the same} metric structure. The canonical
distortion d-tensor $\widehat{\mathbf{Z}}$ in the distortion relation of
type (\ref{whitn}), $\widehat{\mathbf{D}}=\nabla +\widehat{\mathbf{Z}},$ is
an algebraic combination of the coefficients of the corresponding torsion
d-tensor $\widehat{\mathcal{T}}=\{\widehat{\mathbf{T}}_{\ \beta \gamma
}^{\alpha }\}.$ The respective coefficients of the torsions, $\widehat{%
\mathcal{T}}$ and $\ ^{\nabla }\mathcal{T}=0,$ and curvatures, $\widehat{%
\mathcal{R}}=\{\widehat{\mathbf{R}}_{\ \beta \gamma \delta }^{\alpha }\}$
and $\ ^{\nabla }\mathcal{R}=\{R_{\ \beta \gamma \delta }^{\alpha }\},$ of $%
\widehat{\mathbf{D}}$ and $\nabla $ can be defined and computed using
formulas similar to (\ref{torsnmcurv}). We note that the coefficients $%
\widehat{\mathbf{T}}_{\ \beta \gamma }^{\alpha }$ are not trivial but
nonholonomically induced by anholonomy coefficients $W_{\alpha \beta
}^{\gamma }$ (\ref{nonholr}) and certain off-diagonal coefficients of the
metric (\ref{ansatz}).\footnote{%
In the Riemann-Cartan geometry, such a torsion is for a general metric
compatible linear connection, $D$, which is not necessarily completely
defined by the data $(\mathbf{g,N}).$}

The Ricci tensors of $\widehat{\mathbf{D}}$ and $\nabla $ are computed in
the standard form,
\begin{equation*}
\widehat{\mathcal{R}}ic=\{\widehat{\mathbf{R}}_{\ \beta \gamma }:=\widehat{%
\mathbf{R}}_{\ \alpha \beta \gamma }^{\gamma }\}\mbox{ and }Ric=\{R_{\ \beta
\gamma }:=R_{\ \alpha \beta \gamma }^{\gamma }\}.
\end{equation*}%
With respect to N-adapted coframes (\ref{nadif}), the Ricci d-tensor $%
\widehat{\mathcal{R}}ic$ is characterized by four $h$-$v$ N-adapted
coefficients{\small
\begin{equation}
\widehat{\mathbf{R}}_{\alpha \beta }=\{\widehat{R}_{ij}:=\widehat{R}_{\
ijk}^{k},\ \widehat{R}_{ia}:=-\widehat{R}_{\ ika}^{k},\ \widehat{R}_{ai}:=%
\widehat{R}_{\ aib}^{b},\ \widehat{R}_{ab}:=\widehat{R}_{\ abc}^{c}\},
\label{driccic}
\end{equation}%
} and (an alternative to the LC-scalar curvature, $\ R:=\mathbf{g}^{\alpha
\beta }R_{\alpha \beta })$ scalar curvature,
\begin{equation}
\ \widehat{\mathbf{R}}:=\mathbf{g}^{\alpha \beta }\widehat{\mathbf{R}}%
_{\alpha \beta }=g^{ij}\widehat{R}_{ij}+g^{ab}\widehat{R}_{ab}.
\label{sdcurv}
\end{equation}

We emphasize that any (pseudo) Riemannian geometry can be equivalently
described by both geometric data $\left( \mathbf{g,\nabla }\right) $ and $(%
\mathbf{g,N,}\widehat{\mathbf{D}}).$ For instance, there are canonical
distortion relations
\begin{equation*}
\widehat{\mathcal{R}}=\ ^{\nabla }\mathcal{R+}\ ^{\nabla }\mathcal{Z}%
\mbox{
and }\widehat{\mathcal{R}}ic=Ric+\widehat{\mathcal{Z}}ic,
\end{equation*}%
where the respective distortion d-tensors $\ ^{\nabla }\mathcal{Z}$ and $%
\widehat{\mathcal{Z}}ic$ are computed by introducing $\widehat{\mathbf{D}}%
=\nabla +\widehat{\mathbf{Z}}$ into the corresponding formulas (\ref%
{torsnmcurv}) and (\ref{driccic}). The canonical data $(\mathbf{g,N,}%
\widehat{\mathbf{D}})$ provide an example of nonholonomic (pseudo-)
Riemannian manifold which is a standard one but enabled with a nonholonomic
distribution determined by $(\mathbf{g,N}).$ If the coefficients $\Omega
_{ij}^{a}=0$, such a distribution is holonomic, i.e. integrable.

Nevertheless, physical theories formulated in terms of data as $\left(
\mathbf{g,\nabla }\right) $, or $(\mathbf{g,N,}\widehat{\mathbf{D}})$, are
not equivalent if certain additional conditions are not imposed. Let us
consider an explicit example. We can introduce the Einstein d-tensor of $%
\widehat{\mathbf{D}},$
\begin{equation}
\widehat{\mathbf{E}}_{\alpha \beta }:=\widehat{\mathbf{R}}_{\alpha \beta }-%
\frac{1}{2}\mathbf{g}_{\alpha \beta }\ \widehat{\mathbf{R}},  \label{enstdt}
\end{equation}%
and construct a N-adapted energy momentum tensor for a Lagrange density $\
^{m}\mathcal{L}$ of the matter fields, $\widehat{\mathbf{T}}_{\alpha \beta
}:=-\frac{2}{\sqrt{|\mathbf{g}_{\mu \nu }|}}\frac{\delta (\sqrt{|\mathbf{g}%
_{\mu \nu }|}\ \ ^{m}\widehat{\mathcal{L}})}{\delta \mathbf{g}^{\alpha \beta
}}$, performing a N-adapted variational calculus with respect to N-elongated
(co) frames (\ref{nader}) and (\ref{nadif}), and consider that $\widehat{%
\mathbf{D}}$ is used as covariant derivative instead of $\nabla $. In this
way a nonholonomic deformation of Einstein's gravity is constructed, being $%
\nabla \rightarrow $ $\widehat{\mathbf{D}}=\nabla +\widehat{\mathbf{Z}},$
with gravitational field equations
\begin{equation}
\widehat{\mathbf{R}}_{\alpha \beta }=\kappa ^{2}(\widehat{\mathbf{T}}%
_{\alpha \beta }-\frac{1}{2}\mathbf{g}_{\alpha \beta }\widehat{\mathbf{T}})
\label{nheeq}
\end{equation}%
for a conventional gravitational constant $\kappa ^{2}$ and $\widehat{%
\mathbf{T}}:=\mathbf{g}^{\mu \nu }\widehat{\mathbf{T}}_{\mu \nu }.$ Such
equations are different from the standard Einstein equations in GR because,
in general, $\widehat{\mathbf{R}}_{\alpha \beta }\neq R_{\alpha \beta }$ and
$\widehat{\mathbf{T}}_{\alpha \beta }\neq T_{\alpha \beta },$ where $%
T_{\alpha \beta }:=-\frac{2}{\sqrt{|\mathbf{g}_{\mu \nu }|}}\frac{\delta (%
\sqrt{|\mathbf{g}_{\mu \nu }|}\ \ ^{m}\mathcal{L})}{\delta \mathbf{g}%
^{\alpha \beta }}$ for $\ ^{m}\mathcal{L}[\mathbf{g}_{\alpha \beta },\nabla ]%
\mathcal{\neq \ }^{m}\widehat{\mathcal{L}}[\mathbf{g}_{\alpha \beta },%
\widehat{\mathbf{D}}].$

LC-configurations can be extracted from certain classes of solutions of
Eqs.~(\ref{nheeq}) if additional conditions are imposed, resulting in zero
values for the canonical d-torsion, $\widehat{\mathcal{T}}=0$. In N-adapted
coefficient form, such condition is equivalent to {\small
\begin{equation}
\widehat{T}_{\ jk}^{i}=\widehat{L}_{jk}^{i}-\widehat{L}_{kj}^{i},\widehat{T}%
_{\ ja}^{i}=\widehat{C}_{jb}^{i},\widehat{T}_{\ ji}^{a}=-\Omega _{\ ji}^{a},%
\widehat{T}_{aj}^{c}=\widehat{L}_{aj}^{c}-e_{a}(N_{j}^{c}),\widehat{T}_{\
bc}^{a}=\ \widehat{C}_{bc}^{a}-\ \widehat{C}_{cb}^{a}.  \label{dtors}
\end{equation}%
} It should be emphasized that we are able to find generic off-diagonal
solutions of the Einstein equations in GR depending on three and more
coordinates for $\widehat{\mathbf{D}}\rightarrow \nabla ,$ when $\widehat{%
\mathbf{R}}_{\alpha \beta }\rightarrow $ $R_{\alpha \beta }$ and $\widehat{%
\mathbf{T}}_{\alpha \beta }\rightarrow $ $T_{\alpha \beta },$ if the
nonholonomic constraints (\ref{dtors}) are imposed after certain classes of
solutions were found for $\widehat{\mathbf{D}}\neq \nabla .$ But we are not
able to decouple such systems of nonlinear PDEs if the zero torsion
condition for $\nabla $ is imposed from the very beginning.

\subsection{Nonholonomic structures in $f(R,...)$-modified gravity theories}

In general, different models of modified gravity are formulated for
independent metric and linear connection fields with a corresponding
Palatini type variational formulation (see \cite{revfmod,revfmod01,revfmod02,revfmod03,revfmod04,revfmod05,
revfmod06,revfmod07,revfmod08,revfmod09,revfmod10,revfmod11,revfmod12,revfmod13}). The
gravitational and matter field equations in modified gravities consist in
very sophisticate systems of nonlinear PDEs for which finding exact
solutions is a very difficult technical task, even for the simplest diagonal
ans\"{a}tze with the coefficients of the metrics and connections depending
on just one (time or space variable). Nevertheless, the anholonomic frame
deformation method \cite{voffds,voffds1,voffds2,voffds3,voffds4} seems to work efficiently and allows to
construct off-diagonal solutions in modified gravity theories
\cite{voffdmgt,voffdmgt1,voffdmgt2,voffdmgt3}.

\subsubsection{Equivalent modeling of modified gravity}

Consider three classes of equivalent theories of modified gravity defined
for the same metric field $\mathbf{g}=\{g_{\mu \nu }\}$ but with different
actions (and related functionals) for gravity, $\ ^{g}S,$ and matter, $\
^{m}S,$ fields,%
\begin{eqnarray}
\mathcal{S} &=&\ ^{g}\mathcal{S}+\ ^{m}\mathcal{S}=\frac{1}{2\kappa ^{2}}%
\int f(R,T,R_{\alpha \beta }T^{\alpha \beta })\sqrt{|g|}d^{4}u+\int \ ^{m}%
\mathcal{L}\sqrt{|g|}d^{4}u  \notag \\
&=&\ ^{g}\widehat{\mathbf{S}}+\ ^{m}\widehat{\mathbf{S}}=\frac{1}{2\kappa
^{2}}\int \widehat{\mathbf{f}}(\widehat{\mathbf{R}},\widehat{\mathbf{T}},%
\widehat{\mathbf{R}}_{\alpha \beta }\widehat{\mathbf{T}}^{\alpha \beta })%
\sqrt{|\widehat{\mathbf{g}}|}\mathbf{d}^{4}u+\int \ ^{m}\widehat{\mathbf{L}}%
\sqrt{|\widehat{\mathbf{g}}|}\mathbf{d}^{4}u  \notag \\
&=&\ ^{g}\mathbf{\check{S}}+\ ^{m}\mathbf{\check{S}}=\frac{1}{2\kappa ^{2}}%
\int \mathbf{\check{R}}\sqrt{|\mathbf{\check{g}}|}\mathbf{d}^{4}u+\check{%
\Lambda}\int \sqrt{|\mathbf{\check{g}}|}\mathbf{d}^{4}u.  \label{mgts}
\end{eqnarray}%
Here, we use boldface $\mathbf{d}^{4}u$ in order to emphasize that the
integration volume is for N-elongated partial derivatives (\ref{nadif}), $%
\kappa ^{2}$ is the gravitational coupling constant, the values with ``$%
\symbol{94}$'' are computed for a canonical d-connection $\widehat{\mathbf{D}%
}$ and the values with ``$\vee $'' for re-defined geometric data $(\mathbf{%
\check{g},\check{N},\check{D}})$ for certain nonholonomic frame transforms
and nonholonomic deformations $g_{\alpha \beta }\sim \widehat{\mathbf{g}}%
_{\alpha \beta }\sim \mathbf{\check{g}}_{\alpha \beta }.$\footnote{%
We shall give details in Sects.~\ref{sslc} and \ref{sseflrw}.} For
simplicity, we consider matter actions which only depend on the coefficients
of a metric field and not on their derivatives,
\begin{equation*}
\widehat{\mathbf{T}}^{\alpha \beta }=\ ^{m}\widehat{\mathbf{L}}\ \widehat{%
\mathbf{g}}^{\alpha \beta }+2\delta (\ ^{m}\widehat{\mathbf{L}})/\delta
\widehat{\mathbf{g}}_{\alpha \beta }.
\end{equation*}%
Such variations can be performed with respect to coordinate frames for $%
g_{\mu \nu }$ or in various N-adapted forms for $\widehat{\mathbf{g}}_{\mu
\nu }$ and $\mathbf{\check{g}}_{\mu \nu }.$

\subsubsection{Off-diagonal deformations of FLRW metrics}

We assume that the matter content of the universe can be approximated by a
perfect (pressureless) fluid, where
\begin{equation}
\widehat{\mathbf{T}}_{\alpha \beta }=p\widehat{\mathbf{g}}_{\alpha \beta
}+(\rho +p)\widehat{\mathbf{v}}_{\alpha }\widehat{\mathbf{v}}_{\beta }
\label{dsourc}
\end{equation}%
is defined for certain (effective) energy and pressure densities,
respectively, $\widehat{\mathbf{v}}_{\alpha }$ being the four-velocity of
the fluid for which $\widehat{\mathbf{v}}_{\alpha }\widehat{\mathbf{v}}%
^{\alpha }=-1$ and $\widehat{\mathbf{v}}^{\alpha }=(0,0,0,1)$ in N-adapted
comoving frames/coordinates. Here frame deformations/transforms of metrics
of type $\widehat{\mathbf{g}}_{\alpha \beta }=\mathbf{e}_{\ \alpha }^{\alpha
^{\prime }}\mathbf{e}_{\ \beta }^{\beta ^{\prime }}\mathring{g}_{\alpha
^{\prime }\beta ^{\prime }},$ will be studied, being the FLRW diagonalized
element
\begin{eqnarray}
d\mathring{s}^{2} &=&\mathring{g}_{\alpha ^{\prime }\beta ^{\prime
}}du^{\alpha ^{\prime }}du^{\beta ^{\prime }}=\mathring{a}%
^{2}(t)[dr^{2}+r^{2}d\theta ^{2}+r^{2}\sin ^{2}\theta d\varphi ^{2}]-dt^{2},
\label{flrw} \\
&=&\mathring{a}^{2}(t)[dx^{2}+dy^{2}+dz^{2}]-dt^{2}  \notag
\end{eqnarray}%
where the scale factor $\mathring{a}(t)$ (we use also the value $\mathring{H}%
:=\mathring{a}/\mathring{a},$ for $\mathring{a}^{\diamond }:=d\mathring{a}%
/dt)$ with signature $(+,+,+,-)$, and a parametrization of coordinates in
the form $u^{\alpha ^{\prime }}=(x^{1^{\prime }}=r,x^{2^{\prime }}=\theta
,y^{3^{\prime }}=\varphi ,y^{4^{\prime }}=t),$ or as Cartesian coordinates $%
(x^{1^{\prime }}=x,x^{2^{\prime }}=y,y^{3^{\prime }}=z,y^{4^{\prime }}=t).$
For such cosmological metrics, the main issues of the Einstein and modified
Universes are encoded into energy-momentum tensor $\mathring{T}_{\alpha
\beta }=\mathring{p}\mathring{g}_{\alpha \beta }+(\mathring{\rho}+\mathring{p%
})\mathring{v}_{\alpha }\mathring{v}_{\beta }$ (we omit primes or other
distinctions in the coordinate indices if there is no ambiguity) arising
from a matter Lagrangian $\ ^{m}\mathcal{\mathring{L}}$ through
\begin{equation}
\mathring{T}(t)=\mathring{T}_{\ \alpha }^{\alpha }=-\mathring{\rho},%
\mathring{P}(t)=\mathring{R}_{\alpha \beta }\mathring{T}^{\alpha \beta }=%
\mathring{R}_{44}\mathring{T}^{44}=-3\mathring{\rho}(\mathring{H}^{2}+%
\mathring{H}^{\diamond }),  \label{auxaa}
\end{equation}%
for $\mathring{T}_{\ \beta }^{\alpha }=diag[0,0,0,-\mathring{\rho}].$

We will consider nonhomogeneous and locally anisotropic cosmological
solutions of type (\ref{ansatz}) and/or \ (\ref{dm1}) generated by
off-diagonal deformations of \ (\ref{flrw})
\begin{eqnarray}
g_{i} &=&g_{i}{(x^{k})}=\eta _{i}(x^{k},y^{4})\mathring{g}%
_{i}(x^{k},y^{4})=e^{\psi {(x^{k})}},  \label{polarf} \\
g_{a} &=&\omega ^{2}(x^{k},y^{4})h_{a}(x^{k},y^{4})=\omega
^{2}(x^{k},y^{4})\eta _{a}(x^{k},y^{4})\mathring{g}_{a}(x^{k},y^{4}),  \notag
\\
\ N_{i}^{3} &=&n_{i}(x^{k}),N_{i}^{4}=w_{i}(x^{k},y^{4}).  \notag
\end{eqnarray}%
In Eqs.~(\ref{polarf}) there is no summation on repeated indices, $\eta
_{\alpha }=(\eta _{i},\eta _{a})$ are polarization functions, the
N-connection coefficients are determined by $n_{i}$ and $w_{i},$ the
vertical conformal factor $\omega $ may depend on all spacetime coordinates
and $\mathring{g}_{\alpha }=(\mathring{g}_{i},\mathring{g}_{a})$ define the
``prime'' diagonal metric if $\eta _{\alpha }=1$ and $N_{i}^{a}=0.$ The
``target'' off-diagonal metrics are with Killing symmetry on $\partial
/\partial y^{3}$ when the coefficients (\ref{polarf}) do not depend on $%
y^{3}.$\footnote{%
We can consider nonholonomic deformations with non-Killing symmetries when,
for instance, $\omega (x^{k},y^{4})\rightarrow \omega (x^{k},y^{3},y^{4})$,
which results in a more cumbersome calculus and geometric techniques. For
simplicity, we do not study such generalizations in this work (see examples
in \cite{voffds,voffds1,voffds2,voffds3,voffds3}).}
\begin{equation}
ds^{2} =a^{2}(x^{k},t)[\eta _{1}(x^{k},t)(dx^{1})^{2}+\eta
_{2}(x^{k},t)(dx^{2})^{2}]+ a^{2}(x^{k},t)\widehat{h}_{3}(x^{k},t)(\widehat{%
\mathbf{e}}^{3})^{2}+\omega ^{2}(x^{k},t)h_{4}(x^{k},t)(\widehat{\mathbf{e}}%
^{4})^{2},  \label{flrwod}
\end{equation}%
when $a^{2}(x^{k},t)\eta _{i}(x^{k},t)=e^{\psi {(x^{k})}},$ for $i=1,2;$ $%
a^{2}\ \widehat{h}_{3}=\omega ^{2}(x^{k},t)h_{3}(x^{k},t),$ and
\begin{equation*}
\widehat{\mathbf{e}}^{3}=dy^{3}+n_{i}(x^{k})dx^{i},\widehat{\mathbf{e}}%
^{4}=dy^{4}+w_{i}(x^{k},t)dx^{i}.
\end{equation*}%
Functions $\eta _{i},\eta _{a},a,\psi ,\omega ,n_{i},w_{i}$ will be found
such that, via nonholonomic transforms (\ref{polarf}), when $\mathring{g}%
_{\alpha ^{\prime }\beta ^{\prime }}(t)$ (\ref{flrw}) $\rightarrow $ $%
\widehat{\mathbf{g}}_{\alpha \beta }(x^{k},t)$ (\ref{flrwod}), off-diagonal
nonhomogeneous cosmological solutions are generated in a model of modified
gravity (\ref{mgts}). We can consider subclasses of off-diagonal
cosmological solutions but with deformed symmetries when certain nontrivial
limits $\widehat{\mathbf{g}}_{\alpha \beta }(x^{k},t)\rightarrow $ $\widehat{%
\mathbf{g}}_{\alpha \beta }(t)$ can be found and define viable cosmological
models.

\subsubsection{Field equations for nonholonomic modified gravities and FLRW
cosmology}

Applying an N-adapted variational procedure with respect to a nonholonomic
basis (\ref{nader}) and (\ref{nadif}) for the action $\mathcal{S}=\ ^{g}%
\widehat{\mathbf{S}}+\ ^{m}\widehat{\mathbf{S}}$, which is similar to that
in \cite{odgom,odgom1,odgom2} but for $\nabla \rightarrow $ $\widehat{\mathbf{D}}$ and
matter source $\widehat{\mathbf{T}}_{\alpha \beta }$ (\ref{dsourc}), we
obtain the field equations for the corresponding modified gravity theory%
{\small
\begin{eqnarray}
&&\ \widehat{\mathbf{R}}_{\alpha \beta }\ \ ^{1}\widehat{\mathbf{f}}-\frac{1%
}{2}\ \widehat{\mathbf{g}}_{\alpha \beta }\widehat{\mathbf{f}}+(\widehat{%
\mathbf{g}}_{\alpha \beta }\widehat{\mathbf{D}}^{\mu }\widehat{\mathbf{D}}%
_{\mu }-\widehat{\mathbf{D}}_{\alpha }\widehat{\mathbf{D}}_{\beta })\ ^{1}%
\widehat{\mathbf{f}}+(\widehat{\mathbf{T}}_{\alpha \beta }+\mathbf{\Theta }%
_{\alpha \beta })\ ^{2}\widehat{\mathbf{f}}+  \label{mgtfe} \\
&&\mathbf{\Xi }_{\alpha \beta }\ ^{3}\widehat{\mathbf{f}}+\frac{1}{2}(%
\widehat{\mathbf{D}}^{\mu }\widehat{\mathbf{D}}_{\mu }\widehat{\mathbf{T}}%
_{\alpha \beta }\ ^{3}\widehat{\mathbf{f}}+\widehat{\mathbf{g}}_{\alpha
\beta }\widehat{\mathbf{D}}_{\mu }\widehat{\mathbf{D}}_{\nu }\widehat{%
\mathbf{T}}^{\mu \nu }\ ^{3}\widehat{\mathbf{f}})-\widehat{\mathbf{D}}_{\nu }%
\widehat{\mathbf{D}}_{(\alpha }\widehat{\mathbf{T}}_{\beta )}^{\ \nu }\ ^{3}%
\widehat{\mathbf{f}}=\kappa ^{2}\ \widehat{\mathbf{T}}_{\alpha \beta },
\notag
\end{eqnarray}%
}
\begin{equation*}
\mbox{ for }\mathbf{\Theta }_{\alpha \beta }=p\ \widehat{\mathbf{g}}_{\alpha
\beta }-2\widehat{\mathbf{T}}_{\alpha \beta },\ \mathbf{\Xi }_{\alpha \beta
}=2\ \widehat{\mathbf{E}}_{\ (\alpha }^{\nu }\widehat{\mathbf{T}}_{\beta
)\nu }-p\ \widehat{\mathbf{E}}_{\alpha \beta }-\frac{1}{2}\widehat{\mathbf{R}%
}\widehat{\mathbf{T}}_{\alpha \beta },
\end{equation*}%
with respective d-tensors defined by Eqs.~(\ref{driccic}), (\ref{sdcurv})
and (\ref{enstdt}), where $\ ^{1}\widehat{\mathbf{f}}:=\partial \widehat{%
\mathbf{f}}/\partial \widehat{\mathbf{R}},$ $\ \ \ ^{2}\widehat{\mathbf{f}}%
:=\partial \widehat{\mathbf{f}}/\partial \widehat{\mathbf{T}}$ and $\ \ ^{3}%
\widehat{\mathbf{f}}:=\partial \widehat{\mathbf{f}}/\partial \widehat{%
\mathbf{P}},$ when $\widehat{\mathbf{P}}=\widehat{\mathbf{R}}_{\alpha \beta }%
\widehat{\mathbf{T}}^{\alpha \beta }$ and $(\alpha \beta )$ denotes
symmetrization of the indices.

In general, the divergence with $\widehat{\mathbf{D}}$ and/or $\nabla $ of
Eqs.~(\ref{mgtfe}) is not zero. Also Eqs.~(\ref{nheeq}) have a similar
property. In the last case, we can obtain the continuity equations as in GR
and then deform them by using the distortions (\ref{distr}), which for the
canonical d-connections are completely determined by the metric structure.
There are certain types of conservation laws for matter fields with
additional nonholonomic constraints. We can consider the field equations (%
\ref{mgtfe}) and the equations derived by taking the divergence with $%
\widehat{\mathbf{D}}$ as nonholonomic distortions of similar systems of
nonlinear functional PDEs considered in \cite{odgom,odgom1,odgom2} for $\nabla .$
Remarkably, such sophisticate nonholonomic and nonlinear systems can be
solved in very general off-diagonal forms, by applying the anholonomic frame
deformation method. In order to compare these results and to find possible
applications in modern cosmology, we will consider a particular equation of
state (EoS) $p=\varpi \rho $ with $\varpi =const,$ and study the cosmology
of off-diagonal distortions of certain FLRW models considered in the
framework of GR and its modifications. In both cases, by exploring some
particular classes of solutions, the dynamics of the matter sector of
generalized $f(R,T,R_{\mu \nu }T^{\mu \nu })$\ gravity (with respect to
N-adapted frames) may lead to similar cosmological scenarios as GR, but with
nonholonomic constraints and deformations.

\section{ The anholonomic frame deformation method and exact solutions in
modfied gravities}

\label{safdm}

A surprising property of Eqs.~(\ref{nheeq}) and (\ref{mgtfe}) is that they
can be integrated in very general form with generic off-diagonal metrics
when their coefficients depend on all spacetime coordinates via various
classes of generating and integration functions and constants. In
particular, we can consider such generating and integration functions when $%
\widehat{\mathbf{g}}_{\alpha \beta }(x^{k},t)$ (\ref{flrwod}) result in
off-diagonal metrics of type $\widehat{\mathbf{g}}_{\alpha \beta }(t)$
depending on the parameters and possible (non-) commutative Lie algebra or
algebroid symmetries.

\subsection{Off-diagonal FLRW like cosmological models}

We shall study cosmological models with sources of type (\ref{dsourc}) when
the four-velocity $\widehat{\mathbf{v}}_{\alpha }$ is re-parameterized in a
way that for some frame transforms as
\begin{eqnarray}
\widehat{\mathcal{Y}}_{\alpha \beta }&:=&\kappa ^{2}\ (\widehat{\mathbf{T}}%
_{\alpha \beta }-\frac{1}{2}\mathbf{g}_{\alpha \beta }\widehat{\mathbf{T}})
\notag \\
&\rightarrow &diag[\Upsilon _{1}=\Upsilon _{2},\Upsilon _{2}=\ ^{h}\Upsilon
(x^{i}),\Upsilon _{3}=\Upsilon _{4},\Upsilon _{4}=\ ^{v}\Upsilon (x^{i},t)]
\label{dsours1} \\
&\rightarrow &\widehat{\Lambda }\ \mathbf{g}_{\alpha \beta }\
\mbox{
(redefining the generating functions and sources)},  \label{dsours2}
\end{eqnarray}%
for effective $h$- and $v$-polarized sources, respectively, $\ ^{h}\Upsilon
(x^{i})$ and $\Upsilon _{4}=\ ^{v}\Upsilon (x^{i},t),$ or an effective
cosmological constant $\widehat{\Lambda }.$ For simplicity, we can consider
effective matter sources and ``prime'' metrics with Killing symmetry on $%
\partial /\partial _{3},$ i.e. when the effective matter sources and
d-metrics do not depend on the coordinate $y^{3}.$\footnote{%
The method can be extended to account for $y^{3}$ dependence and non-Killing
configurations (see \cite{voffds,voffds1,voffds2,voffds3}). In this paper the local coordinates and
ans\"{a}tze for d-metrics are parameterized in different forms than in
previous works, what is more convenient for the study of cosmological models.%
} In brief, the partial derivatives $\partial _{\alpha }=\partial /\partial
u^{\alpha }$ on a 4-d manifold \ will be written as $s^{\bullet }=\partial
s/\partial x^{1},s^{\prime }=\partial s/\partial x^{2},s^{\ast }=\partial
s/\partial y^{3},s^{\diamond }=\partial s/\partial y^{4}.$

The nontrivial components of the Ricci d-tensor (\ref{driccic}) and
nonholonomic Einstein equations (\ref{enstdt}), with source (\ref{dsours1})
parameterized with respect to N-adapted bases (\ref{nader}) and (\ref{nadif}%
), for a d-metric ans\"{a}tze (\ref{dm1}) with coefficients (\ref{polarf}),
are{\small
\begin{eqnarray}
-\widehat{R}_{1}^{1} &=&-\widehat{R}_{2}^{2}=\frac{1}{2g_{1}g_{2}}%
[g_{2}^{\bullet \bullet }-\frac{g_{1}^{\bullet }g_{2}^{\bullet }}{2g_{1}}-%
\frac{\left( g_{2}^{\bullet }\right) ^{2}}{2g_{2}}+g_{1}^{\prime \prime }-%
\frac{g_{1}^{\prime }g_{2}^{\prime }}{2g_{2}}-\frac{(g_{1}^{\prime })^{2}}{%
2g_{1}}]=\ ^{h}\Upsilon ,  \label{eq1b} \\
-\widehat{R}_{3}^{3} &=&-\widehat{R}_{4}^{4}=\frac{1}{2h_{3}h_{4}}%
[h_{3}^{\diamond \diamond }-\frac{\left( h_{3}^{\diamond }\right) ^{2}}{%
2h_{3}}-\frac{h_{3}^{\diamond }h_{4}^{\diamond }}{2h_{4}}]=\ ^{v}\Upsilon ,
\label{eq2b} \\
\widehat{R}_{3k} &=&\frac{h_{3}}{2h_{4}}n_{k}^{\diamond \diamond }+(\frac{%
h_{3}}{h_{4}}h_{4}^{\diamond }-\frac{3}{2}h_{3}^{\diamond })\frac{%
n_{k}^{\diamond }}{2h_{4}}=0,  \label{eq3b} \\
\widehat{R}_{4k} &=&\frac{w_{k}}{2h_{3}}[h_{3}^{\diamond \diamond }-\frac{%
\left( h_{3}^{\diamond }\right) ^{2}}{2h_{3}}-\frac{h_{3}^{\diamond
}h_{4}^{\diamond }}{2h_{4}}]+\frac{h_{3}^{\diamond }}{4h_{3}}(\frac{\partial
_{k}h_{3}}{h_{3}}+\frac{\partial _{k}h_{4}}{h_{4}})-\frac{\partial
_{k}h_{3}^{\diamond }}{2h_{3}}=0,  \label{eq4b}
\end{eqnarray}%
}where $\widehat{\mathbf{R}}_{\ \beta }^{\alpha }$ are computed for $\omega
=1$ and then the formulas are generalized for $\omega \neq 1$ via
v-conformal transforms (see Refs.~\cite{odgom,odgom1,odgom2} for details),%
\begin{equation}
\mathbf{e}_{i}\omega =\partial _{i}\omega -n_{i}\ \omega ^{\ast
}-w_{i}\omega ^{\diamond }=0.  \label{conf}
\end{equation}
The d-torsion (\ref{dtors}) vanishes if the (Levi-Civita, LC) conditions $%
\widehat{L}_{aj}^{c}=e_{a}(N_{j}^{c}),\widehat{C}_{jb}^{i}=0,\Omega _{\
ji}^{a}=0,$ are satisfied for
\begin{eqnarray}
w_{i}^{\diamond } &=&(\partial _{i}-w_{i}\partial _{4})\ln \sqrt{|h_{4}|}%
,(\partial _{i}-w_{i}\partial _{4})\ln \sqrt{|h_{3}|}=0,  \label{lccondb} \\
\partial _{k}w_{i} &=&\partial _{i}w_{k},n_{i}^{\diamond }=0,\partial
_{i}n_{k}=\partial _{k}n_{i}.  \notag
\end{eqnarray}%
The above system of equations can be integrated in very general situations,
for instance, for d-metrics with Killing symmetry on $\partial _{3}.$

\subsubsection{Decoupling of PDEs for inhomogeneous cosmological metrics}

The system of equations (\ref{eq1b})-(\ref{conf}) has an important
decoupling property. To show this explicitly, we rewrite it as nonlinear PDE
which posses an important decoupling property, allowing integration step by
step of such equations. For $h_{a}^{\diamond }\neq 0,$ $\ ^{h}\Upsilon ,\
^{v}\Upsilon \neq 0,$ Killing symmetry on $\partial _{3}$ and
parameterizations (\ref{polarf}), these equations can be written as
\begin{eqnarray}
\psi ^{\bullet \bullet }+\psi ^{\prime \prime } &=&2~^{h}\Upsilon
\label{eq1m} \\
\phi ^{\diamond }h_{3}^{\diamond } &=&2h_{3}h_{4}~^{v}\Upsilon  \label{eq2m}
\\
n_{i}^{\diamond \diamond }+\gamma n_{i}^{\diamond } &=&0,  \label{eq3m} \\
\beta w_{i}-\alpha _{i} &=&0,  \label{eq4m} \\
\partial _{i}\omega -(\partial _{i}\phi /\phi ^{\diamond })\omega ^{\diamond
} &=&0,  \label{confeq}
\end{eqnarray}%
for
\begin{equation}
\alpha _{i}=h_{3}^{\diamond }\partial _{i}\phi ,\beta =h_{3}^{\diamond }\
\phi ^{\diamond },\gamma =\left( \ln |h_{3}|^{3/2}/|h_{4}|\right) ^{\diamond
},  \label{abc}
\end{equation}%
where
\begin{equation}
{\phi =\ln |h_{3}^{\diamond }/\sqrt{|h_{3}h_{4}|}|,\mbox{ and/ or }}\Phi
:=e^{{\phi }},  \label{genf}
\end{equation}%
is considered as a generating function. Eq.~(\ref{confeq}) is just Eq.~(\ref%
{conf}) for a nontrivial solution of (\ref{eq4m}) with coefficients (\ref%
{abc}), when
\begin{equation}
w_{i}=\partial _{i}\phi /\phi ^{\diamond }.  \label{aux4}
\end{equation}%
The decoupling property of the above system of equations follows from the
facts that: 1) integrating the 2-d Laplace equation (\ref{eq1m}) one finds
solutions for the $h$-coefficients of the d-metric, and 2) the solutions for
the coefficients of the d-metric can be found from (\ref{eq2m}) and (\ref%
{genf}). \ 3) Then the N-connection coefficients $w_{i}$ and $n_{i}$ can be
found from (\ref{eq3m}) and (\ref{eq4m}), respectively.

\subsubsection{Cosmological solutions with nonholonomically induced torsion}

The equations (\ref{eq1m}) and (\ref{eq4m}) can be solved, respectively, for
any source $~^{h}\Upsilon (x^{k})$ and generating function ${\phi (x}^{k},t{%
).}$ The system (\ref{eq2m}) and (\ref{genf}) \ can be written under the
form
\begin{equation*}
h_{3}h_{4} =\phi ^{\diamond }h_{3}^{\diamond }/2~^{v}\Upsilon \mbox{ and }
|h_{3}h_{4}| = ({h_{3}^{\diamond })}^{2}e^{-2\phi },
\end{equation*}
for any nontrivial source $~^{v}\Upsilon (x^{i},t)$ in (\ref{eq2m}).
Introducing the first equation into the second, one finds $|h_{3}^{\diamond
}|=\frac{(e^{2\phi })^{\diamond }}{4|~^{v}\Upsilon |}=\frac{\Phi ^{\diamond
}\Phi \ }{2|~^{v}\Upsilon |}$, i.e. $h_{3}=\ ^{0}h_{3}(x^{k})+\frac{\epsilon
_{3}\epsilon _{4}}{4}\int dt\frac{(\Phi ^{2})^{\diamond }}{~^{v}\Upsilon }$,
where $\ ^{0}h_{3}(x^{k})$ and $\epsilon _{3},\epsilon _{4}=\pm 1.$ Using
again the first equation, we obtain
\begin{equation}
h_{4}=\frac{\phi ^{\diamond }(\ln \sqrt{|h_{3}|})^{\diamond }}{%
2~^{v}\Upsilon }=\frac{1}{2~^{v}\Upsilon }\frac{\Phi ^{\diamond }}{\Phi }%
\frac{h_{3}^{\diamond }}{h_{3}}.  \label{aux1}
\end{equation}%
We can simplify such formulas for $h_{3}$ and $h_{4}$ if we redefine the
generating function, $\Phi \rightarrow \widehat{\Phi },$ where $(\Phi
^{2})^{\diamond }/|~^{v}\Upsilon |=(\widehat{\Phi }^{2})^{\diamond }/\Lambda
,$ i.e.%
\begin{equation}
\Phi ^{2}=\Lambda ^{-1}\left[ \widehat{\Phi }^{2}|~^{v}\Upsilon |+\int dt\
\widehat{\Phi }^{2}|~^{v}\Upsilon |^{\diamond }\right] ,  \label{aux2}
\end{equation}%
for an effective cosmological constant $\Lambda $ which may take positive or
negative values. We can integrate on $t,$ include the integration function $%
\ ^{0}h_{3}(x^{k})$ in $\widehat{\Phi }$ and write
\begin{equation}
h_{3}[\widehat{\Phi }]=\widehat{\Phi }^{2}/4\Lambda .  \label{h3}
\end{equation}%
Introducing this formula and (\ref{aux2}) into (\ref{aux1}), we compute
\begin{equation}
h_{4}[\widehat{\Phi }]=\frac{(\ln |\Phi |)^{\diamond }}{4|~^{v}\Upsilon |}=%
\frac{(\widehat{\Phi }^{2})^{\diamond }}{8}\left[ \widehat{\Phi }%
^{2}|~^{v}\Upsilon |+\int dt\ \widehat{\Phi }^{2}|~^{v}\Upsilon |^{\diamond }%
\right] ^{-1}.  \label{h4}
\end{equation}

As next step, we need solve Eq.~(\ref{eq3m}) by integrating on $t$ twice. We
obtain
\begin{equation}
n_{k}=\ _{1}n_{k}+\ _{2}n_{k}\int dt\ h_{4}/(\sqrt{|h_{3}|})^{3},
\label{n1b}
\end{equation}%
where $\ _{1}n_{k}(x^{i}),\ _{2}n_{k}(x^{i})$ are integration functions and $%
h_{a}[\widehat{\Phi }]$ are given by formulas (\ref{h3}) and (\ref{h4}). If
we fix $\ _{2}n_{k}=0,$ we shall be able to find $n_{k}=\ _{1}n_{k}(x^{i})$
which have zero torsion limits (see examples in subsection \ref{sslc}).

The solutions of (\ref{eq4m}) are given by (\ref{aux4}), which for different
types of generating functions are parameterized as
\begin{equation}
w_{i}=\frac{\partial _{i}\Phi }{\Phi ^{\diamond }}=\frac{\partial _{i}(\Phi
^{2})}{(\Phi ^{2})^{\diamond }},  \label{w1b}
\end{equation}%
where the integral functional $\Phi \lbrack \widehat{\Phi },~^{v}\Upsilon ]$
is given by (\ref{aux2}).

We can introduce certain polarization functions $\eta _{\alpha }$ in order
to write the d-metric of such solutions in the form (\ref{flrwod}). Let us
fix $\omega ^{2}=|h_{4}|^{-1}$ to satisfy the condition (\ref{confeq}),
which for a generating function $\Phi \lbrack \phi ]$ is equivalent to
\begin{equation}
\Phi ^{\diamond }\partial _{i}h_{4}-\partial _{i}\Phi \ h_{4}^{\diamond }=0.
\label{aux5}
\end{equation}%
These first order PDE equations impose certain conditions on the class of
generating function $\Phi $ and source $~^{v}\Upsilon .$ For instance, we
can choose such a system of coordinates where $~^{v}\Upsilon =\frac{1}{4}%
(e^{-\phi })^{\diamond }$ which transforms Eq.~(\ref{h4}) into $h_{4}=\Phi ,$
i.e. this coefficient of the d-metric is considered as a generating
function, and Eqs.~(\ref{aux5}) are solved. In general, the integral
varieties of such equations cannot be expressed in explicit holonomic form.

A modification of the scale factor $\mathring{a}(t)\rightarrow a(x^{k},t),$
for the FLRW metric (\ref{flrw}) (with for $\mathring{g}_{1}=\mathring{g}%
_{2}=\mathring{g}_{3}=\mathring{a}^{2},\mathring{g}_{4}=-1$, has to be
chosen in order to explain observational cosmological data. For any
prescribed functions $a(x^{k},t)$ and $\omega ^{2}=|h_{4}|^{-1}$ and
solutions \ $e^{\psi {(x^{k})}},$ (see (\ref{eq1m})) and $h_{a}[\widehat{%
\Phi }],n_{k}(x^{i}),w_{i}[\widehat{\Phi }]$ (given respectively by formulas
(\ref{h3})-(\ref{w1b})), we can compute the polarization functions $\eta
_{i}=a^{-2}e^{\psi },\eta _{3}=\mathring{a}^{-2}h_{3},\eta _{4}=1$ and
function $\widehat{h}_{3}=h_{3}/a^{2}|h_{4}|.$ Such coefficients (see the
data (\ref{polarf})), define off-diagonal metrics of type (\ref{flrwod}),
\begin{equation}
ds^{2} =a^{2}(x^{k},t)[\eta _{1}(x^{k},t)(dx^{1})^{2}+\eta
_{2}(x^{k},t)(dx^{2})^{2}]+ a^{2}(x^{k},t)\widehat{h}%
_{3}(x^{k},t)[dy^{3}+n_{i}(x^{k})dx^{i}]^{2}-[dt+\frac{\partial _{i}\Phi
\lbrack \widehat{\Phi },~^{v}\Upsilon ]}{\Phi ^{\diamond }[\widehat{\Phi }%
,~^{v}\Upsilon ]}dx^{i}]^{2}.  \label{odfrlwtors}
\end{equation}%
Choosing any generating functions $a^{2}(x^{k},t),\psi (x^{i})$ and $\Phi
\lbrack \widehat{\Phi },~^{v}\Upsilon ]$ and integration functions $%
n_{i}(x^{k}),$ we generate a nonhomogeneous cosmological model with
nonholonomically induced torsion (\ref{dtors}). More general torsions can be
induced if $n_{i}(x^{k},t)$ is taken with two types of integration functions
$\ _{1}n_{i}(x^{k})$ and $\ _{2}n_{i}(x^{k})$ (see Eqs.~(\ref{n1b})). Having
constructed this solution, we can now consider certain subclasses of
generating and integration functions where $a(x^{k},t)\rightarrow a(t)\neq
\mathring{a}(t),w_{i}\rightarrow w_{i}(t),n_{i}\rightarrow const$, etc. In
this way generic off-diagonal cosmological metrics are generated (because
there are nontrivial anholonomy coefficients $W_{ia}^{b}$ in (\ref{nonholr}%
)).

\subsubsection{Levi-Civita off-diagonal cosmological configurations}

\label{sslc}

The LC-conditions (\ref{lccondb}) are given by a set of nonholonomic
constraints which cannot be solved in explicit form for arbitrary data $%
(\Phi ,\Upsilon )$ and integration functions $\ _{1}n_{k}$ and $\ _{2}n_{k}.$
However, some subclasses of off-diagonal solutions can still be constructed
where via frame and coordinate transforms we can chose $\ _{2}n_{k}=0$ and $%
\ _{1}n_{k}=\partial _{k}n$ with a function $n=n(x^{k}).$ It should be noted
that $(\partial _{i}-w_{i}\partial _{4})\Phi \equiv 0$ for any $\Phi
(x^{k},y^{4})$ if $w_{i}$ is defined by (\ref{w1b}). Introducing a new
functional $B(\Phi ),$ we find that $(\partial _{i}-w_{i}\partial _{4})B=%
\frac{\partial B}{\partial \Phi }(\partial _{i}-w_{i}\partial _{4})\Phi =0.$
Using Eq.~(\ref{h3}) for functionals of type $h_{3}=B(|\tilde{\Phi}(\Phi
)|), $ we solve Eqs.~$(\partial _{i}-w_{i}\partial _{4})h_{3}=0,$ what is
equivalent to the second system of equations in (\ref{lccondb}), because $%
(\partial _{i}-w_{i}\partial _{4})\ln \sqrt{|h_{3}|}\sim (\partial
_{i}-w_{i}\partial _{4})h_{3}.$

We can use a subclass of generating functions $\Phi =\check{\Phi}$ for which
\begin{equation}
(\partial _{i}\check{\Phi})^{\diamond }=\partial _{i}\check{\Phi}^{\diamond }
\label{aux4a}
\end{equation}%
and get for the left part of the second equation in (\ref{lccondb}), $%
(\partial _{i}-w_{i}\partial _{4})\ln \sqrt{|h_{3}|}=0.$ The first system of
equations in (\ref{lccondb}) can be solved in explicit form if $w_{i}$ are
determined by formulas (\ref{w1b}), and $h_{3}[\tilde{\Phi}]$ and $h_{4}[%
\tilde{\Phi},\tilde{\Phi}^{\diamond }]$ are chosen respectively for any $%
\Upsilon \rightarrow \Lambda .$ We can consider $\tilde{\Phi}=\tilde{\Phi}%
(\ln \sqrt{|h_{4}|})$ for a functional dependence $h_{4}[\tilde{\Phi}[\check{%
\Phi}]].$ This allows us to obtain $w_{i}=\partial _{i}|\tilde{\Phi}|/|%
\tilde{\Phi}|^{\diamond }=\partial _{i}|\ln \sqrt{|h_{4}|}|/|\ln \sqrt{%
|h_{4}|}|^{\diamond }.$ Taking the derivative $\partial _{4}$ on both sides
of these equations, we get
\begin{equation*}
w_{i}^{\diamond }=\frac{(\partial _{i}|\ln \sqrt{|h_{4}|}|)^{\diamond }}{%
|\ln \sqrt{|h_{4}|}|^{\diamond }}-w_{i}\frac{|\ln \sqrt{|h_{4}|}%
|^{^{\diamond }}}{|\ln \sqrt{|h_{4}|}|^{^{\diamond }}}.
\end{equation*}%
If the mentioned conditions are satisfied, we can construct in explicit form
generic off-diagonal configurations with $w_{i}^{^{\diamond }}=(\partial
_{i}-w_{i}\partial _{4})\ln \sqrt{|h_{4}|},$ which is necessary for the zero
torsion conditions.

We need als solve for the conditions $\partial _{k}w_{i}=\partial _{i}w_{k}$
from the second line in (\ref{lccondb}). We find in explicit form the
solutions for such coefficients if
\begin{equation}
\check{w}_{i}=\partial _{i}\check{\Phi}/\check{\Phi}^{^{\diamond }}=\partial
_{i}\widetilde{A},  \label{w1c}
\end{equation}%
with a nontrivial function $\widetilde{A}(x^{k},y^{4})$ depending
functionally on the generating function $\check{\Phi}.$

Finally, we conclude that we generate LC-configurations for a class of
off-diagonal cosmological metric type (\ref{dm1}) for $\Upsilon =\breve{%
\Upsilon}=\Lambda ,$ $\Phi =\check{\Phi}=\tilde{\Phi}$ and $\ _{2}n_{k}=0$
in (\ref{n1b}) which are parameterized by quadratic elements
\begin{equation}
ds^{2}=e^{\psi (x^{k})}[(dx^{1})^{2}+(dx^{2})^{2}]+\frac{\check{\Phi}^{2}}{%
4|\Lambda |}[dy^{3}+(\partial _{k}n(x^{i}))dx^{k}]^{2}-\frac{(\check{\Phi}%
^{^{\diamond }})^{2}}{|\Lambda |\check{\Phi}^{2}}[dt+(\partial _{i}%
\widetilde{A}[\check{\Phi}])dx^{i}]^{2}.  \label{qelgen}
\end{equation}%
We can re-write such solutions in the form (\ref{odfrlwtors}), which
provides us a general procedure of off-diagonal deformations with $\mathring{%
a}(t)\rightarrow a(x^{k},t)$ (see the FLRW metric (\ref{flrw})), resulting
in nonhomogeneous cosmological metrics in GR. Prescribing a function $%
a(x^{k},t),$ a generating function $\check{\Phi}(x^{k},t)$ satisfying the
condition (\ref{aux4a}) and a solution $e^{\psi {(x^{k})}}$ (see (\ref{eq1m}%
)), we respectively compute the v-conformal factor and the polarization
functions for
\begin{equation*}
\widehat{h}_{3}=h_{3}/a^{2}|h_{4}|=\check{\Phi}^{4}/4a^{2}(\check{\Phi}%
^{^{\diamond }})^{2},\ \omega ^{2}=|h_{4}|^{-1}=|\Lambda |\check{\Phi}^{2}/(%
\check{\Phi}^{^{\diamond }})^{2},\ \eta _{i}=a^{-2}e^{\psi },\eta _{3}=%
\mathring{a}^{-2}h_{3}=\check{\Phi}^{2}/4|\Lambda |\mathring{a}^{2},\eta
_{4}=1.
\end{equation*}%
Such coefficients (see data (\ref{polarf})) transform the off-diagonal
cosmological solutions (\ref{qelgen}) into metrics of type (\ref{flrwod}),
\begin{equation}
ds^{2}=a^{2}(x^{k},t)\{[\eta _{1}(x^{k},t)(dx^{1})^{2}+\eta
_{2}(x^{k},t)(dx^{2})^{2}]+\widehat{h}_{3}(x^{k},t)[dy^{3}+(\partial
_{k}n(x^{i}))dx^{k}]^{2}\}-[dt+(\partial _{i}\widetilde{A}[\check{\Phi}%
])dx^{i}]^{2}.  \label{qelgenofd}
\end{equation}%
The dependence on the source $\Lambda $ is contained in explicit form in the
polarization $\eta _{3}$, for instance. This class of effective Einstein
off-diagonal metrics $\mathbf{g}_{\alpha \beta }(x^{k},t)$ define new
nonhomogeneous cosmological solutions in GR as off-diagonal deformations of
the FLRW cosmology. For certain well-defined conditions, one can find limits
$\mathbf{g}_{\alpha \beta }\rightarrow \mathbf{g}_{\alpha \beta }(t,a(t),%
\widehat{h}_{3}(t),\check{\Phi}(t),\eta _{i}(t)).$ This provides explicit
geometric models of nonlinear off-diagonal anisotropic cosmological
evolution which, with respect to N-adapted frames, describe $a(t)$ with
modified re-scaling factors.

\subsection{Effective FLRW cosmology for $f$-modified gravity}

\label{sseflrw}

The anholonomic frame deformation method outlined in previous subsections
can be applied for the generation of off-diagonal cosmological solutions of
field equations of modified gravities, see (\ref{mgtfe}). Redefining the
generating functions via the transforms (\ref{aux2}) and (\ref{aux4a}), $%
\Phi \rightarrow \check{\Phi}\rightarrow \tilde{\Phi},$ we can generate
off-diagonal cosmological configurations with $\ \widehat{\mathbf{R}}%
=4\Lambda ,$ see (\ref{dsours1}) and (\ref{dsours2}). Such parameterizations
of geometric data and sources are possible for certain general conditions
via transforms of N-adapted frames when the action functional functionally
depends on $\Lambda $ and on the effective sources, $\widehat{\mathbf{f}}[%
\widehat{\mathbf{R}}(\Lambda ),\widehat{\mathbf{T}}(\Lambda ),\widehat{%
\mathbf{P}}],$ with $\widehat{\mathbf{P}}(t)=\widehat{\mathbf{R}}_{\alpha
\beta }\widehat{\mathbf{T}}^{\alpha \beta }=-3\mathring{\rho}%
(H^{2}+H^{\diamond })\ $ and $H=a^{\diamond }/a~\ $with scaling factor $a(t)$
taken for some limits of a solution (\ref{odfrlwtors}), or (\ref{qelgenofd}%
). It should be noted that in the variables corresponding to the Levi-Civita
connection $\nabla $ the functional $\widehat{\mathbf{f}}\rightarrow
f(R,T,R_{\mu \nu }T^{\mu \nu })$ describe very general modifications of GR
which in our approach are encoded into a very sophisticated off-diagonal
effective vacuum structure with nontrivial vacuum constants.

We assume that the density of matter $\rho =\mathring{\rho}$ in $\widehat{%
\mathbf{T}}_{\alpha \beta }$ (\ref{dsourc}) is the same as for a standard
FLRW metric (\ref{flrw}) and does not change under off-diagonal deformations
with respect to N-adapted frames. For such configurations, the functions $%
\mathbf{\Theta }_{\alpha \beta }$ and $\mathbf{\Xi }_{\alpha \beta }$ are
parameterized, respectively, as
\begin{equation*}
\mathbf{\Theta }_{\ \beta }^{\alpha }=(p\ -2\Lambda )\delta _{\ \beta
}^{\alpha },\ \mathbf{\Xi }_{\ \beta }^{\alpha }=(2\Lambda ^{2}-p\ \Lambda -%
\frac{1}{2}4\Lambda ^{2})\delta _{\ \beta }^{\alpha }=-p\Lambda \delta _{\
\beta }^{\alpha },
\end{equation*}%
where terms with $\Lambda ^{2}$ compensate each other in 4-d. We can write $%
\widehat{\mathbf{D}}_{\mu }\widehat{\mathbf{T}}_{\alpha \beta }=0,$ $%
\widehat{\mathbf{D}}_{\mu }\ ^{1}\widehat{\mathbf{f}}\sim \partial ^{2}%
\widehat{\mathbf{f}}/\partial \widehat{\mathbf{R}}_{\ldots }^{2}$ $\mathbf{e}%
_{\mu }\Lambda \sim 0,$ and (similarly) $\widehat{\mathbf{D}}_{\mu }\ ^{2}%
\widehat{\mathbf{f}}\sim 0,$ $\widehat{\mathbf{D}}_{\mu }\ ^{3}\widehat{%
\mathbf{f}}\sim 0,$ for \ $\widehat{\mathbf{R}}_{\alpha \beta }\sim $ $%
\widehat{\mathbf{T}}_{\alpha \beta }\sim \Lambda \delta _{\alpha \beta },$ $%
\Lambda =const,$ with respect to corresponding classes of N-adapted frames.
Eqs.~(\ref{mgtfe}) transform into a system of nonholonomic nonlinear PDEs of
type (\ref{eq1b})-(\ref{eq4b}), $\ \widehat{\mathbf{R}}_{\ \ \beta }^{\alpha
}=\ \widehat{\mathbf{\Upsilon }}\delta _{\ \beta }^{\alpha }$, with
effective diagonalized source
\begin{equation}
\ \widehat{\mathbf{\Upsilon }}=\frac{\Lambda }{\ ^{1}\widehat{\mathbf{f}}}+%
\frac{\widehat{\mathbf{f}}}{2\ ^{1}\widehat{\mathbf{f}}}+(2\Lambda -\kappa
^{-2}\Lambda -p)\frac{\ ^{2}\widehat{\mathbf{f}}}{\ ^{1}\widehat{\mathbf{f}}}%
+p\Lambda \frac{\ ^{3}\widehat{\mathbf{f}}}{\ ^{1}\widehat{\mathbf{f}}},
\label{effsource}
\end{equation}%
which can be parameterized with dependencies on $(x^{i},t),$ or on $t.$
These equations can be solved for very general off-diagonal forms, depending
on generating and integration functions, following the procedure outlined in
previous subsections. Redefining the generation function as in (\ref{aux2}),
when an effective cosmological constant $\check{\Lambda}$ is generated from $%
\widehat{\mathbf{\Upsilon }}(x^{i},t)$, one has
\begin{equation}
\check{\Phi}^{2}=\check{\Lambda}^{-1}\left[ \widehat{\Phi }^{2}|\ \widehat{%
\mathbf{\Upsilon }}|+\int dt\ \widehat{\Phi }^{2}|\ \widehat{\mathbf{%
\Upsilon }}|^{\diamond }\right] .  \label{aux2b}
\end{equation}%
Such a generating function defines off-diagonal cosmological solutions of
type (\ref{odfrlwtors}), or (\ref{qelgen}), as solutions of field equations
for an effective (nonholonomic) Einstein space $\mathbf{\check{R}}_{\ \beta
}^{\alpha }=\check{\Lambda}\delta _{\ \beta }^{\alpha }$. In this way, a
geometric method is provided when the (effective or modified) matter sources
transform as $\ \widehat{\mathbf{\Upsilon }}$ (\ref{dsours1}) $\rightarrow $
$\check{\Lambda}$ (\ref{dsours2}) and the gravitational field equations in
modified gravity can be effectively expressed as nonholonomic Einstein
spaces when the d-metric coefficients encode the contributions of $\widehat{%
\mathbf{f}},\ ^{1}\widehat{\mathbf{f}},\ ^{2}\widehat{\mathbf{f}}$ and $\
^{3}\widehat{\mathbf{f}}$ and of the matter sources.

We can consider inverse transforms with $\ \check{\Lambda}\rightarrow
\widehat{\mathbf{\Upsilon }}$ for (\ref{aux2b}) and state that for certain
well-defined conditions of type (\ref{aux4a}) and (\ref{w1c}) we can mimic
both $f$-functional contributions and/or massive gravitational theories \cite%
{voffdmgt,voffdmgt1,voffdmgt2,voffdmgt3}. Here we emphasize that off-diagonal configurations (of vacuum
and non-vacuum types) are possible even if the effective sources from
modified gravity are constrained to be zero.

\section{Off-diagonal modeling of cosmological modified gravity theories}

\label{s4} This section has three goals. The first is to provide a
reconstruction procedure for off-diagonal effective Einstein and modified
gravity cosmological scenarios. The second is to apply these methods in
practice and provide explicit examples related to $f(R)$ gravity and
cosmology. The third goal is to analyze how matter stability problems for $%
f(R)$-theories can be solved by nonholonomic frame transforms and
deformations and imposing non-integrable constraints.

\subsection{Reconstructing nonholonomic general $f(R)$-models}

Let us construct an effective Einstein space which models a quite general
modified gravity theory with $f(R,T,R_{\alpha \beta }T^{\alpha \beta
})=R+F(R_{\alpha \beta }T^{\alpha \beta })+G(T).$ This theory admits a
reconstruction procedure which does not affect the observational constraints
when a realistic evolution is studied \cite{odgom,odgom1,odgom2}. Following the
anholonomic frame deformation method with an auxiliary canonical
d-connection $\widehat{\mathbf{D}},$ the modified gravity (\ref{mgtfe}) is
formulated for
\begin{equation}
\widehat{\mathbf{f}}(\widehat{\mathbf{R}},\widehat{\mathbf{T}},\widehat{%
\mathbf{R}}_{\alpha \beta }\widehat{\mathbf{T}}^{\alpha \beta })=\widehat{%
\mathbf{R}}+\widehat{\mathbf{F}}(\widehat{\mathbf{P}})+\widehat{\mathbf{G}}(%
\widehat{\mathbf{T}}).  \label{aux10}
\end{equation}%
We can self-consistently embed this model into a nonholonomic background
determined by N-adapted frames (\ref{nader}) and (\ref{nadif}) for a generic
off-diagonal solution (\ref{qelgenofd}) with limits $\widehat{\mathbf{D}}%
\rightarrow \nabla $ and $\mathbf{g}_{\alpha \beta }\rightarrow \mathbf{g}%
_{\alpha \beta }(t,a(t),\widehat{h}_{3}(t),\check{\Phi}(t),\eta _{i}(t)).$
With respect to such frames, the nonholonomic FLRW equations are similar to
those found in section III B of \cite{odgom,odgom1,odgom2} (see the second paper for
details on methods of constructing solutions and speculations on the problem
of matter instability).\footnote{%
In section III A of that work, a model with $G(T)=0$ was investigated in
detail. The conclusion was that in order to elaborate a realistic evolution
it is necessary to consider nontrivial values for $G(T).$ In nonholonomic
variables, such term $\widehat{\mathbf{G}}(\widehat{\mathbf{T}})$ allows to
encode $f(R)$ modified theories and related into certain off-diagonal
configurations in GR, which simplifies the solution of the problem of matter
instability (see subsection \ref{ssminst}).}

\subsubsection{$f$-modified off-diagonal FLRW equations}

The effective function $a(t)$ defines in our case off-diagonal cosmological
evolution scenarios which are different from those where $\mathring{a}(t)$
stands for a standard diagonal FLRW cosmology. For $H:=a^{\diamond }/a,$ $\
^{1}\widehat{\mathbf{G}}:=d\widehat{\mathbf{G}}/d\widehat{\mathbf{T}}$ and $%
\ ^{1}\widehat{\mathbf{F}}:=d\widehat{\mathbf{F}}/d\widehat{\mathbf{P}},$ we
have%
\begin{eqnarray}
3H^{2}+\frac{1}{2}\left[ \widehat{\mathbf{f}}+\widehat{\mathbf{G}}%
-3(3H^{2}-H^{\diamond })\ \rho \ ^{1}\widehat{\mathbf{F}}\right] -\rho
(\kappa ^{2}-\ ^{1}\widehat{\mathbf{G}}) &=&0,  \label{ceq} \\
-3H^{2}-2H^{\diamond }-\frac{1}{2}[\widehat{\mathbf{f}}+\widehat{\mathbf{G}}%
-\left( \rho \ ^{1}\widehat{\mathbf{F}}\right) ^{\diamond \diamond
}-4H\left( \rho \ ^{1}\widehat{\mathbf{F}}\right) ^{\diamond }-\left(
3H^{2}+H^{\diamond }\right) \ \rho \ ^{1}\widehat{\mathbf{F}}] &=&0.  \notag
\end{eqnarray}%
An observer is here in a nonholonomic basis determined by $%
N_{i}^{a}=\{n_{i},w_{i}(t)\}$ for a nontrivial off-diagonal vacuum with
effective polarizations $\eta _{\alpha }(t)$, and can test cosmological
scenarios in terms of the redshift $1+z=a^{-1}(t)$ for $P=P(z)$ and $T=T(z),$
with a new \textquotedblleft shift\textquotedblright\ derivative when (for
instance, for a function $s(t)$) $s^{\diamond }=-(1+z)H\partial _{z}.$

The system of two equations (\ref{ceq}) simplifies by extending it to a set
of three equations for four unknown functions $\{\widehat{\mathbf{f}}(z),%
\widehat{\mathbf{G}}(z),\rho (z),\varsigma (z)\}$ with a new variable $%
\varsigma (z):=\ \rho \ ^{1}\widehat{\mathbf{F}},$
\begin{eqnarray}
3H^{2}+\frac{1}{2}[\widehat{\mathbf{f}}(z)+\widehat{\mathbf{G}}(z)]-\frac{3}{%
2}[3H^{2}-(1+z)H(\partial _{z}H)]\ \varsigma (z) \frac{3}{2}%
H^{2}(1+z)\partial _{z}\varsigma (z)-\kappa ^{2}\rho (z)&=&0,  \notag \\
-3H^{2}+(1+z)H(\partial _{z}H)-\frac{1}{2}\{\widehat{\mathbf{f}}(z)+\widehat{%
\mathbf{G}}(z)-[3H^{2}-(1+z)H(\partial _{z}H)]\varsigma (z) &&  \notag \\
+[3(1+z)H^{2}-(1+z)H(\partial _{z}H)]\partial _{z}\varsigma
(z)+(1+z)^{2}\partial _{zz}^{2}\varsigma (z)\}&=&0,  \notag \\
(\partial _{z}\ ^{1}\widehat{\mathbf{F}})\ \varsigma (z)-\rho (z)\ (\partial
_{z}\ \widehat{\mathbf{f}})&=&0.  \label{ceq1}
\end{eqnarray}%
Here, by re-scaling the generating function, we have fixed the condition $%
\partial _{z}\ ^{1}\widehat{\mathbf{G}}(z)=0$. Such a nontrivial term must
be considered if one wants to transform $\widehat{\mathbf{f}}$ into a
standard theory $f(R,T,R_{\alpha \beta }T^{\alpha \beta })$. The functional $%
\widehat{\mathbf{G}}(\widehat{\mathbf{T}})$, in both holonomic and
nonholonomic forms, encodes a new degree of freedom for the evolution of the
energy-density of type
\begin{equation}
\rho =\rho _{0}a^{-3(1+\varpi )}=\rho _{0}(1+z)a^{3(1+\varpi )},
\label{densas}
\end{equation}%
which is taken for the dust matter approximation $\varpi $ when the
evolution reduces to $\rho \sim (1+z)^{3}.$ For the assumption that such an
evolution can be considered with respect to N-adapted frames, the solutions
of (\ref{ceq1}) are determined by data $\{\widehat{\mathbf{f}}(z),\widehat{%
\mathbf{G}}(z),\varsigma (z)\}$ by replacing the second and third equations
into the first one and obtaining a single fourth-order equation for $%
\widehat{\mathbf{f}}(z).$

\subsubsection{Reconstructing $\widehat{\mathbf{f}}$-models and effective
Einstein spaces}

The reconstruction procedure is restricted to fluids without pressure when
such approximation is considered locally with N-adapted frames and the
expressions (\ref{auxaa}) for $(\mathring{a},\mathring{H},\mathring{\rho})$
are re-defined in terms of $(a,H,\rho )$; data are written with a script
\textquotedblleft 0\textquotedblright\ if $z=z_{0},$ with $\xi =\kappa
^{2}\rho _{0}/3H_{0}^{2}$. One should not confused, e.g., $\mathring{H}$ and
$H_{0}$, because these values are computed for \textit{different} FLRW
solutions, with $\mathring{a}(z)$ determined for a diagonal configuration
and $a(z)$ for an off-diagonal one, respectively. We can express
\begin{equation*}
\widehat{\mathbf{T}}=\widehat{\mathbf{T}}_{\ \alpha }^{\alpha }=-\xi \frac{%
3H_{0}^{2}}{\kappa ^{2}}(1+z)^{3}\mbox{ and }\widehat{\mathbf{P}}=\widehat{%
\mathbf{R}}_{\alpha \beta \ }\widehat{\mathbf{T}}_{\ }^{\alpha \beta }=-3\xi
\frac{3H_{0}^{2}}{\kappa ^{2}}(1+z)^{3}[H^{2}-(1+z)H(\partial _{z}H)].
\end{equation*}%
Following the approach outlined in Sect.~IIIB of \cite{odgom,odgom1,odgom2}, we introduce
the parameterizations
\begin{equation}
\widehat{\mathbf{F}}(\widehat{\mathbf{P}})=H_{0}^{2}\mathbf{\check{F}}(%
\mathbf{\check{P}})\mbox{ and }\widehat{\mathbf{G}}(\widehat{\mathbf{T}}%
)=H_{0}^{2}\mathbf{\check{G}}(\mathbf{\check{T}}),  \label{aux9}
\end{equation}%
where $\mathbf{\check{P}=}\widehat{\mathbf{P}}/P_{0}$ and $\mathbf{\check{T}=%
}$ $\widehat{\mathbf{T}}/T_{0},$ for $P_{0}=-9H_{0}^{4}\xi /\kappa ^{2}$ and
$T_{0}=-3H_{0}^{2}\xi /\kappa ^{2}.$ In correspondingly N-adapted variables,
the off-diagonal cosmological solutions can be associated with a class of de
Sitter (dS) solutions with effective cosmological constant $\check{\Lambda}$
(see (\ref{aux2b})), where $H(z)=\check{H}_{0}$ results in $\mathbf{\check{P}%
}=\mathbf{\check{T}}=(1+z)^{3}$ for the energy-density (\ref{densas}). In
these variables, the solutions of (\ref{ceq1}) can be written as
\begin{eqnarray}
\mathbf{\check{F}} &=&c_{1}\mathbf{\check{P}}^{b_{1}}+\mathbf{\check{P}}%
^{b_{2}/3}[c_{2}\cos (\frac{b_{3}}{3}\ln \mathbf{\check{P}})+c_{3}\sin (%
\frac{b_{3}}{3}\ln \mathbf{\check{P}})]+c_{4}+3\xi \mathbf{\check{P},}
\label{aux8} \\
\mathbf{\check{G}} &=&\tilde{c}_{1}\mathbf{\check{T}}^{b_{1}}+\mathbf{\check{%
T}}^{b_{2}/3}[\tilde{c}_{2}\cos (\frac{b_{3}}{3}\ln \mathbf{\check{T}})+%
\tilde{c}_{3}\sin (\frac{b_{3}}{3}\ln \mathbf{\check{P}})]+\tilde{c}%
_{4}-3\xi \mathbf{\check{T},}  \notag
\end{eqnarray}%
being the constants $b_{1}=-1.327,b_{2}=3.414$ and $b_{3}=1.38.$ The values $%
c_{1},c_{2},c_{3}$ and $c_{4}$ are integration constants, and the second set
of constants $\tilde{c}_{1},\tilde{c}_{2},\tilde{c}_{3}$ and $\tilde{c}_{4}$
can be expressed via such integration constants, and $b_{1},b_{2}$ and $%
b_{3}.$ We omit explicit formulas because for general solutions they can be
included in certain generating or integration functions for the modified
gravity equations and ultimately related to real observation data for the
associated cosmological models.

For off-diagonal configurations, the $\widehat{\mathbf{f}}(\widehat{\mathbf{R%
}},\widehat{\mathbf{T}},$ $\widehat{\mathbf{R}}_{\alpha \beta }\widehat{%
\mathbf{T}}^{\alpha \beta })$ gravity positively allows for dS solutions in
presence of non-constant fluids, not only due to the term $\widehat{\mathbf{P%
}}=\widehat{\mathbf{R}}_{\alpha \beta }\widehat{\mathbf{T}}^{\alpha \beta }$
in (\ref{mgts}), and respective gravitational field and cosmological
equations. This is possible also because of the off-diagonal nonlinear
gravitational interactions in the effective gravitational models. It should
be emphasized that the reconstruction procedure elaborated in \cite{odgom,odgom1,odgom2},
see also references therein, can be extended to more general classes of
modified gravity theories, to Finsler like theories and the ensuing
cosmological models \cite{voffdmgt,voffdmgt1,voffdmgt2}. Introducing (\ref{aux8}) and (\ref{aux9}%
) into (\ref{aux10}), we reconstruct a function $\widehat{\mathbf{f}}=%
\widehat{\mathbf{R}}+\widehat{\mathbf{F}}(\widehat{\mathbf{P}})+\widehat{%
\mathbf{G}}(\widehat{\mathbf{T}}).$ As a result, we can associate an
effective matter source $\ \widehat{\mathbf{\Upsilon }}$ (\ref{effsource}),
which allows the definition of a corresponding generating function $\check{%
\Phi}$ (\ref{aux2b}) (see also $\Phi $ and (\ref{aux2})). Finally, we can
reconstruct an off-diagonal cosmological solution with nonholonomically
induced torsion of type (\ref{odfrlwtors}) or to model a similar
cosmological metric for LC configurations (\ref{qelgen}) (equivalently, (\ref%
{qelgenofd})).

\subsection{How specific $f(R)$ gravities and the FLRW cosmology are encoded
in nonholonomic deformations?}

It is well known that any FLRW cosmology can be realized in a specific $f(R)$
gravity (see Ref.~\cite{odintsplb,odintsplb1} and, for further generalizations, \cite%
{odgom,odgom1,odgom2}).\footnote{%
We use a system of notations different from that article;\ here, e.g., $N$
in used for the N-connection and we work with nonholonomic geometric objects.%
} In this subsection we analyze two examples of reconstruction of $f(R)$%
-gravities where the "e-folding'' variable $\zeta :=\ln a/a_{0}=-\ln (1+z) $
is used instead of the cosmological time $t$ and in related nonholonomic
off-diagonal deformations. For such models, we consider $\widehat{\mathbf{f}}%
=\widehat{\mathbf{f}}(\widehat{\mathbf{R}})$ in (\ref{mgts}), use $\
\widehat{\mathbf{\Upsilon }}(x^{i},\zeta )=\Lambda /\ ^{1}\widehat{\mathbf{f}%
}+\widehat{\mathbf{f}}/2\ ^{1}\widehat{\mathbf{f}}$ instead of (\ref%
{effsource}) and introduce these values in Eq.~(\ref{aux2b}), which can be
parameterized with dependencies on $(x^{i},\zeta )$ (in particular, only on $%
\zeta $), $\check{\Phi}^{2}=\check{\Lambda}^{-1}[\widehat{\Phi }^{2}|\
\widehat{\mathbf{\Upsilon }}|+\int d\zeta \ \widehat{\Phi }^{2}\partial
_{\zeta }|\ \widehat{\mathbf{\Upsilon }}|],$ when $\partial _{\zeta
}=\partial /\partial \zeta $ with $s^{\diamond }=H\partial _{\zeta }s$ for
any function $s.$ The matter energy density $\rho $ is taken as in (\ref%
{ceq1}).

We restrict ourselves to N-adapted frames, (\ref{nader}) and (\ref{nadif}),
determined by an off-diagonal cosmological solution of the (modified)
gravitational field equations, and can repeat all computations leading to
Eqs.~(2)-(7) in \cite{odintsplb,odintsplb1} and prove that a modified gravity with $%
\widehat{\mathbf{f}}(\widehat{\mathbf{R}})$ realizes the FLRW cosmological
model. Such solutions depend on the above source type $\widehat{\mathbf{%
\Upsilon }}(x^{i},\zeta )$ and generating function $\check{\Phi}(x^{i},\zeta
)$; also the nonholonomic background can be modeled to be nonhomogeneous
(via $w_{i}$ and $n_{i}$ depending respectively on $x^{i}$ and $\zeta ,$ or
only on $\zeta ).$ The off-diagonal analog of the field equation
corresponding to the first FLRW equation is {\small \
\begin{equation*}
\widehat{\mathbf{f}}(\widehat{\mathbf{R}})=(H^{2}+H\ \partial _{\zeta
}H)\partial _{\zeta }[\widehat{\mathbf{f}}(\widehat{\mathbf{R}})]-36H^{2}%
\left[ 4H+(\partial _{\zeta }H)^{2}+H\partial _{\zeta \zeta }^{2}H\right]
\partial _{\zeta \zeta }^{2}[\widehat{\mathbf{f}}(\widehat{\mathbf{R}}%
\mathbf{)]+}\kappa ^{2}\rho .
\end{equation*}%
} In terms of an effective quadratic Hubble rate, $q(\zeta ):=H^{2}(\zeta ),$
and considering that $\zeta =\zeta (\widehat{\mathbf{R}})$ for certain
parameterizations, this equation yields {\small
\begin{equation}
\widehat{\mathbf{f}}(\widehat{\mathbf{R}})=-18q(\zeta (\widehat{\mathbf{R}}%
))[\partial _{\zeta \zeta }^{2}q(\zeta (\widehat{\mathbf{R}}))+4\partial
_{\zeta }q(\zeta (\widehat{\mathbf{R}}))]\frac{d^{2}\widehat{\mathbf{f}}(%
\widehat{\mathbf{R}})}{d\widehat{\mathbf{R}}^{2}}+6[q(\zeta (\widehat{%
\mathbf{R}}))+\frac{1}{2}\partial _{\zeta }q(\zeta (\widehat{\mathbf{R}}))]
\frac{d\widehat{\mathbf{f}}(\widehat{\mathbf{R}})}{d\widehat{\mathbf{R}}}%
+2\rho _{0}a_{0}^{-3(1+\varpi )}a^{-3(1+\varpi )\zeta (\widehat{\mathbf{R}}%
)}.  \label{f1gen}
\end{equation}%
} We can construct an off-diagonal cosmological model with metrics of type (%
\ref{odfrlwtors}) and nonholonomically induced torsion (when $t\rightarrow
\zeta )$ if a solution $\widehat{\mathbf{f}}(\widehat{\mathbf{R}})$ is used
for computing $\widehat{\mathbf{\Upsilon }}$ and $\check{\Phi}.$ Modeling
such nonlinear systems we can consider solutions of the field equations for
an effective (nonholonomic) Einstein space $\mathbf{\check{R}}_{\ \beta
}^{\alpha }=\check{\Lambda}\delta _{\ \beta }^{\alpha }$, when certain terms
of type $d\widehat{\mathbf{f}}(\mathbf{\check{R}})/d\mathbf{\check{R}}$ and
higher derivatives vanish for a functional dependence $\widehat{\mathbf{f}}(%
\check{\Lambda})$ with $\partial _{\zeta }\check{\Lambda}=0.$ The
nonholonomic cosmological evolution is determined by off-diagonal
coefficients of the metrics and by certain non-explicit relations for the
functionals variables, like $q(\zeta (\widehat{\mathbf{R}}(\check{\Lambda}%
))) $ and (effective/modified) matter sources transform as $\ \widehat{%
\mathbf{\Upsilon }}$ (\ref{dsours1}) $\rightarrow $ $\check{\Lambda}$ (\ref%
{dsours2}).

LC-configurations can be modeled by off-diagonal cosmological metrics of
type (\ref{qelgen}) when the zero torsion conditions (\ref{lccondb}) are
satisfied. We obtain a standard expression (see \cite{odintsplb,odintsplb1}) for the
curvature of $\nabla ,$
\begin{equation}
R=3\partial _{\zeta }q(\zeta )+12q(\zeta ),  \label{cscurv}
\end{equation}
if the polarization or generating functions for (\ref{qelgen}) and the
solutions of (\ref{f1gen}) are taken for diagonal configurations.

\subsubsection{Off-diagonal encoding of $f(R)$ gravity and reproduction of
the ${\Lambda}$CDM-era}

We here provide an example of reconstruction of models of $f(R)$ gravity and
nonholonomically deformed GR when both reproduce the ${\Lambda}$CDM era. For
simplicity, we do not consider a real matter source (if such a source
exists, it can be easily encoded into a nontrivial vacuum structure with
generic off-diagonal contributions).

With respect to correspondingly N-adapted frames and for $a(\zeta )$ and $%
H(\zeta )$ determined by an off-diagonal solution (\ref{odfrlwtors}), with
nonholonomically induced torsion, or (\ref{qelgenofd}), for
LC-configurations, the FLRW equation for ${\Lambda}$CDM cosmology is given
by
\begin{equation}
3\kappa ^{-2}H^{2}=3\kappa ^{-2}H_{0}^{2}+\rho _{0}a^{-3}=3\kappa
^{-2}H_{0}^{2}+\rho _{0}a_{0}^{-3}e^{-3\zeta }.  \label{aux11}
\end{equation}%
This equation looks similar to the one for Einstein gravity for diagonal
configurations but contains values determined, in general, for other classes
of models with off-diagonal interactions. Thus, $H_{0} $ and $\rho _{0}$ are
fixed to be certain constant values, after the coefficients of off-diagonal
solutions are found, and for an approximation were the dependencies on $%
(x^{i},\zeta )$ are changed into dependencies on $\zeta $ (via a
corresponding re-definition of the generating functions and the effective
sources). We can relate the first term on the rhs to an effective
cosmological constant $\Lambda $ (\ref{dsours2}), which in our approach
appears via a re-definition (\ref{aux2}). The second term in the formula
describes, in general, an inhomogeneous distribution of cold dark mater
(CDM) with respect to N-adapted frames. In order to keep the similarity with
the diagonalizable cosmological models in GR we can choose these integration
constants for $\Lambda =12H_{0}^{2}$ to survive in the limit $%
w_{i},n_{i}\rightarrow 0.$ It should be noted that such limit must be
computed for ``nonlinear'' nonholonomic constraints via generating functions
and effective sources.

Using (\ref{aux11}), the effective quadratic Hubble rate and the modified
scalar curvature, $\widehat{\mathbf{R}}$, are computed to be, respectively,
\begin{equation*}
q(\zeta ) := H_{0}^{2}+\kappa ^{2}\rho _{0}a_{0}^{-3}e^{-3\zeta }\mbox{ and }
\widehat{\mathbf{R}} = 3\partial _{\zeta }q(\zeta )+12q(\zeta
)=12H_{0}^{2}+\kappa ^{2}\rho _{0}a_{0}^{-3}e^{-3\zeta }.
\end{equation*}
These functional formulas can be used for the dependencies on $\widehat{%
\mathbf{R}}$ if a necessary re-definition of the generation functions, or an
approximation $(x^{i},\zeta )\rightarrow \zeta $ is performed. Expressing
\begin{equation*}
3\zeta =-\ln [\kappa ^{-2}\rho _{0}^{-1}a_{0}^{3}(\widehat{\mathbf{R}}%
-12H_{0}^{2})]\mbox{ and }X:=-3+\widehat{\mathbf{R}}/3H_{0}^{2},
\end{equation*}%
we obtain from Eq.~(\ref{f1gen})%
\begin{equation}
X(1-X)\frac{d^{2}\widehat{\mathbf{f}}}{dX^{2}}+[\chi _{3}-(\chi _{1}+\chi
_{2}+1)X]\frac{d\widehat{\mathbf{f}}}{dX}-\chi _{1}\chi _{2}\widehat{\mathbf{%
f}}=0,  \label{gauss}
\end{equation}%
for certain constants, for which $\chi _{1}+\chi _{2}=\chi _{1}\chi
_{2}=-1/6 $ and $\chi _{3}=-1/2.$ The solutions of this equation with
constant coefficients and for $R$ (\ref{cscurv}) were found in \cite%
{odintsplb,odintsplb1} as Gauss hypergeometric function, denoted there by $\widehat{%
\mathbf{f}}=F(X):=F(\chi _{1},\chi _{2},\chi _{3};X),$ as (for some
constants $A$ and $B $) \
\begin{equation*}
F(X)=AF(\chi _{1},\chi _{2},\chi _{3};X)+BX^{1-\chi _{3}}F(\chi _{1}-\chi
_{3}+1,\chi _{2}-\chi _{3}+1,2-\chi _{3};X).
\end{equation*}%
This provides a proof of the statement that $f(R)$ gravity can indeed
describe ${\Lambda}$CDM scenarios without the need of an effective
cosmological constant. Working with auxiliary connections of the type $%
\widehat{\mathbf{D}},$ we can generalize the constructions to off-diagonal
configurations and various classes of modified gravity theories, where $A,B$
and $\chi _{1},\chi _{2},\chi _{3}$ are appropriate functions of the $h $
coordinates. For instance, reconstruction procedures for Finsler like
theories and cosmology models on tangent and Lorentz bundles, and
bi-metric/massive gravity models are given in \cite{voffdmgt,voffdmgt1,voffdmgt2,voffdmgt3,massgr,massgr1}.

Having chosen $\widehat{\mathbf{f}}=F(X)$ for a modified gravity, we can go
further and mimic an off-diagonal configuration when $\widehat{\mathbf{f}}=%
\widehat{\mathbf{f}}(\widehat{\mathbf{R}})$ is introduced in (\ref{mgts})
and the source $\ \widehat{\mathbf{\Upsilon }}(x^{i},\zeta )=\Lambda /\ ^{1}%
\widehat{\mathbf{f}}+\widehat{\mathbf{f}}/2\ ^{1}\widehat{\mathbf{f}}$ is
considered instead of (\ref{effsource}) and (\ref{aux2b}) for $\check{\Phi}%
^{2}=\check{\Lambda}^{-1}[\widehat{\Phi }^{2}|\ \widehat{\mathbf{\Upsilon }}%
|+\int d\zeta \ \widehat{\Phi }^{2}\partial _{\zeta }|\ \widehat{\mathbf{%
\Upsilon }}|].$ Nevertheless, recovering nonhomogeneous modified
cosmological models cannot be completed for general re-parameterized
dependencies on $(x^{i},\zeta )$ (in particular, only on $\zeta $). This
distinguishes explicitly the modified gravity theories of type $f(R)$ from
those generated by nonholonomic deformations. For certain homogeneity
conditions, we can state an equivalence of some classes of gravities and
cosmological models, or analyze their alternative physical implications. But
a complete recovering is only possible if all generating and integration
functions and the effective sources are correlated with certain observable
cosmological effects and further approximations and re-definitions in terms
of constant parameters and functionals depending on a time-like coordinate
can be effectively performed.

In general, a modified gravity theory is not transformed completely into a
nonholonomic off-diagonal Einstein manifold; an overlap between certain
classes of solutions and cosmological and quantum gravity models may exist
(see the constructions and discussion in \cite{vhl,vhl1,vhl2,vhl3}, related to \cite{covhl,covhl1,covhl2,covhl3}%
9. A rigorous theoretical analysis of various types of classical and quantum
corrected solutions is necessary and new experimental data are compulsory in
order to conclude that an orthodox paradigm with nonholonomic off-diagonal
sophisticate (non) vacuum configurations in GR may be enough for elaborating
a final, viable cosmological model and to perform a variant of the effective
covariant anisotropic quantization, what is indeed missing, up to now. In a
more radical case, we will have to modify substantially the GR theory and a
number of additional issues may arise on the status of off-diagonal
solutions, on methods of quantization and recovering formalism, on stability
issues and some other, in the search for a matching solution.

\subsubsection{Nonholonomic configurations mimicking phantom and non-phantom
matter in $f(R)$ gravity}

The anholonomic frame deformation method allows to reconstruct off-diagonal
configurations modeling $f(R)$ gravity and cosmology with non-phantom or
phantom matter in GR. With respect to N-adapted frames in an off-diagonal
(modified, or not) gravitational background, the FLRW equations can be
written as
\begin{equation}
3\kappa ^{-2}H^{2}=\ _{s}\rho (x^{k})a^{-c(x^{k})}+\ _{p}\rho
(x^{k})a^{c(x^{k})},  \label{aux11a}
\end{equation}%
where $a(x^{k},\zeta )$ and $H(x^{k},\zeta )$ are determined by a solution (%
\ref{odfrlwtors}), or (\ref{qelgenofd}). For re-parameterizations or
approximations with $(x^{i},\zeta )\rightarrow \zeta ,$ we assume that the
positive functions$\ _{s}\rho (x^{k}),\ _{p}\rho (x^{k})$ and $c(x^{k})$ can
be considered. The first term on the rhs dominates for small $a$ in the
early universe, as in GR with non-phantom matter described by an EoS
parameter $w=-1+c/3>-1.$ Introducing $q(x^{k},\zeta ):=H^{2}(x^{k},\zeta )$
and the respective functionals $\ _{s}q:=\frac{\kappa ^{2}}{3}\ _{s}\rho
a_{0}^{-c}$ and $\ _{p}q:=\frac{\kappa ^{2}}{3}\ _{p}\rho a_{0}^{c},$ for $%
q=\ _{s}qe^{-c\zeta }+$ $\ _{p}qe^{c\zeta },$ in $\widehat{\mathbf{R}}%
=3\partial _{\zeta }q(\zeta )+12q(\zeta ),$ we find
\begin{equation}
e^{c\zeta }=\left\{
\begin{array}{cc}
\lbrack \widehat{\mathbf{R}}\pm \sqrt{\widehat{\mathbf{R}}^{2}-4(144-9c^{2})}%
]/6(4+c), & \mbox{ for }c\neq 4; \\
\widehat{\mathbf{R}}/24, & \mbox{ for }c=4.%
\end{array}%
\right.  \label{aux12}
\end{equation}
The non-phantom matter may correspond to the case $c=4$ in (\ref{aux12}),
including radiation with $w=1/3.$ Eq.~(\ref{aux11a}) transform into a
functional equation on $Y$ determined by changing the functional variable $%
\widehat{\mathbf{R}}^{2}=-576\ _{s}q\ \ _{p}q\ Y,$ $4Y(1-Y)\frac{d^{2}%
\widehat{\mathbf{f}}}{dY^{2}}+(3+Y)\frac{d\widehat{\mathbf{f}}}{dY}-2%
\widehat{\mathbf{f}}=0$. This is again a functional variant (if we consider
dependencies on $x^{k}$) of the generating Gauss' hypergeometric function,
similarly to (\ref{gauss}), which can be solved in explicit form.

For the case $c\neq 4$ in (\ref{aux12}), we come to models with phantom-like
dominant components. A similar procedure as for deriving Eqs.~(22) and (23)
in \cite{odintsplb,odintsplb1}, results in a functional generalization of the Euler
equation, namely%
\begin{equation*}
\widehat{\mathbf{R}}^{2}\frac{d^{2}\widehat{\mathbf{f}}(\widehat{\mathbf{R}})%
}{d\widehat{\mathbf{R}}^{2}}+A\widehat{\mathbf{R}}\frac{d\widehat{\mathbf{f}}%
(\widehat{\mathbf{R}})}{d\widehat{\mathbf{R}}}+B\widehat{\mathbf{f}}(%
\widehat{\mathbf{R}})=0,
\end{equation*}%
for some coefficients $A=-H_{0}(1+H_{0})$ and $B=(1+2H_{0})/2,$ for $%
H_{0}=1/3(1+\ _{ph}w)$. Here we consider, for simplicity, homogenous limits
and approximations $H^{2}(t)=\frac{\kappa ^{2}}{3}_{ph}\rho $ for the
phantom EoS fluid-like states, $_{ph}p=\ _{ph}w\ _{ph}\rho ,$ with $\
_{ph}w<-1.$ In both cases, with a trivial or a nontrivial nonholonomically
induced torsion, there are solutions of the nonholonomic Euler equations
above which can be expressed in the form $\widehat{\mathbf{f}}(\widehat{%
\mathbf{R}})=C_{+}\widehat{\mathbf{R}}^{m_{+}}+C_{-}\widehat{\mathbf{R}}%
^{m_{-}}$, for some integration constants $C_{\pm }$ and $2m_{\pm }=1-A\pm
\sqrt{(1-A)^{2}-4B}.$ This reproduces with respect to N-adapted frames the
phantom dark energy cosmology with a behavior of the type $%
a(t)=a_{0}(t_{s}-t)^{-H_{0}},$ where $t_{s} $ is the so-called Rip time. If
the generating functions for the off-diagonal cosmological solutions are
chosen in a way such that the N-connection coefficients $w_{i}$ and $n_{i}$
transform to zero, the solutions describe universes which end at a Big Rip
singularity during $t_{s}. $ Additionally to the former result that in the $%
f(R)$ theory no phantom fluid is needed, we conclude that for off-diagonal
configurations we can effectively model such locally anisotropic
cosmological configurations.

\subsubsection{On nonholonomic constraints and non-conservation of the
effective energy-momentum tensor}

One can encode and effectively model various types of cosmological solutions
for modified gravity theories with $f(R)$ and/or $f(R,T,R_{\alpha \beta
}T^{\alpha \beta })$ functionals and their nonholonomic deformations. The
cosmological reconstruction procedures can be elaborated for various types
of viable modified gravity which may pass, or not, local gravitational tests
and explain observational data for accelerating cosmology, dark energy and
dark matter interactions \cite{odgom,odgom1,odgom2,odintsplb,odintsplb1,massgr,massgr1,voffdmgt,voffdmgt1,voffdmgt2,voffdmgt3}.
Nevertheless, these theories exhibit certain specific problems as
non-conservation of the energy-momentum tensors for the effective or
physical matter fields.

Let us discuss the ``non-conservation'' issue which is related to the
nonholonomic deformations of GR. Even in the case $\nabla \rightarrow $ $%
\widehat{\mathbf{D}}=\nabla +\widehat{\mathbf{Z}},$ we have the condition $%
\widehat{\mathbf{D}}_{\gamma }\widehat{\mathbf{T}}_{\alpha \beta }\neq 0$,
which is a typical one for nonholonomic (subjected to non-integrable
constraints) mechanical or classical field theories. In Lagrange mechanics,
for instance, the issue of nonholonomic restrictions is solved by
introducing additional integration constants. In such cases, the
conservation laws should be re-considered by taking into account various
classes of non-dynamical functions. Because the distortion tensor $\widehat{%
\mathbf{Z}}$ is completely defined by the data $(\widehat{\mathbf{g}},%
\mathbf{N}),$ we can compute in unique form the value $\widehat{\mathbf{D}}%
_{\gamma }\widehat{\mathbf{T}}_{\alpha \beta }$ and relate this to the fact
that, in general, $\widehat{\mathbf{R}}_{\alpha \beta }$ is not symmetric.
This is a consequence of a nonholonomically induced torsion. It is
convenient to work with such nonholonomic variables $(\widehat{\mathbf{g}},%
\mathbf{N,}\widehat{\mathbf{D}})$ in order to apply the anholonomic frame
deformation method and decouple certain modified gravitational equations and
generate off-diagonal solutions of (\ref{nheeq}). After this general,
integral variety of solutions has been found, one can re-define the
generating functions and sources in order to generate LC-configurations, as
was shown in section \ref{sslc}. This way, the problem of
``non-conservation'' of effective and physical $\widehat{\mathbf{T}}_{\alpha
\beta }$ can be solved by encoding into generic off-diagonal configurations
with effective conservation laws which are similar to GR.

In explicit form, we explain how the ``non-conservation'' problem can be
solved for off-diagonal solutions with one Killing symmetry in the framework
of $f(R,T)$ theories generalizing certain constructions from \cite{alvar}.
Following a similar procedure as in Sect.~II of that work, but using the
operator $\widehat{\mathbf{D}}$ instead of $\nabla ,$ for $\widehat{\mathbf{f%
}}=\widehat{\mathbf{f}}(\widehat{\mathbf{R}},\widehat{\mathbf{T}})$, and
considering an N-adapted parametrization of the effective source $\widehat{%
\mathbf{\Upsilon }}=const$ $,$ we prove that
\begin{equation}
(1+\frac{\kappa ^{2}}{\ ^{2}\widehat{\mathbf{f}}})\widehat{\mathbf{D}}%
^{\alpha }\widehat{\mathbf{T}}_{\alpha \beta }=\frac{1}{2}\mathbf{g}_{\alpha
\beta }\widehat{\mathbf{D}}^{\alpha }\widehat{\mathbf{T}}-(\widehat{\mathbf{T%
}}_{\alpha \beta }+\widehat{\Theta }_{\alpha \beta })\widehat{\mathbf{D}}%
^{\alpha }\ln (\ ^{2}\widehat{\mathbf{f}})-\widehat{\mathbf{D}}^{\alpha }%
\widehat{\Theta }_{\alpha \beta }.  \label{divemt}
\end{equation}%
In these equations the values $\ ^{2}\widehat{\mathbf{f}}:=\partial \widehat{%
\mathbf{f}}/\partial \widehat{\mathbf{T}}$ and $\widehat{\Theta }_{\alpha
\beta }=-2\widehat{\mathbf{T}}_{\alpha \beta }-p\mathbf{g}_{\alpha \beta }$
are used, with an energy-momentum tensor (\ref{dsourc}) for nonholonomic
flows of a perfect fluid. In general, $\widehat{\mathbf{D}}^{\alpha }%
\widehat{\mathbf{T}}_{\alpha \beta }\neq 0$ even for nonholonomic
deformations of GR. Nevertheless, we can consider a subclass of off-diagonal
configurations in $\widehat{\mathbf{f}}(\widehat{\mathbf{R}},\widehat{%
\mathbf{T}})$ gravity when $\ \widehat{\mathbf{\Upsilon }}$ (\ref{dsours1}) $%
\rightarrow $ $\check{\Lambda}$ (\ref{dsours2}) and $\check{\Phi}^{2}=\check{%
\Lambda}^{-1}[\widehat{\Phi }^{2}|\ \widehat{\mathbf{\Upsilon }}|+\int
d\zeta \ \widehat{\Phi }^{2}\partial _{\zeta }|\ \widehat{\mathbf{\Upsilon }}%
|]$ (\ref{aux2b}) result in $\widehat{\mathbf{f}}\rightarrow \mathbf{\check{f%
}=\check{R}}$ and effective $\mathbf{\check{R}}_{\ \beta }^{\alpha }=\check{%
\Lambda}\delta _{\ \beta }^{\alpha }$ which admit LC-solutions with zero
torsion. For such nonholonomic distributions with $\widehat{\mathbf{D}}%
\rightarrow \nabla ,$ $\widehat{\mathbf{D}}^{\alpha }\widehat{\mathbf{T}}%
_{\alpha \beta }\rightarrow \check{\nabla}\check{\Lambda}=0$ and all terms
on the lhs of (\ref{divemt}) vanish, because they are nonholonomically
equivalent to functionals o f the effective cosmological constant $\check{%
\Lambda}.$ Such conditions are satisfied in correspondingly N-adapted frames
and for canonical d-connections. The equations (\ref{divemt}) generalize to
nonholonomic forms the similar ones derived for the Levi-Civita connection $%
\nabla$ (see Eq.~(10) in Ref.~\cite{alvar}).

The already mentioned problem of ``non-conservation'' becomes worse for
general $f(R,T,R_{\alpha \beta }T^{\alpha \beta })$ theories. Even if there
are certain special cases where it can be solved \cite{odgom,odgom1,odgom2,odintsplb,odintsplb1}, it
is the case that no solution can be found in general form. Surprisingly, the
anholonomic frame deformation method also suggests a procedure for selecting
off-diagonal configurations which can mimic modified gravity theories in a
self-consistent way, admitting effective torsions completely determined by
observational data $(\widehat{\mathbf{f}},\ \widehat{\mathbf{\Upsilon }},%
\widehat{\mathbf{g}},\mathbf{N,}\widehat{\mathbf{D}}),$ and/or by
constraints to LC-configurations. By redefining the generating functions in
the form $\ \widehat{\mathbf{\Upsilon }}$ (\ref{dsours1}) $\rightarrow $ $%
\check{\Lambda}$ (\ref{dsours2}), we get the possibility to consider
effective sources and off-diagonal Einstein spaces for which the divergence
and non-conservation problem become similar to those in GR or (for more
geometrically complex configurations) to those in viable $f(R)$ models.
Finally, we emphasize the fact that in this way we do not find a general
solution for all $f(R,...)$-modified theories but only for those models
which admit an encoding in an effective GR system and a general decoupling
via the nonholonomic deformations.

\subsection{Nonholonomic constraints and matter instability}

\label{ssminst} There is another serious problem in modified gravities which
is related to possible matter instabilities related to modifications of the
gravitational actions. Even tiny modifications of GR may make the new model
to posses unstable interior solutions (see, e.g., \cite{minst,minst1}). It was
demonstrated however that there are viable $f(R)$ theories (with
appropriated choices of the functional) where such instabilities may not
occur \cite{revfmod,revfmod01,revfmod02,revfmod03,revfmod04,revfmod05,
revfmod06,revfmod07,revfmod08,revfmod09,revfmod10,revfmod11,revfmod12,revfmod13,stabodin}. In this section, we apply the anholonomic
frame deformation method to more general $f(R,T,R_{\alpha \beta }T^{\alpha
\beta })$ theories. The corresponding field equations are very difficult to
solve even in a linear approximation \cite{odgom,odgom1,odgom2}, if we work in coordinate
frames and general functionals. In the nonholonomic variable formalism, the
gravitational field equations in modified gravity theories posses the
decoupling property exhibited above, which allows to encode $f(R,...)$%
-modifications into off-diagonal nonholonomic configurations for the
effective Einstein manifolds.

For a stability analysis, the trace equations where (\ref{mgtfe}) are
multiplied by $\mathbf{g}^{\mu \nu }$ are to be considered, namely%
\begin{equation}
-2\widehat{\mathbf{f}}+(\widehat{\mathbf{R}}\ +3\widehat{\mathbf{D}}^{\mu }%
\widehat{\mathbf{D}}_{\mu })\ ^{1}\widehat{\mathbf{f}}+(\widehat{\mathbf{T}}+%
\mathbf{\Theta })\ ^{2}\widehat{\mathbf{f}}+(\frac{1}{2}\widehat{\mathbf{D}}%
^{\mu }\widehat{\mathbf{D}}_{\mu }\widehat{\mathbf{T}}+\widehat{\mathbf{D}}%
_{\mu }\widehat{\mathbf{D}}_{\nu }\widehat{\mathbf{T}}^{\mu \nu }+\mathbf{%
\Xi )}\ ^{3}\widehat{\mathbf{f}}=\kappa ^{2}\ \widehat{\mathbf{T}},
\label{lineardef}
\end{equation}
where $\ ^{1}\widehat{\mathbf{f}}:=\partial \widehat{\mathbf{f}}/\partial
\widehat{\mathbf{R}},$ $\ ^{2}\widehat{\mathbf{f}}:=\partial \widehat{%
\mathbf{f}}/\partial \widehat{\mathbf{T}}$ and $\ \ ^{3}\widehat{\mathbf{f}}%
:=\partial \widehat{\mathbf{f}}/\partial \widehat{\mathbf{P}},$ when $%
\widehat{\mathbf{P}}=\widehat{\mathbf{R}}_{\alpha \beta }\widehat{\mathbf{T}}%
^{\alpha \beta }.$ Let us envisage a trace configuration in the interior of
a celestial body, when $\widehat{\mathbf{T}}=\widehat{\mathbf{T}}_{0}$ and $%
-2\widehat{\mathbf{f}}+\widehat{\mathbf{R}}_{0}\ (\ ^{1}\widehat{\mathbf{f}}%
)=\kappa ^{2}\ \widehat{\mathbf{T}}_{0}.$ Imposing nonholonomic constraints,
we parameterize a LC-configuration in GR and model an interior solution in
the presence of some gravitational objects (for instance, the Sun or the
Earth). The $f$-modifications (in general with strong coupling for the
curvature and the energy-momentum tensor) may result in a worsening of the
stability problems and may prevent $\widehat{\mathbf{T}}_{0}$ to be a
solution of any suitable background equation. It is difficult to find
solutions of (\ref{lineardef}) even for very much simplified cases in the
nonlinear situation if we work in coordinate frames for the connection $%
\widehat{\mathbf{D}}=\nabla .$

A rigorous study of the problem of matter instability for $f(R)$ and more
generally $f(R,T,R_{\alpha \beta }T^{\alpha \beta })$ gravities, for certain
illustrative cases when $\ ^{1}\widehat{\mathbf{f}}=R$, and for restrictive
conditions where there is a qualitative description via additional
functionals on $T$ and $P$ shows that the appearance of a large instability
can actually be avoided. Using the anholonomic frame deformation method, we
can consider modified gravity theories with $f$-modifications which are
effectively described by $\mathbf{\check{R}}_{\ \beta }^{\alpha }=\check{%
\Lambda}\delta _{\ \beta }^{\alpha }$ when the modifications are encoded
into polarization functions and N-coefficients. For models generated by
\begin{equation*}
\widehat{\mathbf{f}}(\widehat{\mathbf{R}},\widehat{\mathbf{T}},\widehat{%
\mathbf{R}}_{\alpha \beta }\widehat{\mathbf{T}}^{\alpha \beta })=\widehat{%
\mathbf{f}}_{1}(\widehat{\mathbf{R}})+\widehat{\mathbf{F}}(\widehat{\mathbf{P%
}})+\widehat{\mathbf{G}}(\widehat{\mathbf{T}}),
\end{equation*}%
we take a constant interior solution with $\widehat{\mathbf{T}}_{0}=const$
and $\widehat{\mathbf{P}}_{0}=const,$ and denote by $\widehat{\mathbf{f}}%
_{1}^{(s)}:=\partial ^{s}\widehat{\mathbf{f}}_{1}/\partial \widehat{\mathbf{R%
}}^{s}$ and $\widehat{\mathbf{F}}^{(s)}:=\partial ^{s}\widehat{\mathbf{F}}%
/\partial \widehat{\mathbf{P}}^{s}$ for $s=1,2,...$ We can repeat, with
respect to the N-frames (\ref{nader}) and (\ref{nadif}), the computations
presented in detail for Eqs.~(45)-(48) in \cite{odgom,odgom1,odgom2} (see also references
therein), and prove that Eqs.~(\ref{lineardef}) for linear pertubations can
be written in the form%
\begin{equation*}
\lbrack \mathbf{\check{D}}^{\mu }\mathbf{\check{D}}_{\mu }+2\frac{\mathbf{%
\check{T}}_{0}^{\mu \nu }}{\mathbf{\check{T}}_{0}}\mathbf{\check{D}}_{\mu }%
\mathbf{\check{D}}_{\nu }+2\frac{\mathbf{\Xi }_{0}}{\mathbf{\check{T}}_{0}}+4%
\frac{\mathbf{\check{P}}_{0}}{\mathbf{\check{T}}_{0}}\frac{\widehat{\mathbf{f%
}}_{1}^{(1)}}{\widehat{\mathbf{F}}^{(2)}}]\ \delta \mathbf{\check{P}}=[\frac{%
2}{\mathbf{\check{T}}_{0}}\frac{\widehat{\mathbf{f}}_{1}^{(1)}}{\widehat{%
\mathbf{F}}^{(2)}}-\frac{\mathbf{\check{P}}_{0}}{\mathbf{\check{T}}_{0}}%
\frac{\widehat{\mathbf{F}}^{(1)}}{\widehat{\mathbf{F}}^{(2)}}\left( 2\ \ ^{m}%
\widehat{\mathcal{L}}-\mathbf{\check{T}}_{0}\right) ]\ \delta \mathbf{\check{%
R}}.
\end{equation*}%
The values of type $\delta \widehat{\mathbf{P}}$ and $\delta \widehat{%
\mathbf{R}}$ are considered for a small perturbations in the curvature where
$\widehat{\mathbf{R}}=\widehat{\mathbf{R}}_{0}+\delta \widehat{\mathbf{R}}$
and $\widehat{\mathbf{P}}=\widehat{\mathbf{P}}_{0}+\delta \widehat{\mathbf{P}%
}.$ No instability appears if $\delta \mathbf{\check{P}=}\delta \mathbf{%
\check{R}=0}$ which is a particular solution of the above equation. We can
in fact model a damped oscillator with additional nonholonomic constraints
if $\mathbf{\Xi }_{0}+2\mathbf{\check{P}}_{0}\widehat{\mathbf{f}}_{1}^{(1)}/%
\widehat{\mathbf{F}}^{(2)}\geq \mathbf{\check{T}}_{0},$ which allows to
avoid large instabilities in the interior of a spherical body. For some
specific functionals $f(R)$ and appropriate $G(T),$ the same behavior as in
GR results (with mass stability in the cosmological context), although there
are possible large perturbations $\delta R$ and $\delta P$ remaining. The
ideas how to circumvent the mass instability problem for holonomic
configurations has been studied in \cite{massinst,massinst1,massinst2,massinst3,massinst4,massinst5,massinst6}. Redefining the
generating functions and sources in a $f$-modified model into an effective
Einsteinian theory, with $\mathcal{S}[\mathbf{\check{R},}\check{\Lambda}],$
one can consider a nonholonomically deformed Hilbert-Einstein action with $%
\widehat{\mathbf{f}}\rightarrow \mathbf{\check{f}=\check{R}.}$ In such
cases, $\delta \widehat{\mathbf{R}}=\delta \mathbf{\check{R}=0}$ and
instabilities are not produced, indeed, if we impose the zero torsion
conditions (see (\ref{dtors})), we get back to the GR theory. Even if Eq.~(%
\ref{lineardef}) involves not only perturbations of the Ricci scalar $%
\widehat{\mathbf{R}}$ but also of the Ricci d-tensor $\widehat{\mathbf{R}}%
_{\alpha \beta }$ (through $\delta \widehat{\mathbf{P}}$), via nonholonomic
transforms to effective $\mathbf{\check{R}}_{\ \beta }^{\alpha }=\check{%
\Lambda}\delta _{\ \beta }^{\alpha },$ the stability of the system is
obtained via off-diagonal interactions and the nonholonomic constraints used
for an effective modeling of a subclass of $\widehat{\mathbf{f}}$-theories
to certain nonholonomic deformations of the Einstein equations with
effective cosmological constant $\check{\Lambda}.$ This is indeed a
remarkable result.

\section{Effective field theory for off--diagonal cosmological configurations%
}

\label{s5} Both in particle and condensed matter physics, effective field
theory (EFT) methods have proven so far to provide a quick and economic way
in order to connect experimental data and phenomenological results with
certain fundamental theories (see the reviews \cite{eft,eft1}). Recently methods
of that sort have been applied in cosmology, in particular to inflation \cite%
{eftinfl,eftinfl1,eftinfl2}, late--time acceleration \cite{eftacc,eftacc1,eftacc2} and dark energy physics
\cite{eftde,eftde1} (details and references can be found in \cite{bloom,bloom1}). The goal
of this section is to construct an EFT describing perturbations both over
diagonal and off--diagonal cosmological background solutions in modified
gravities, in cases where an effective Einsteinian manifold can be
associated, and when the matter sector obeys the weak equivalence principle
and all modifications of gravity and the matter fields can be encoded into
an effective cosmological constant.

\subsection{Off--diagonal background evolution and $\Lambda$CDM}

We shall consider here configurations with redefined generating functions
and sources and where the third effective action in (\ref{mgts}) is taken to
be of the form
\begin{equation}
\mathcal{S}=\ ^{g}\mathbf{S}+\ ^{m}\mathbf{S}=\int [\frac{1}{2\overline{%
\kappa }^{2}}\Omega (t)\mathbf{R}+\widehat{\Lambda }(t)-c(t)\delta \mathbf{g}%
^{44}]\sqrt{|\mathbf{g}|}\mathbf{d}^{4}u+\ ^{m}\mathbf{S[g}_{\alpha \beta }%
\mathbf{]}.  \label{effbact}
\end{equation}%
This is related to a background FLRW solution constrained, on its turn, by
observation to be in close agreement with $\Lambda $CDM.\footnote{%
In effective field theories the mass scale $\overline{m}^2=\overline{\kappa }%
^{-2}$ can be different from the Plank mass of GR. It is used to render $%
\Omega $ dimensionless.} In this formula, the value $\delta \mathbf{g}^{44}$
is the perturbation of the time--time like component of the d--metric, and
the Ricci scalar $\mathbf{R}$ is defined by a canonical d--connection $%
\mathbf{D} $ which can be obtained by a finite chain of redefinitions,
resulting in $\lbrack \widehat{\mathbf{g}},\widehat{\mathbf{N}}\mathbf{,}%
\widehat{\mathbf{D}},\widehat{\Upsilon }]\rightarrow \lbrack \mathbf{g},%
\mathbf{N,D,T}_{\alpha \beta },\widehat{\Lambda }(t)]$ where $\widehat{%
\Lambda }(t)$ takes a constant value $\Lambda /2\overline{\kappa }^{2}$ for
the $h$--equations while it is a function of $t$ for the $v$--equations. The
energy--momentum tensor $\mathbf{T}_{\alpha \beta }$ determined by $\ ^{m}%
\mathbf{S[g}_{\alpha \beta }\mathbf{]}$ does not satisfy, in general, the
condition $\mathbf{D}_{\alpha }\mathbf{T}^{\alpha \beta }\neq 0,$ but $%
\mathbf{\nabla }_{\alpha }\mathbf{T}^{\alpha \beta }\rightarrow 0$ for $%
\mathbf{D}_{\alpha }\rightarrow \mathbf{\nabla }_{\alpha }.$ For simplicity,
effective matter is treated as a perfect fluid, which in the N--adapted
model is described by a time dependent average energy density $\overline{%
\rho }(t)$ and pressure $\overline{p}(t).$ For LC--configurations one has
the usual continuity equation
\begin{equation*}
\overline{\rho }^{\diamond }+3H(\overline{\rho }+\overline{p})=0,
\end{equation*}%
where $H$ is determined by a scaling factor $a(t)$ for a generic
off--diagonal solution of type (\ref{odfrlwtors}), or (\ref{qelgenofd}). We
note that in this section $\overline{\rho }^{\diamond }$ is the derivative
with respect to the physical time (the formalism works also in conformal
time). The function $a(t)$ can be considered as the limiting result for $%
\varepsilon \rightarrow 0$ of an integration in a metric of type (\ref%
{ans1bb}) (where nonlinear interactions are encoded by f--modifications), as
we explained before for that formula.

\subsubsection{ Off--diagonal background evolution}

The background evolution is in general off--diagonal (values $\varepsilon
\approx 10^{-20}$ do not contradict present experimental data \cite{appleby}%
). For simplicity, we can consider a diagonal background with a FLRW metric
with zero spatial curvature. We can chose $a(t),\overline{\rho }(t)$ and $%
\overline{p}(t)$ to be close to $\Lambda$CDM evolution if $\varepsilon
\rightarrow 0.$ As in standard (diagonal) cosmology we can use the Friedmann
equations to eliminate the functions $\widehat{\Lambda }(t)$ and $c(t)$
(which can again be considered as a redefinition of the generating functions
in our approach) but keep $\Omega (t)$ as a free function, similarly to \cite{bloom,bloom1} and \cite{starob}. For diagonal configurations, such theory can be
formulated in the Jordan frame and dealt with a non--minimal coupling
between an effective scalar field and metric. In the Einstein frame, we get
a coupling of matter to the effective scalar field. Nevertheless, the
scaling factor $a(t)$ has some ``memory'' of the genuine nonholonomic
redefinition of the integration functions and corresponding contributions of
the modified gravity theory.

Varying (\ref{effbact}) with respect to N--adapted frames, we get an
effective Friedman equation, which allows to express%
\begin{eqnarray*}
-2c(t) &=&\overline{\rho }+\overline{p}+(2\Omega H^{\diamond }+\Omega
^{\diamond \diamond }/2-H\Omega ^{\diamond })/\overline{\kappa }^{2} \\
-\widehat{\Lambda }(t) &=&\overline{p}+(3\Omega H^{2}+2\Omega H+\Omega
^{\diamond \diamond }+H\Omega ^{\diamond })/\overline{\kappa }^{2}
\end{eqnarray*}%
for any given data $(\Omega ,a,\overline{\rho },\overline{p}).$\footnote{%
This can be rewritten in a more conventional form in terms of the dark
energy density, $\rho _{DE},$ and pressure, $p_{DE},$ (cf.~(\ref{demeq})),
when
\begin{equation*}
3\Omega H^{2}=\overline{\kappa }^{2}(\rho _{DE}+\overline{\rho })\mbox{ and }%
2\Omega H^{\diamond }=-\overline{\kappa }^{2}(\rho _{DE}+p_{DE}+\overline{%
\rho }+\overline{p}),
\end{equation*}%
for $2c(t)=\rho _{DE}+p_{DE}+(H\Omega ^{\diamond }-\Omega ^{\diamond
\diamond })/\overline{\kappa }^{2}$ and $\widehat{\Lambda }%
(t)=p_{DE}-(H\Omega ^{\diamond }+\Omega ^{\diamond \diamond })/\overline{%
\kappa }^{2}.$} These equations describe how the background energy and
pressure of the DE component evolve over cosmological history, corresponding
to the evolution of the N--coefficients $w_{i}(t).$

\subsubsection{N--adapted perturbations}

Assuming that for $\mathbf{D\rightarrow \nabla }$ the weak equivalence
principle is true, one can always introduce a conformal (Jordan) frame, when
the matter fields couples only to the d--metric $\mathbf{g}_{\alpha \beta } $
and not to the scalar field.

Starting from an N--adapted form for (\ref{effbact}), a procedure similar to
that for the construction of effective field theories (see Eq.~(2.1) of \cite{bloom,bloom1}) leads to the general effectively unitary gauge action%
\begin{eqnarray}
\mathcal{S} &=&\int \sqrt{|\mathbf{g}|}\mathbf{d}^{4}u[\frac{1}{2\overline{%
\kappa }^{2}}\Omega (t)\mathbf{R}+\widehat{\Lambda }(t)-c(t)\delta \mathbf{g}%
^{44}+\frac{M_{2}^{4}(t)}{2}(\delta \mathbf{g}^{44})^{2}+\frac{M_{3}^{4}(t)}{%
2}(\delta \mathbf{g}^{44})^{3}+...  \label{pertact} \\
&&-\frac{\overline{M}_{1}^{3}(t)}{2}\delta \mathbf{g}^{44}\ \delta \mathbf{K}%
_{\ \alpha }^{\alpha }-\frac{\overline{M}_{2}^{2}(t)}{2}(\delta \mathbf{K}%
_{\ \alpha }^{\alpha })^{2}-\frac{\overline{M}_{3}^{2}(t)}{2}\delta \mathbf{K%
}_{\ \beta }^{\alpha }\delta \mathbf{K}_{\ \alpha }^{\beta }+...+\
^{1}\lambda (t)(\delta \mathbf{R})^{2}+\ ^{2}\lambda (t)\delta \mathbf{R}%
_{\alpha \beta }\ \delta \mathbf{R}^{\alpha \beta }  \notag \\
&&+\ ^{1}\gamma (t)\mathbf{C}_{\mu \nu \sigma \lambda }\mathbf{C}^{\mu \nu
\sigma \lambda }+\ ^{2}\gamma (t)\epsilon ^{\mu \nu \sigma \lambda }\mathbf{C%
}_{\mu \nu }^{\qquad ^{\alpha \beta }}\mathbf{C}_{\sigma \lambda \alpha
\beta }+...  \notag \\
&&+\ ^{1}m^{2}(t)\mathbf{n}^{\mu }\mathbf{n}^{\nu }(\mathbf{e}_{\mu }\mathbf{%
g}^{44})(\mathbf{e}_{\nu }\mathbf{g}^{44})+\ ^{2}m^{2}(t)(\mathbf{g}^{\mu
\nu }+\mathbf{n}^{\mu }\mathbf{n}^{\nu })(\mathbf{e}_{\mu }\mathbf{g}^{44})(%
\mathbf{e}_{\nu }\mathbf{g}^{44})+...]+\ ^{m}\mathbf{S[g}_{\alpha \beta }%
\mathbf{].}  \notag
\end{eqnarray}%
The value $\mathbf{C}_{\mu \nu \sigma \lambda }$ is the Weyl tensor
determined by $\mathbf{D}$, while $\mathbf{n}^{\mu }$ is the vector normal
to the surfaces of constant time. Each term in (\ref{pertact}) can have a
time-dependent coefficient because the background solutions is, in general,
off--diagonal and breaks time translation symmetry. The matter action $\ ^{m}%
\mathbf{S[g}_{\alpha \beta }\mathbf{]}$ can be arbitrary, with sources to be
re--defined by integration functions. We will fix below the quantities $%
\delta \mathbf{g}^{44},\delta \mathbf{R,}\delta \mathbf{R}_{\alpha \beta }$
and $\delta \mathbf{K}_{\alpha \beta }$ as perturbations, the dots
corresponding to various terms which we do not specify of quadratic and
higher-order perturbations. N--elongated partial derivatives $\mathbf{e}%
_{\mu }$ (\ref{nader}) are used instead of $\partial _{\mu }.$

We involve and additional 3+1 splitting (the variant 2+2 is convenient for
constructing off--diagonal, exact solutions), where the Ricci d--tensor of $%
\mathbf{D}$ decomposes as
\begin{equation*}
\mathbf{R}_{\alpha \beta }\mathbf{n}^{\alpha }\mathbf{n}^{\beta }=\mathbf{K}%
_{\alpha \beta }\mathbf{K}^{\alpha \beta }-\mathbf{K}_{\alpha }^{\ \beta }%
\mathbf{K}_{\beta }^{\ \alpha }+\mathbf{D}_{\alpha }^{\ }(\mathbf{n}^{\gamma
}\mathbf{D}_{\gamma }^{\ }\mathbf{n}^{\alpha })-\mathbf{D}_{\alpha }^{\ }(%
\mathbf{n}^{\alpha }\mathbf{D}_{\gamma }^{\ }\mathbf{n}^{\gamma }).
\end{equation*}
As a result of the N--adapted construction, the value $\mathbf{n}^{\alpha
}\delta \mathbf{K}_{\alpha \beta },$ where $\delta \mathbf{K}_{\alpha \beta
}=\mathbf{K}_{\alpha \beta }-\ ^{0}\mathbf{K}_{\alpha \beta }:=\mathbf{K}%
_{\alpha \beta }+3H(\mathbf{g}^{\mu \nu }+\mathbf{n}^{\mu }\mathbf{n}^{\nu
}) $ vanishes.

\subsubsection{Effective field theory in terms of St\"{u}ckelberg d--fields}

In effective field theories and cosmological models, the St\"{u}kelberg\
technique \cite{eftinfl,eftinfl1,eftinfl2,bloom,bloom1} is used when explicit functions of time are
modified according to
\begin{equation}
\varphi (t)\rightarrow \varphi (t+\pi (u^{\alpha }))  \label{aux61}
\end{equation}
and then Taylor--expanded in $\pi (u^{\alpha }).$ Such procedure is applied
to the action (\ref{pertact}) when the ansatz for off--diagonal solutions (%
\ref{odfrlwtors}), or (\ref{qelgenofd}), is reparameterized as a $3+1$ form,%
\begin{eqnarray}
ds^{2} &=&\underline{g}_{\alpha \beta }du^{\alpha }du^{\beta }=a^{2}(\tilde{g%
}_{\underline{i}\underline{j}}+\varsigma _{\underline{i}\underline{j}})d%
\widetilde{x}^{\underline{i}}d\widetilde{x}^{\underline{j}}-d\widetilde{t}%
^{2},\mbox{\  synchronous gauge };  \label{synchr} \\
&=&a^{2}(\mathbf{\tilde{g}}_{\underline{i}\underline{j}}+\tilde{\varsigma}_{%
\underline{i}\underline{j}})d\widetilde{\mathbf{e}}^{\underline{i}}d%
\widetilde{\mathbf{e}}^{\underline{j}}-(\delta \widetilde{t})^{2},%
\mbox{\
N-adapted synchronous gauge };  \notag \\
ds^{2} &=&a^{2}(1-2\widetilde{\phi })\tilde{g}_{\underline{i}\underline{j}}d%
\widetilde{x}^{\underline{i}}d\widetilde{x}^{\underline{j}}-(1+2\psi )d%
\widetilde{t}^{2},\mbox{\  Newtonian gauge };  \label{newtong} \\
&=&a^{2}(1-2\widetilde{\phi })\mathbf{\tilde{g}}_{\underline{i}\underline{j}%
}d\widetilde{\mathbf{e}}^{\underline{i}}d\widetilde{\mathbf{e}}^{\underline{j%
}}-(1+2\psi )(\delta \widetilde{t})^{2},\mbox{\ N--adapted  Newtonian gauge }%
,  \notag
\end{eqnarray}%
where $d\widetilde{\mathbf{e}}^{\underline{i}}=(d\widetilde{x}^{i},\delta
\tilde{e}^{3}=dy^{3}+\tilde{n}_{i}d\widetilde{x}^{i})$ and $\delta
\widetilde{t}=d\widetilde{t}+\tilde{w}_{i}d\widetilde{x}^{i},$ see (\ref%
{nadif}). The basic premise for this is that we can perform coordinate
transformations $\widetilde{t}=t+\pi (u^{\alpha }),\widetilde{x}^{\underline{%
i}}=x^{\underline{i}},$ were the convention for indices is $\underline{i},%
\underline{j},...=1,2,3$ and the coordinates $u^{\alpha
}=(x^{i},y^{3},t)\rightarrow \widetilde{u}^{\alpha }=(\widetilde{x}^{%
\underline{i}},\widetilde{t}).$ We consider $\tilde{g}_{\underline{i}%
\underline{j}}$ to be a time--independent background metric but $a(%
\widetilde{t})$ and $\varsigma _{\underline{i}\underline{j}%
}[N_{i}^{a},g_{i},h_{a}]$ are certain nonlinear functionals determined by a
cosmological off--diagonal solution in the modified theory. The metric (\ref%
{synchr}) is written in a form which allows to describe perturbations in a
synchronous gauge. This parametrization can be obtained for any d--metric (%
\ref{dm1}) which with respect to coordinate frames is rewritten in the form (%
\ref{m1}) with coefficients (\ref{ansatz}). In the synchronous gauge, the
coordinate transforms $u^{\alpha }\rightarrow \widetilde{u}^{\alpha }$ are
chosen in a for to be satisfied the conditions $\delta g^{44}=g^{4\underline{%
i}}=0$ (we note that here we do not use boldface symbols because, in
general, such conditions can be imposed for not N--adapted frames). It will
be convenient to discuss some phenomena in the Newtonian gauge (\ref{newtong}%
). For small $\varepsilon $--deformations (see (\ref{ans1bb})),
off--diagonal extensions of cosmological metrics can be treated as effective
fluctuations which may be nonholonomically constrained, or not, to be
parameterized in N--adapted form. In both cases, we can use the formulas
derived in \cite{bloom,bloom1} but keeping in mind that we work with a
d--connection $\mathbf{D}$ which only at the end will be additionally
constrained, when $\mathbf{D\rightarrow \nabla .}$\footnote{%
For simplicity, we do not dub, in an N--adapted form, the proofs from those
works but use directly the synchronous gauge representation which is more
convenient for studying both perturbations and $\varepsilon $--deformations
all included in a term $\varsigma _{\underline{i}\underline{j}}(\varepsilon
,x^{\underline{i}},t);$ we omit the tilde on spacetime coordinates when this
does not result in ambiguities, and use in brief the term perturbations both
for $\varepsilon $--deformations and for fluctuations; in a more general
context, perturbations with respect to a fixed N--adapted background can be
considered.}

We can decompose
\begin{equation}
\varsigma _{\underline{i}\underline{j}}=\frac{\varsigma }{3}\tilde{g}_{%
\underline{i}\underline{j}}+(\widetilde{\mathbf{D}}_{\underline{i}}%
\widetilde{\mathbf{D}}_{\underline{j}}-\frac{\tilde{g}_{\underline{i}%
\underline{j}}}{3}\widetilde{\mathbf{D}}^{2})\eta \mbox{ and }\tilde{%
\varsigma}_{\underline{i}\underline{j}}=\frac{\tilde{\varsigma}}{3}\mathbf{%
\tilde{g}}_{\underline{i}\underline{j}}+(\widetilde{\mathbf{D}}_{\underline{i%
}}\widetilde{\mathbf{D}}_{\underline{j}}-\frac{\mathbf{\tilde{g}}_{%
\underline{i}\underline{j}}}{3}\widetilde{\mathbf{D}}^{2})\tilde{\eta},
\label{aux62}
\end{equation}%
where tildes refer to quantities associated with the spatial metric, and
express the metric determinant $\sqrt{|\underline{g}|}=a^{3}\sqrt{|\tilde{g}|%
}(1+\varsigma /2)$ and $\sqrt{|\mathbf{g|}}=a^{3}\sqrt{|\mathbf{\tilde{g}}|}%
(1+\tilde{\varsigma}/2).$ It is convenient to use a definition in mommentum
space $k^{\alpha }=(\overrightarrow{k},\tilde{k}),$ where
\begin{equation}
\eta (\overrightarrow{k},t)=-k^{-2}[\varsigma (\overrightarrow{k},t)+6%
\mathring{\eta}(\overrightarrow{k},t)]  \label{aux63}
\end{equation}
(see details on similar conventions in \cite{bloom,bloom1}), where the effective
field constructions are also performed in the massive case.

\subsection{Linking off--diagonal perturbations to observations}

The goal is here to identify and study the full set of perturbations which
result in off--diagonal deformations of the FRW background. Linearized
equations of motion for certain effective scalar perturbations will be
considered, which will allow to determine: 1) the speed of sound; 2)
Poisson's equation; 3) the anisotropic stress; 4) the effective Newton
constant; and 5) Caldwell's parameter \cite{cald}. \ Reparametrizations of
the metric in the respective forms (\ref{synchr}) and (\ref{newtong}) allow
for this association with the standard phenomenological functions and
parameters appearing in literature, connecting in this way observational
features with the off--diagonal cosmological solutions, for different
modified gravities and effective Einstein spaces. In what follows, and as a
hint to explicitly show these possibilities, we will consider in brief 1),
4) and 5) above, as obtained for certain classes of cosmological models.

\subsubsection{Effective speed of sound in off--diagonal media}

The data $\pi ,\varsigma $ and $\mathring{\eta}$ defined by Eqs.~(\ref{aux61}%
), (\ref{aux62}) and (\ref{aux63}) are used for writing the so--called $\pi$%
--equation of motion along with the time--time and space--space trace
components of the Einstein equation (in our case, modified in terms of $%
\nabla \rightarrow \mathbf{D}$) via the kinetic matrix $\gamma _{X,Y}$
(namely, the coefficient of $Y$ in the $X$ equation of motion). In its turn,
this allows to compute the speed of sound of scalar perturbations in the
sub--horizon limit using the synchronous gauge for this calculation. The
differential operators are transformed into matrix components by changing $%
z^{\diamond }\rightarrow -i\omega .$ This allows to compute the coefficients
of (\ref{pertact}), following the same procedure as in Table 1 and Appendix
D (see also Eqs.~(4.1)--(4.5)) of \cite{bloom,bloom1}, where the determinant of the
kinetic matrix is set to be zero. Such equations are parameterized in the
form%
\begin{equation*}
\left(
\begin{array}{ccc}
\gamma _{\pi ,\pi } & \gamma _{\pi ,\varsigma } & \gamma _{\pi ,\mathring{%
\eta}} \\
\gamma _{\widetilde{t}\widetilde{t},\pi } & \gamma _{\widetilde{t}\widetilde{%
t},\varsigma } & \gamma _{\widetilde{t}\widetilde{t},\mathring{\eta}} \\
\gamma _{ss,\pi } & \gamma _{ss,\varsigma } & \gamma _{ss,\mathring{\eta}}%
\end{array}%
\right) \left(
\begin{array}{c}
\pi \\
\varsigma \\
\mathring{\eta}%
\end{array}%
\right) =0
\end{equation*}
The resulting expression yields, in general, a number of nonlinear
dispersion relations. For instance, the presence of the operator $%
M_{2}^{4}(t)$ results in the dispersion formula for the speed of sound, $c,$
\begin{equation*}
c_{s}^{-2}=1+2M_{2}^{4}\left[ c+\frac{3(\Omega ^{\diamond })^{2}}{4\overline{%
\kappa }^{2}\Omega }\right] ^{-1}.
\end{equation*}%
The value $c$ is linked to various stability issues, as analyzed in \cite{eftde,eftde1}, where it was concluded that ghost--free conditions are satisfied if
\begin{equation*}
2\left[ c+\frac{3(\Omega ^{\diamond })^{2}}{4\overline{\kappa }^{2}\Omega }%
\right] =\rho _{DE}+p_{DE}+\overline{\kappa }^{-2}\left[ H\Omega ^{\diamond
}+\frac{3}{2}\frac{(\Omega ^{\diamond })^{2}}{\Omega }-\Omega ^{\diamond
\diamond }\right] >0.
\end{equation*}%
For certain off--diagonal configurations, the effective EoS of DE can become
of phantom type, $w_{DE}<-1,$ even if the nonholonomic
deformations/perturbations do not seem to host one (it may depend on the
choice of generating function $\Omega (t)$).

\subsubsection{The effective gravitational constant and Caldwell's parameter}

In the Newtonian limit, we look at off--diagonal deformations and modified
theories of gravity that share the horizon $k^{2}/a^{2}\gg H^{2}.$ Such a a
horizon is indeed shared by different types of solutions in different
models, and may show up, for instance, in $\varepsilon $--deformed
nonholonomic backgrounds. Similar considerations as for subsection 4.4 in
\cite{bloom,bloom1} result in an expression for the effective Newton constant $4\pi
\ ^{eff}G:=\ ^{eff}\kappa /2$ (note here $\pi$ is the mathematical constant,
and not the function $\pi (u^{\alpha })$ from (\ref{aux61}))%
\begin{equation*}
\ ^{eff}\kappa ^{2}=\overline{\kappa }^{2}\Omega ^{-1}\left[ 1+(\Omega
^{\diamond })^{2}/4\Omega (c\overline{\kappa }^{2}+\frac{3}{4}\frac{(\Omega
^{\diamond })^{2}}{\Omega }+a^{2}\frac{\overline{\kappa }^{2}}{k^{2}}M^{2})%
\right] .
\end{equation*}%
This expression becomes more complicated if more terms, in addition to $%
M^{2} $, are included. There are modifications of $^{eff}G$ even if $\Omega
^{\diamond }=0.$ We note here the explicit dependence on $a^{2}$, which can
be considered as the generating function for off--diagonal solutions.

Let us now consider the parameter $\ ^{c}\varpi $ considered by Caldwell et
al. \cite{cald} as the cosmological analog of Eddington's PPN parameter $%
\gamma ,$ when $\ ^{c}\varpi \approx 1-\gamma $ (we use here a notation
which differs from those in other works on the subject, in order not to
interfere with $\varpi $ from Eq.~(\ref{dssol}), etc.).\footnote{%
There are bounds of the type $\ ^{c}\varpi $ $=0.02\pm 0.07(2\sigma ),$ for
scales of about 10 Kpc, and =$0.03\pm 0.10(2\sigma ),$ for hundreds of Kpc,
but as of now there are not known limits at the Mpc scale yet.} Using
background terms, one obtains%
\begin{equation*}
\ ^{c}\varpi =\frac{(\Omega ^{\diamond })^{2}}{2\Omega }\left[ c\overline{%
\kappa }^{2}+\frac{(\Omega ^{\diamond })^{2}}{2\Omega }+a^{2}\frac{\overline{%
\kappa }^{2}}{k^{2}}M^{2}\right] ^{-1}.
\end{equation*}%
Such formula becomes more complicated if further terms are included. In both
cases of this and the previous formulas, only the terms $\overline{M}_{3}$
and $\widehat{\mathbf{M}}^{2}=\delta \mathbf{g}^{44}\delta \mathbf{R}^{(3)}$
(for perturbations of the spacial scalar curvature $\mathbf{R}^{(3)}$ of $%
\mathbf{D}$) contribute, if $\Omega ^{\diamond }=0.$ Both $\ ^{eff}\kappa
^{2}(a)$ and $\ ^{c}\varpi (a)$ are functionals of the generating functions.
In GR, $\ ^{eff}\kappa ^{2}=1/m_{P}$ and $\ ^{c}\varpi =0.$ Nevertheless,
these values are nonzero if the configurations are off--diagonal with an
effective cosmological constant. A kind of nonlinear classical polarization
of the gravitational constant is there obtained at certain scales.

\subsubsection{Confronting an off--diagonal cosmology which encodes modified
gravity with actual observational results}

In Sect.~4.6 of \cite{bloom,bloom1} (see, in particular, Tables 4 and 5 there), the
contributions from various operators in a generalized model with an
effective field theory action were calculated. We will here combine this
knowledge on the dark energy models (in our case with off--diagonal
interactions modeling the $\Lambda$CDM, $f(\mathbf{R})$, and Ho\v{r}%
ava--Lifshitz \cite{horava} theories) for the nontrivial N--adapted
coefficients in (\ref{pertact}):%
\begin{equation*}
\begin{array}{ccccccc}
\mbox{\bf Operator } &  & \Omega & \Lambda & c & \overline{\mathbf{M}}%
_{2}^{2} & \ ^{2}m^{2} \\
\mbox{\bf  MGTs } &  & \mathbf{R} &  & \delta \mathbf{g}^{44} & (\delta
\mathbf{K}_{\ \alpha }^{\alpha })^{2} & a^{-2}\mathbf{g}^{\underline{i}%
\underline{j}}(\mathbf{e}_{\underline{i}}\mathbf{g}^{44})(\mathbf{e}_{%
\underline{j}}\mathbf{g}^{44}) \\
&  &  &  &  &  &  \\
\Lambda CDM &  & 1 & \curlyvee & 0 & - & - \\
f(\mathbf{R}) &  & \curlyvee & \curlyvee & 0 & - & - \\
\mbox{ Ho\v rava--Lifshitz } &  & 1 & \curlyvee & 0 & \curlyvee & \curlyvee%
\end{array}%
\end{equation*}%
In the above matrix, the symbols for the respective operators mean that $%
\curlyvee $ is necessary, the $-$ is not included, and $1,0$ is for unity or
exactly vanishing, respectively.

We conclude that off--diagonal modeling of modified gravities can be
performed as small perturbations of $\Lambda$CDM, when dark energy exists in
the form of a cosmological constant. The off--diagonal redefinitions of
generating functions result in a modification of $a(t)$ which is different
from that in diagonal cosmological models. Nevertheless, even very small
values of $\varepsilon $--off--diagonal deformations may introduce a certain
speed of sound and clumping of DE (this is different from diagonal
configurations in when such effects are zero and the effective Newtonian
constant is just that for GR). Nevertheless, the models can be characterized
by different physical values if the off--diagonal modeling is performed for
different modified gravities, indeed:
\begin{equation*}
\begin{array}{ccccc}
\mbox{\bf Operator } &  & \Omega & \overline{\mathbf{M}}_{2}^{2} & \
^{2}m^{2} \\
\mbox{\bf  MGTs } &  & \mathbf{R} & (\delta \mathbf{K}_{\ \alpha }^{\alpha
})^{2} & a^{-2}\mathbf{g}^{\underline{i}\underline{j}}(\mathbf{e}_{%
\underline{i}}\mathbf{g}^{44})(\mathbf{e}_{\underline{j}}\mathbf{g}^{44}) \\
&  &  &  &  \\
\mbox{ Speed of sound } &  & 1 & \curlyvee +k^{4} & \ast \\
\mbox{ Effective Newtonian constant } &  & \curlyvee & \curlyvee +k^{2} &
\curlyvee \\
\mbox{Caldwell's parameter }\ ^{c}\varpi &  & \curlyvee & \curlyvee & -%
\end{array}%
\end{equation*}%
Here additional labels have been used, for instance, $\curlyvee +k^{4}$ is
for a new scale dependence, and $\ast $ is set for an operator behaving
unusually.

Finally, we note that the effective field theory can be generalized in a
form which allows to describe dark energy and modified gravities in the late
universe when off--diagonal nonlinear parametric gravitational interactions
encode contributions of such modified theories. The action for perturbations
(\ref{pertact}) of the effective models (\ref{effbact}) includes positive
small $\varepsilon$--deformations of the $\Lambda$CDM models with
nonholonomically constrained fluctuations which contribute to the background
cosmological evolution. A systematic investigation of the physical effects
of N--adapted operators was undertaken which emphasizes the importance of
off--diagonal effects when modeling modified gravities with the aim to
constrain the possible nature of DE.


\section{Short summary of new geometric ideas and methods for the construction of off--diagonal solutions}

\label{s6} In the modern era of precision cosmology, a series of evidences have
been found pointing towards deviations from the standard big bang cosmology. The
universe may be slightly non-homogeneous and anisotropic at very large
scales and the accelerating expansion phases are determined by dark energy
and dark matter effects. A number of alternative modifications of gravity
have been proposed already, with the aim to elaborate realistic cosmological models. Some of
the most intensively exploited are the so--called $f(R,...)$--modified gravity
theories  \cite{revfmod,revfmod01,revfmod02,revfmod03,revfmod04,revfmod05,revfmod06,revfmod07,revfmod08,revfmod09,revfmod10,revfmod11,revfmod12,revfmod13}. The gravitational field equations
in GR and the modified gravities consist of very involved systems of nonlinear PDE. A
rigorous study of the cosmological evolution in the frames of the different theories
requires  new
analytic, geometric and numerical methods for constructing exact and
approximate solutions, reconstructing procedures and effective field models.
In scenarios closely related to the standard $\Lambda $CDM universe, the
ansatz for the alternative metric is usually taken of FLRW diagonal type and the
interactions are modeled by effective matter, exotic fluids and
modifications of GR. The advantage of the anholonomic frame deformation
method, AFDM, \cite{odgom,odgom1,odgom2} is that it provides a geometric formalism for
constructing exact generic off--diagonal solutions encoding nonlinear
parametric effects which mimic scenarios of anisotropic cosmology via
generating and integration functions. Such nonlinear gravitational
interactions, and the possible cosmological implications thereof, are not considered if we
fix from the very beginning the diagonal ansatz, which results in a system of
ordinary differential equations. For generic off--diagonal configurations,
we can always derive an effective field theory and confront cosmological
theories with existing observational data as we showed in Sect.~\ref{s5}.


For the equivalent modeling of exact solutions and cosmological scenarios in
different classes of modified and GR theories, we consider three different
parametrization of the action for gravitational and matter fields (\ref{mgts}),
with respective Lagrange functionals%
\begin{eqnarray}
\mathcal{L} &=&\ ^{g}\mathcal{L}[f(R,T,R_{\alpha \beta }T^{\alpha \beta
})]+\ ^{m}\mathcal{L}[g_{\alpha \beta },\nabla ,\ ^{m}\varphi ]  \label{a61}
\\
&=&\ ^{g}\widehat{\mathbf{L}}[\widehat{\mathbf{f}}(\widehat{\mathbf{R}},%
\widehat{\mathbf{T}},\widehat{\mathbf{R}}_{\alpha \beta }\widehat{\mathbf{T}}%
^{\alpha \beta })]+\ ^{m}\widehat{\mathbf{L}}[\widehat{\mathbf{g}}_{\alpha
\beta },\widehat{\mathbf{D}},\ ^{m}\widehat{\varphi }]  \label{a62} \\
&=&\ ^{g}\mathbf{\check{L}}+\ ^{m}\mathbf{\check{L}=\check{R}+}\check{\Lambda%
}.  \label{a63}
\end{eqnarray}%
In these formulas, $\ ^{m}\varphi $ are some (effective) matter fields, which
can be approximated by the components of perfect (pressureless) fluids, for
instance, with an energy--momentum tensor of e thtype (\ref{dsourc}), and the linear
connections $\nabla ,\widehat{\mathbf{D}}$ and $\mathbf{\check{D}}$ are
related via distortion relations of the type (\ref{distr}), which are completely determined
by the metric field $g_{\alpha \beta }\simeq \widehat{\mathbf{g}}_{\alpha
\beta }\simeq \mathbf{\check{g}}_{\alpha \beta }$ up to frame transformations.
The functionals $\ ^{g}\mathcal{L[}f(...)$ $]$ and $\ ^{m}\mathcal{L}[...]$
determine the corresponding model of $f$--modified theory of gravity. The Lagrange densities
are written in different geometric variables because ofthis allows us to find
exact solutions and model physical effects by means  the same solutions but in
different theories of gravity.

\subsection{Decoupling of the generalized Einstein equations and modified
cosmological solutions}

We can decouple the gravitational and matter field equations (\ref{mgtfe})
in a modified theory of gravity, derived from a Lagrangian density (\ref{a61}), and construct generic
off--diagonal solutions depending (in general) on all spacetime variables, if
we work with geometric data $(\widehat{\mathbf{g}},\widehat{\mathbf{D}},%
\widehat{\mathbf{N}})$. The generalized Einstein equations are written in the
N--adapted form $\widehat{\mathbf{R}}_{\alpha \beta }=\widehat{\mathbf{%
\Upsilon }}_{\alpha \beta }$ with an effective source $\widehat{\mathbf{%
\Upsilon }}_{\alpha \beta }=diag[~^{h}\Upsilon ,~^{v}\Upsilon ]$ determined
by the matter fields and the nonholonomic and $f$--deformations. For a very general
off--diagonal ansatz for metrics with one Killing, or non--Killing,
symmetries, the gravitational field equations transform into a system of
nonlinear PDEs (\ref{eq1m})--(\ref{confeq}), which can be solved in their general
form for nonhomogeneous and locally anisotropic cosmological configurations
(\ref{odfrlwtors}). In the coordinate frame $u^{\underline{\alpha }%
}=(x^{k},y^{3},y^{4}=t)$ (where $t$ is the time like coordinate and $k=1,2),$
such metrics $ds^{2}=g_{\underline{\alpha }\underline{\beta }}du^{\underline{%
\alpha }}du^{\underline{\beta }}$ are of the type (\ref{ans1a}), or (\ref{polarf}%
), and can be written in a form involving a generalized scaling factor  $%
a^{2}(x^{k},t)$,
\begin{equation}
g_{\underline{\alpha }\underline{\beta }}=\left[
\begin{array}{cccc}
a^{2}\eta _{1}+\omega ^{2}[(n_{1})^{2}a^{2}\widehat{h}_{3}-(w_{1})^{2}] &
\omega ^{2}[n_{1}n_{2}a^{2}\widehat{h}_{3}-w_{1}w_{2}] & \omega
^{2}n_{1}a^{2}\widehat{h}_{3} & \omega ^{2}w_{1} \\
\omega ^{2}[n_{1}n_{2}a^{2}\widehat{h}_{3}-w_{1}w_{2}] & a^{2}\eta
_{2}+(n_{2})^{2}a^{2}\widehat{h}_{3}-(w_{2})^{2} & \omega ^{2}n_{2}a^{2}%
\widehat{h}_{3} & \omega ^{2}w_{2} \\
\omega ^{2}n_{1}a^{2}\widehat{h}_{3} & \omega ^{2}n_{2}a^{2}\widehat{h}_{3}
& \omega ^{2}a^{2}\widehat{h}_{3} & 0 \\
\omega ^{2}w_{1} & \omega ^{2}w_{2} & 0 & -\omega ^{2}%
\end{array}%
\right] .  \label{goffds}
\end{equation}%
By fixing the data for the generating functions: $\widehat{\Phi }(x^{k},t),\widehat{%
\Phi }^{\diamond }=\partial _{t}\widehat{\Phi }\neq 0;$ $\omega
^{2}(x^{k},y^{3},t)$ as a solution of $\partial _{i}\omega -(\partial
_{i}\Phi /\Phi ^{\diamond })\omega ^{\diamond }=0;e^{\psi
(x^{i})}=a^{2}(x^{k},t)\eta _{1}(x^{k},t)=a^{2}(x^{k},t)\eta _{2}(x^{k},t)$
as a solution of \ $\psi ^{\bullet \bullet }+\psi ^{\prime \prime
}=2~^{h}\Upsilon$, for the nontrivial sources $~^{h}\Upsilon
(x^{k}),~^{v}\Upsilon (x^{k},t)$ and for the effective cosmological constant $%
\Lambda \neq 0,$ we can the express the coefficients of this metric in a certain
general form. One further obtains [see the related formulas (\ref{h3}), (\ref{h4}), (%
\ref{n1b}) and (\ref{w1b})] $\widehat{h}_{3}=h_{3}/a^{2}|h_{4}|$, with
\begin{eqnarray*}
h_{3}(x^{k},t) &=&\widehat{\Phi }^{2}/4\Lambda ,\ h_{4}=a^{-2}(x^{k},t)=%
\frac{(\widehat{\Phi }^{2})^{\diamond }}{8}\left[ \widehat{\Phi }%
^{2}|~^{v}\Upsilon |+\int dt\ \widehat{\Phi }^{2}|~^{v}\Upsilon |^{\diamond }%
\right] ^{-1}, \\
w_{i}(x^{k},t) &=&\partial _{i}\Phi \lbrack \widehat{\Phi },~^{v}\Upsilon
]/\Phi ^{\diamond }[\widehat{\Phi },~^{v}\Upsilon ],\ n_{k}=\ _{1}n_{k}+\
_{2}n_{k}\int dt\ h_{4}/(\sqrt{|h_{3}|})^{3},
\end{eqnarray*}%
where $\ _{1}n_{k}(x^{i}),\ _{2}n_{k}(x^{i})$ are integration functions. The
solutions (\ref{goffds}) include, in general, a nonholonomically induced
torsion (\ref{dtors}), which is important for constructing exact solutions in
modified theories with nontrivial torsion fields.

We can always put additional zero--torsion constraints (\ref{lccondb}) and
construct Levi--Civita configurations. Such solutions are extracted by
choosing $\ _{2}n_{k}=0$ and $\ _{1}n_{k}=\partial _{k}n$ with a function $%
n=n(x^{k})$ \ and for a subclass of generating functions $\Phi =\check{\Phi}%
(x^{k},t)$ for which $(\partial _{i}\check{\Phi})^{\diamond }=\partial _{i}%
\check{\Phi}^{\diamond },$ see the details for Eq.~(\ref{aux4a}). One finds $%
\ \check{w}_{i}=\partial _{i}\check{\Phi}/\check{\Phi}^{^{\diamond
}}=\partial _{i}\widetilde{A}$ for a nontrivial function $\widetilde{A}%
(x^{k},y^{4})$ taken to be a solution of a first order PDE with effective a
sources depending functionally on $\check{\Phi}.$ The $v$--components of
geometric and physical object are generated by couples of data $(\Phi
,~^{v}\Upsilon )$ and $(\widehat{\Phi },\Lambda ),$ or related by formulas
of the type (\ref{aux2}), or (for zero torsion), of the type (\ref{aux2b}), for a
fixed value of the effective cosmological constant $\Lambda =\check{\Lambda},$
with
\begin{equation}
\check{\Lambda}\check{\Phi}^{2}=\widehat{\Phi }^{2}|~^{v}\Upsilon |+\int dt\
\widehat{\Phi }^{2}|~^{v}\Upsilon |^{\diamond }.  \label{redefgenf}
\end{equation}%
For the data $(\check{\Phi}[\widehat{\Phi },~^{v}\Upsilon ],\check{\Lambda}),$ a
metric (\ref{goffds}) is equivalent to the d--metric (\ref{qelgenofd}).%
\footnote{%
We emphasize that such locally anisotropic cosmological solutions are
generically off--diagonal (because, in general, the anholonomy coefficients $%
W_{ia}^{b},$ see (\ref{nonholr}), are non-zero). Some of the six independent
coefficients of the metric depend on all spacetime coordinates. If we fix $%
\omega ^{2}=1,$ we generate solutions with Killing symmetry on $\partial_{3}.$}

The fact that the AFDM setting allows us to integrate in this general, off--diagonal
(with Killing or non--Killing symmetries), both the gravitational field
equations of GR and of modified theories, is certainly a very important result in mathematical
relativity. This result has also fundamental physical implications
for modern standard and modified gravity theories, particle physics and cosmology.
The first one is that off--diagonal solutions of (generalized)
Einstein equations, depending generically on three or four spacetime
coordinates, can be generated by general classes of generating and
integration functions. This reflects a specific property of nonlinear and
nonholonomic off--diagonal gravitational interactions, where re--definitions
of generating functions (for instance, of type (\ref{redefgenf})) of modified
gravity theories with
sources $\widehat{\mathcal{Y}}_{\alpha \beta }$ [see (\ref{dsours1}) and (%
\ref{dsours2})], allow us to describe a large class of these modified gravity
interactions as effective Einstein spaces. The effective equations $\mathbf{%
\check{R}}_{\ \beta }^{\alpha }=\check{\Lambda}\delta _{\ \beta }^{\alpha }$
are derived for the gravitational Lagrangian (\ref{a63}) with a
correspondingly nonholonomically deformed linear connection $\mathbf{\check{D%
}}$. We have found that we can mimic certain classes of solutions of modified theories as
off--diagonal configurations in ordinary GR, and inversely, but using different
nonholonomic variables and deformed geometric/physical objects. Such
constructions cannot be realized at all if we chose from the very beginning the ubiquitous
diagonal ansatz for the metric and consider holonomic configurations. It is not
clear yet how to prove the stability of the solution for $f$--modified theories, with the
exception of some vary special cases. Nevertheless, for a very general class of
such nonlinear nonholonomic systems re--defined as  $\widehat{\mathbf{f}}$%
--theories in terms of  $\mathbf{\check{R}}_{\ \beta }^{\alpha }$ and  $%
\mathbf{\check{R}}$, the stability can be proven indeed, as for the Einstein
spacetime manifolds (see \ref{ssminst}).

Another important physical implication is that, for correspondingly fixed
data, for generating and integration functions solutions of the type $\mathbf{g}%
_{\alpha \beta }(x^{k},t)$ (\ref{goffds}) determine new classes of
nonhomogeneous cosmological metrics in GR as off-diagonal deformations of
the FLRW cosmology, in the limit $\mathbf{g}_{\alpha \beta }\rightarrow
\mathbf{g}_{\alpha \beta }(t,a(t),\widehat{h}_{3}(t),\check{\Phi}(t),$ $\eta
_{i}(t)).$ We can model accelerating cosmology and dark energy and dark
matter effects via nonlinear off--diagonal interactions and nonholonomic
constraints induced by corresponding transforms of type (\ref{redefgenf})
and modified re-scaling factor $a(t)$ as in d--metrics (\ref{odfrlwtors}),
or (\ref{qelgenofd}). The possibility to mimic modified gravities as analogous models in
GR via nonlinear transformations of generating functions and sources (\ref%
{redefgenf}) reflects a fundamental property of the class of off--diagonal solutions
of the gravitational field equations. This property holds true for solutions
with one Killing symmetry and/or non--Killing symmetries when metrics are
generically off--diagonal and depend (in 4 dimensions) on three/four spacetime
coordinates. It reflects a specific nonlinear dynamics of the gravitational
 and matter fields when off--diagonal interactions are taken into
consideration for maximally possible six degrees of freedom of the metric and with
 certain classes of nonholonomic constraints imposed. Working only
with diagonalizable, two Killing symmetries and stationary configurations,
such nonlinear gravitational physics and cosmologycal effects cannot be
encountered. A  radically ``orthodox'' interpretation of this class of
nonlinear and nonholonomic configurations and cosmological evolution
scenarios is that they may explain the bulk of accelerating cosmology data
and related dark energy and dark matter effects. It might be the case that, in order to understand the observed Universe it is not enough to modify GR in a simplistic way but, rather, to bring into consideration, within the standard GR theory, of a richer class of off--diagonal solutions and then take convenient limits leading to effective theories, in the end. This issue still need rigorous theoretical and observational consideration.

\subsection{Small off--diagonal $f$--deformations and effective FLRW like
cosmologies}

We can consider the subclass of generating functions, effective
sources and cosmological constant \newline
$(\check{\Phi}[\widehat{\Phi }%
,~^{v}\Upsilon ],\check{\Lambda})$ as in (\ref{redefgenf}), where the
cosmological evolution of spacetime regions is approximated by off--diagonal
deformations with polarization functions $\eta _{\alpha }\simeq
1+\varepsilon \chi _{\alpha }(x^{k},t)$ and the N--coefficients $n_{i}(x^{k})
$ and $w_{i}(x^{k},t)$ are proportional to a small parameter $\varepsilon $
when $0\leq \varepsilon \ll 1.$ This is motivated by the fact that although possible
anisotropic cosmological effects are very small, the modifications of the
scale factor $\mathring{a}(t)\rightarrow a(x^{k},t)$ [for a FLRW metric (\ref%
{flrw}) with for $\mathring{g}_{1}=\mathring{g}_{2}=\mathring{g}_{3}=%
\mathring{a}^{2},\mathring{g}_{4}=-1]$ can be substantial for some intervals
of time when the generating functions are of type (\ref{redefgenf}).
Prescribing any $a(x^{k},t)$ and the solution $e^{\psi {(x^{k})}}$
compatible with the observations, and fixing, for simplicity, $\omega ^{2}=1$, as
for the d--metric (\ref{qelgenofd}), when the polarization functions can be
approximated as $1+\varepsilon \chi _{i}=a^{-2}e^{\psi },$ arbitrary $\chi
_{3}(x^{k},t)$ but $\chi _{4}=1$, and the function $\widehat{h}_{3}=\eta
_{3}/a^{2},$ we obtain, up to $\varepsilon ^{2} $,
\begin{equation}
g_{\underline{\alpha }\underline{\beta }}=\left[
\begin{array}{cccc}
a^{2}(1+\varepsilon \chi _{1})+\varepsilon ^{2}[(n_{1})^{2}a^{2}-(w_{1})^{2}]
& \varepsilon ^{2}[n_{1}n_{2}a^{2}-w_{1}w_{2}] & \varepsilon n_{1}a^{2} &
\varepsilon w_{1} \\
\varepsilon ^{2}[n_{1}n_{2}a^{2}-w_{1}w_{2}] & a^{2}(1+\varepsilon \chi
_{2})+\varepsilon ^{2}[(n_{2})^{2}a^{2}\widehat{h}_{3}-(w_{2})^{2}] &
\varepsilon n_{2}a^{2} & \varepsilon w_{2} \\
\varepsilon n_{1}a^{2} & \varepsilon n_{2}a^{2} & a^{2} & 0 \\
\varepsilon w_{1} & \varepsilon w_{2} & 0 & -1%
\end{array}%
\right] .  \label{gofdse}
\end{equation}%
This class of locally anisotropic metrics is of type (\ref{goffds}), with
Killing symmetry on $\partial _{3}$ and small off--diagonal deformations on
an anisotropy parameter $\varepsilon $ which has to be fixed by experimental
data. We can consider the limit $\varepsilon \rightarrow 0$ in (\ref{gofdse}%
) but even in such cases we have an anisotropic scaling factor $a(x^{k},t)$,
or $a(t)$, which is different from a the standard $\mathring{a}(t)$ FLRW one.
This is a consequence of the nonlinear off--diagonal and nonholonomic
gravitational interactions, with generating functions and possible modified gravity
sources related by transforms (\ref{redefgenf}).

Having constructed a class of LC configurations (\ref{gofdse}), we can
extract certain subclasses of cosmological evolution scenarios with
generating and integration functions when $a(x^{k},t)\rightarrow a(t)\neq
\mathring{a}(t),w_{i}\rightarrow w_{i}(t),n_{i}\rightarrow$ const., etc. In this
way we reproduce an effective FLRW like cosmology when the scaling factor $%
a(t)$ is defined not by exotic dark matter and dark energy interactions
but by certain off--diagonal gravitational interactions which mimic
contributions of the modified gravity type. In a more general context, metrics of type (\ref%
{gofdse}) may encode certain nonholonomic torsion configurations if the
LC--constraints (\ref{lccondb}) are not imposed. \ Such solutions also
contain a small parameter $\varepsilon $ but a scaling factor $%
a(x^{k},t)\rightarrow a(t)\neq \mathring{a}(t)$ is generated by data $(\Phi
\lbrack \widehat{\Phi },~^{v}\Upsilon ],\Lambda )$ related by formula (\ref%
{aux2}) instead of (\ref{redefgenf}).

\section{Concluding remarks and discussion}

We have proven in this paper that a wide class of $f(R,...)$ modified
gravity theories can be encoded into effective off-diagonal Einstein spaces
if nonholonomic deformations and constraints are considered for the
nonlinear dynamics of gravity and matter fields. A special attention has
been paid to a new version of modified gravity theory which includes strong
coupling of the fields \cite{odgom,odgom1,odgom2}. Such modified gravity theories have
physical motivations from the covariant Ho\v{r}ava-Lifshitz like gravity
models, with dynamical breaking of the Lorentz invariance \cite{covhl,covhl1,covhl2,covhl3},
which provides also an example of a covariant, power-counting renormalizable
theory and is represented by a simplest power-law $f(R,T,R_{\alpha \beta
}T^{\alpha \beta })$ gravity.

We have demonstrated that the gravitational field equations in such modified
gravity theories admit a decoupling property with respect to certain classes
of nonholonomic frames, which allows us to generate exact solutions for very
general off-diagonal forms. The corresponding integral varieties of
solutions are parameterized by generating and integration functions and
various classes of commutative and noncommutative symmetry parameters. For
certain nonholonomic configurations, it is possible to re-define the
generating functions and effective sources of matter fields in such a way
that the $f(R)$-terms are equivalently encoded into effective Einstein
spaces with complex parametric nonlinear structure for the gravitational
vacuum. We argue that certain nonholonomic configurations model also
covariant gravity theories with nice ultraviolet behaviors and seem to be
(super-)renormalizable in the sense of Ho\v rava-Lifshitz gravity
\cite{vhl,vhl1,vhl2,vhl3,covhl,covhl1,covhl2,covhl3,voffdmgt,voffdmgt1,voffdmgt2,voffdmgt3}.

Notwithstanding the fact that the various $f(R)$ modified theories and
general relativity are actually very different theories, the off-diagonal
configurations and nonlinear parametric interactions considered in GR may
encode various classes of such modified gravity effects and explain
alternatively observational data for accelerating cosmology and certain
effects in dark energy and dark matter physics. In both cases, it is
possible to find cosmological solutions and reconstruct the corresponding
action. In the already mentioned classes of modified gravity theories with $%
f $-modifications \cite{revfmod,revfmod01,revfmod02,revfmod03,revfmod04,revfmod05,revfmod06,revfmod07,revfmod08,
revfmod09,revfmod10,revfmod11,revfmod12,revfmod13,covhl,covhl1,covhl2,covhl3,odgom,odgom1,odgom2}, the dynamics of the matter
sector is modeled by a perfect fluid. This is necessary to satisfy the
continuity equation and guarantee an evolution which is similar to that in
GR. For the alternative models with nonholonomic configurations \cite%
{vhl,vhl1,vhl2,vhl3,voffdmgt,voffdmgt1,voffdmgt2,voffdmgt3,voffds,
voffds1,voffds2,voffds3,voffds4}, the behavior of modified gravity theories is
determined by the off-diagonal terms and non-integrable constraints. In
general, we cannot distinguish the effects of $f$-modifications from the
off-diagonal ones because nonholonomic frame transforms mix different
classes of nonlinear interactions and parametric constraints. Nevertheless,
for certain well-defined parameterizations, we can work with effective $%
a(\zeta )$ and $H(\zeta )$ which, with respect to N-adapted frames and for
appropriate types of nonholonomic constraints, mimic ${\Lambda}$CDM
cosmology when the gravitational background is generically off-diagonal and
with a nontrivial gravitational vacuum structure (which may be a
nonholonomically induced torsion) with an effective cosmological constant.

Off-diagonal cosmological solutions can be described by a realistic Hubble
parameter but with an anomalous behavior for the barionic dark matter as it
is shown in the first subsection of Sect.~\ref{s4}. This kind of nonlinear
parametric evolution allows to reproduce de Sitter like universes modeled on
nonholonomic backgrounds of certain forms, encoding $f(R,T,R_{\alpha \beta
}T^{\alpha \beta })$ gravity. For a corresponding class of generating
functions, we can model nonholonomic deformations of the ${\Lambda}$CDM
universe with a standard evolution for dust matter. It is possible to
distinguish corrections with $f(T)$ and/or $f(R_{\alpha \beta }T^{\alpha
\beta })$ terms. The priority of the anholonomic frame deformation method
(AFDM) is that in such way we can generate analytical and exact formulas for
the field and cosmological evolution equations, to formulate equivalent
modeling criteria, etc.

Another priority of the AFDM is that we can study the issue of matter
instability with various classes of modified gravity theories using
geometric methods. The equations for the perturbations are complicated
fourth-order differential equations involving linear perturbations of the
Ricci scalar and tensor for different classes of linear connections.
Nevertheless, we were able to consider specific nonholonomic transforms and
constraints which allowed us to avoid matter instabilities. In this
approach, some viable effective off-diagonal Einstein models and $f(R)$
gravities could be elaborated to encode the $f(R_{\alpha \beta }T^{\alpha
\beta })$ contributions.

An effective field theory approach to off--diagonal cosmological
cosmological configurations can be elaborated in terms of St\"{u}ckelberg
fields adapted to nonlinear connection structures, see Sect.~\ref{s5}. The
constructions for  generic off--diagonal perturbations of MGTs  (in terms of
effective speed of sound, gravitational constant, Caldwell's parameter etc.)
are linked and confronted with actual observational data.

We note, finally, that modified gravity theories in general contain ghosts,
due to the higher-derivative terms in the action. However, we can select
certain ghost-free configurations determined by corresponding classes of
nonholonomic deformations or constraints. Such models of bi-metric and
massive graviton gravities were recently studied in \cite{massgr,massgr1,voffdmgt,voffdmgt1,voffdmgt2,voffdmgt3}.
Together with the results in \cite{covhl,covhl1,covhl2,covhl3,vhl,vhl1,vhl2,vhl3},
the conclusion is reached
that some $f(R,T,R_{\alpha \beta }T^{\alpha \beta })$ models, and their
off-diagonal nonholonomic equivalents, may possess nice ultraviolet
properties and that interesting connections can be established with viable
theories of quantum gravity.

\vskip5pt

\textbf{Acknowledgments: } This work has been partially supported by the
Program IDEI, PN-II-ID-PCE-2011-3-0256, by an associated visiting research
position at CERN, by a DAAD fellowship for Munich and Hannover,  by MINECO (Spain), grant PR2011-0128 and project FIS2010-15640, by the CPAN Consolider Ingenio Project, and by AGAUR
(Generalitat de Ca\-ta\-lu\-nya), contract 2009SGR-994. We thank S.
Capozziello, N. Mavromatos, S. D. Odintsov, E. Saridakis, D. Singleton, and
P. Stavrinos for important discussions and support.

\end{document}